\documentclass[twocolumn]{aastex71}

\usepackage[T1]{fontenc}
\usepackage{textcomp}

\usepackage{amsmath,color,textcomp,url,graphicx,subfigure}
\usepackage{listings}
\usepackage{booktabs}
\usepackage{ifpdf}

\newcommand{\nyoungsubstellar}{1620}

\newcommand{\nyoungucd}{455}
\newcommand{\nnewyoungucd}{196}
\newcommand{\ntotalplanemos}{101}
\newcommand{\nnewplanemos}{53}
\newcommand{\nrejectedplanemos}{64}

\newcommand{\totalassociations}{10259}
\newcommand{\bsigmaassociations}{8125}

\newcommand{\nmocatables}{147}

\newcommand{\newcandidates}{11535}

\newcommand{\nyoungexo}{134}
\newcommand{\nnewyoungexo}{46}
\newcommand{\ntessexo}{121}
\newcommand{\nnewtessexo}{53}
\newcommand{\nnewallexo}{99}

\newcommand{\nspectra}{2943} 

\newcommand{\db}{MOCAdb}

\newcommand{\kms}{\hbox{km\,s$^{-1}$}}
\newcommand{\pckms}{\hbox{pc\,km$^{-1}$\,s}}
\newcommand{\mjup}{$M_{\mathrm{Jup}}$}

\newcommand{\logrhk}{$\log R^\prime_{\rm HK}$}
\newcommand{\msol}{$M_{\odot}$}

\newcommand{\asyr}{$\mathrm{as}\,\mathrm{yr}^{-1}$}
\newcommand{\masyr}{$\mathrm{mas}\,\mathrm{yr}^{-1}$}
\newcommand{\teff}{$T_{\rm eff}$}

\newcommand{\li}{\ion{Li}{1}~$\lambda$6708\,\AA}
\newcommand{\halpha}{H$\alpha$~$\lambda$6563\,\AA}

\newcommand{\amnh}{Department of Astrophysics, American Museum of Natural History, Central Park West, New York, NY, USA}

\newcommand{\planetarium}{Plan\'etarium de Montr\'eal, Espace pour la Vie, 4801 av. Pierre-de Coubertin, Montr\'eal, Qu\'ebec, Canada}
\newcommand{\irex}{Trottier Institute for Research on Exoplanets, Universit\'e de Montr\'eal, D\'epartement de Physique, C.P.~6128 Succ. Centre-ville, Montr\'eal, QC H3C~3J7, Canada}
\newcommand{\openu}{School of Physical Sciences, The Open University, Milton Keynes, MK7 6AA, UK}
\newcommand{\uwo}{Department of Physics and Astronomy, The University of Western Ontario, 1151 Richmond St, \\London, Ontario, N6A 3K7, Canada}
\newcommand{\iese}{Institute for Earth and Space Exploration, The University of Western Ontario, 1151 Richmond St, \\London, Ontario, N6A 3K7, Canada}
\newcommand{\jpl}{Jet Propulsion Laboratory, California Institute of Technology, 4800 Oak Grove Drive, Pasadena, CA 91109, USA}
\newcommand{\columbia}{Department of Astronomy, Columbia University, 550 West 120th\ Street, New York, NY 10027, USA}

\submitjournal{ApJS}

\shorttitle{MOCA\lowercase{db}: A Census of Nearby Associations and Open Clusters}
\shortauthors{Gagn\'e et al.}

\begin{document}

\title{THE MONTREAL OPEN CLUSTERS AND ASSOCIATIONS (MOCA) DATABASE: A CENSUS OF NEARBY ASSOCIATIONS, OPEN CLUSTERS, AND YOUNG SUBSTELLAR OBJECTS WITHIN 500 \lowercase{pc} OF THE SUN}

\author[0000-0002-2592-9612]{Jonathan Gagn\'e}
\affiliation{\planetarium}
\affiliation{\irex}
\email{gagne@astro.umontreal.ca}

\author[0000-0001-7171-5538]{Leslie Moranta}
\affiliation{\planetarium}
\affiliation{\irex}
\affiliation{\amnh}
\email{leslie.moranta@umontreal.ca}

\author[0000-0001-6251-0573]{Jacqueline K. Faherty}
\affiliation{\amnh}
\email{jfaherty@amnh.org}

\author[0000-0002-2792-134X]{Jason Lee Curtis}
\affiliation{\columbia}
\affiliation{\amnh}
\email{jasoncurtis.astro@gmail.com}

\author[0000-0003-2235-761X]{Thomas P. Bickle}
\affil {\openu}
\email{tombicklecrypto@gmail.com}

\author[0000-0003-2604-3255]{Dominic Couture}
\affiliation{\planetarium}
\affiliation{\irex}
\email{couture@astro.umontreal.ca}

\author[0009-0007-4005-2218]{Am\'elie Chiasson David}
\affiliation{\planetarium}
\affiliation{\irex}
\email{amelie.chiassondavid@mail.mcgill.ca}

\author[0009-0005-9903-6752]{Katie Christie}
\affiliation{\planetarium}
\affiliation{\irex}
\email{christie.katiebeth@gmail.com}

\author[0000-0003-3219-4818]{Samantha Lambier}
\affiliation{\uwo}
\affiliation{\iese}
\email{slambier@uwo.ca}

\author[0009-0000-4260-1662]{Elise Leclerc}
\affiliation{\planetarium}
\affiliation{\irex}
\email{elise.leclerc.1@umontreal.ca}

\author[0009-0008-5861-0449]{Livia Poliquin}
\affiliation{\planetarium}
\affiliation{\irex}
\email{livpolik@gmail.com}

\author[0009-0007-6298-9802]{Danika Belzile}
\affiliation{\planetarium}
\email{danika.belzile@umontreal.ca}

\author[0000-0003-2008-1488]{Eric E. Mamajek}
\affiliation{\jpl}
\email{eric.mamajek@jpl.nasa.gov}

\begin{abstract}

We present the Montreal Open Clusters and Associations database (MOCAdb), a public MySQL database with a Python interface. MOCAdb provides a census of memberships for \totalassociations\ associations and open clusters, with a comprehensive compilation of literature measurements such as spectral types, kinematics, rotation periods, activity indices, spectral indices, and photometry. All known substellar objects are cataloged in MOCAdb, along with \nspectra\ public spectra, to enable the characterization of substellar association members. MOCAdb also features periodically updated calculations such as Galactic $UVW$ space velocities. We use this compilation to construct mappings between independent association definitions, and to update the BANYAN~$\Sigma$ membership classification tool, which now includes \bsigmaassociations\ associations. The BANYAN~$\Sigma$ model construction is improved to account for heterogeneous and correlated errors and to capture complex association shapes using Gaussian mixture models. Combined with Gaia~DR3, this enabled us to identify \newcandidates\ yet unrecognized candidate members of young associations within 500\,pc, mostly M dwarfs. Our results corroborate a recent observation that systematics up to $\approx$\,4\,\kms\ remain in Gaia~DR3 radial velocities for A-type stars. We present an updated census of age-calibrated exoplanets and substellar objects: \nyoungexo\ age-calibrated exoplanet systems (plus \ntessexo\ TESS exoplanet candidates), \nnewallexo\ of which did not appear to have known memberships so far, and \nyoungucd\ substellar (L0 or later) candidate members of young associations, \nnewyoungucd\ of which appear newly recognized. We bring the total of candidate isolated planetary-mass objects to \ntotalplanemos, \nnewplanemos\ of which are newly recognized candidate members.

\end{abstract}


\section{INTRODUCTION}\label{sec:intro}

Young stellar associations near the Sun are valuable laboratories to study stellar evolution and refine age-dating methods because they contain groups of stars spanning a wide range of masses that formed from the same molecular cloud within a short time period (e.g., \citealp{2004ARAA..42..685Z,2008hsf2.book..757T}). The associations closest to the Sun (within $\approx$\,70\,pc) are particularly valuable because their members appear brighter, but it also causes them to spread over larger areas of the sky, making their initial identification less straightforward than distant associations \citep{2022ApJ...939...94M}. Obtaining reliable lists of members with low contamination by unrelated field stars is challenging and typically requires measuring the six-dimensional position and space velocity of each member. As these stars typically formed from a single molecular cloud, they share the same velocities typically within $\lesssim$\,1\,\kms, allowing us to distinguish them from most field stars.

The study of nearby young stars recently benefited from a tremendous influx of data, in large part due to the highly successful Gaia mission \citep{2016AA...595A...1G} which measured high-precision proper motions and parallaxes for more than a billion stars and heliocentric radial velocities for a subset of 33.8 million stars as of the third data release \citep{2023AA...674A...1G}. This makes it possible to calculate the three-dimensional Galactic position and space velocities for a sample of stars much larger than was previously possible. These new data have led to the discovery of many new nearby young associations (e.g., see \citealp{2017AJ....153..257O,2019AJ....158..122K,2020AJ....160..279K,2022ApJ...939...94M,2019AA...622L..13M}), new coronae around known open clusters (e.g., see \citealp{2019AA...621L...3M,2021AA...645A..84M}), and a significant refinement in the shape and substructure of many star-forming complexes (e.g., see \citealp{2021ApJ...917...23K,2022ApJ...941...49K,2022ApJ...941..143K,2021ApJS..254...20L,2023ApJ...954..134K,2023AA...677A..59R}).

The coalescing of available data has also become a gradually more demanding task given the increasing number of available data sets. The construction of young star samples that benefit from the best available measurements is becoming ever more powerful, yet also a longer process; this situation will only become more pronounced with upcoming data releases from existing missions and imminent releases from the Vera Rubin Observatory \citep{2019ApJ...873..111I} and the EUCLID mission \citep{2011arXiv1110.3193L}.

In this paper, we present the Montreal Open Clusters and Associations (MOCA) database, whose objective is to coalesce currently available data for (1) all nearby associations within 500\,pc of the Sun and (2) all currently known nearby substellar objects. This database was built with the intended goal of facilitating the age-dating of astrophysical objects in the Solar neighborhood, and the identification of age-calibrated exoplanet systems and substellar objects down into the planetary-mass regime.

Section~\ref{sec:sample} describes the starting samples of members of young associations that were used to construct a list of stars, associations, and substellar objects. Section~\ref{sec:dbs} describes the structure of the MOCA database and its main tables. Section~\ref{sec:calcs} describes a set of calculated quantities that are included in the MOCA database and will be periodically updated as new data come in. Section~\ref{sec:discussion} presents a discussion based on the contents of the MOCA database. This includes an overview of the hierarchical relationships between nearby young associations, a refreshed version of the BANYAN~$\Sigma$ tool for membership classification based on improved models, empirical age-calibrated color magnitude diagrams, a search for new candidate members of known associations using Hipparcos and Gaia~DR3 and more. A presentation of the user-accessible MOCAdb website is presented in Section~\ref{sec:website}, and this work is concluded in Section~\ref{sec:conclusion}.

\section{SAMPLE}\label{sec:sample}

This section describes the input sample of young associations and their members that were used to build the MOCA database.

\subsection{Young Associations and open Clusters}\label{sec:associations}

We extended the initial list of members in nearby young associations, compiled in the BANYAN paper series \citep{2013ApJ...762...88M,2014ApJ...783..121G,2018ApJ...856...23G} with a series of recent young association surveys based on the Gaia mission. These include papers from the SPYGLASS series and other studies of substructures in young stellar formation regions \citep{2021ApJ...917...23K,2022ApJ...941...49K,2022ApJ...941..143K,2023ApJ...954..134K,2023AA...677A..59R,2021ApJS..254...20L}, clustering searches for new associations \citep{2017AJ....153..257O,2019AJ....158..122K,2020AJ....160..279K,2022ApJ...939...94M,2019AA...622L..13M}, recent discoveries of nearby open cluster coronae \citep{2019AA...621L...3M,2021AA...645A..84M}, and recent open cluster catalogs and surveys \citep{2005AA...440..403K,1998AJ....116.2423P,2020AA...635A..45C,2020AA...633A..99C,2022AA...664A.175P,2023AA...673A.114H,2023ApJS..265...12Q}. We have also included associations that are currently not thought to be coeval or even physical (e.g., Hercules-Lyra of \citealp{1998PASP..110.1259G}, and the Castor moving group of \citealp{1998AA...339..831B}, see \citealp{2015IAUS..314...21M} for more details) for completeness, but those are marked with a special flag \texttt{is\_real=0}.

\added{The \totalassociations\ associations} that we have considered are listed in Table~\ref{tab:associations} with some of their basic properties, and are stored in the database table \texttt{moca\_associations} as described in Section~\ref{sec:dbs}. 

\subsection{Brown Dwarfs}\label{sec:bds}

We included all objects from the Ultracoolsheet \citep{Best20US}\footnote{\added{A persistent and versioned copy of the Ultracoolsheet is available on Zenodo at \href{https://doi.org/10.5281/zenodo.4169084}{doi:10.5281/zenodo.4169084}}} in the database, with their spectral types, memberships to young associations when available, and their kinematics in the \db. The Ultracoolsheet systematically includes all ultracool dwarfs (spectral types L0 and later) at distances up to $\approx$\,100\,pc. Although most of the brown dwarfs in this list show no signs of youth or known memberships in young associations, we decided to include them all to facilitate future searches for brown dwarfs of intermediate ages ($\approx 200$, Myr and older) and cold, young brown dwarfs (spectral types T0 and later, ie, $\approx 1300$, K and cooler) for which the signatures of low surface gravity (a consequence of youth) are not immediately obvious from low resolution spectroscopy.

\section{DATABASE STRUCTURE}\label{sec:dbs}

The MOCA database compiles not only young associations, their members, and substellar objects, but also a number of literature properties of individual stars, as well as calculations that are performed using a set of custom \texttt{IDL} and \texttt{Python} libraries which automatically update measurements when new data becomes available.

The MOCA database currently includes \nmocatables tables, divided into six categories, from which the table name prefixes are determined.
\begin{itemize}
    \item \texttt{moca\_}: Tables containing unique identifiers used across the MOCA database along with their basic information (e.g., \texttt{moca\_associations}),
    \item \texttt{cat\_}: Tables containing subsets of astronomical catalogs, usually imported from an external database, linked with a MOCA object identifier (e.g., \texttt{cat\_gaiadr3}),
    \item \texttt{data\_}: Tables containing raw data, usually imported from the scientific literature or a \texttt{cat\_} table (e.g., \texttt{data\_radial\_velocities}),
    \item \texttt{calc\_}: Tables containing calculations performed as part of the MOCAdb infrastructure (e.g., \texttt{calc\_radial\_velocities\_combined}),
    \item \texttt{mechanics\_}: Other tables containing quantities necessary to the functioning of MOCAdb features (e.g., \texttt{mechanics\_memberships\allowbreak\_propagated}),
    \item \texttt{summary\_}: Summary tables \added{(or views)} allowing users to quickly access a number of useful quantities (e.g., \texttt{summary\_all\_members}). \added{Some summary tables are precomputed for efficiency.}
\end{itemize}

Most of the foreign keys connecting the MOCAdb tables have column names starting with \texttt{moca\_}, to indicate a database-wide identifier. Some of these identifiers are listed below:

\begin{itemize}
    \item \texttt{moca\_oid}: Unique identifier for a single star or substellar object,
    \item \texttt{moca\_aid}: Unique identifier for a young association,
    \item \texttt{moca\_pid}: Unique identifier for a publication's bibliographic entry,
    \item \texttt{moca\_psid}: Unique identifier for a photometric system,
\end{itemize}

The \texttt{moca\_aid} keys are defined with a unique string identifier that is only composed of alphanumeric characters and underscore symbols (e.g., the $\beta$~Pic Moving Group is linked with the keyword \texttt{BPMG}).

All MOCAdb tables are listed in Table~\ref{tab:mocadb_tables} with their total number of columns and approximate numbers of rows at the time of publication, and detailed column descriptions can be found at \url{https://mocadb.ca/schema}.

\begin{figure*}
 	\centering
 	\includegraphics[width=0.965\textwidth]{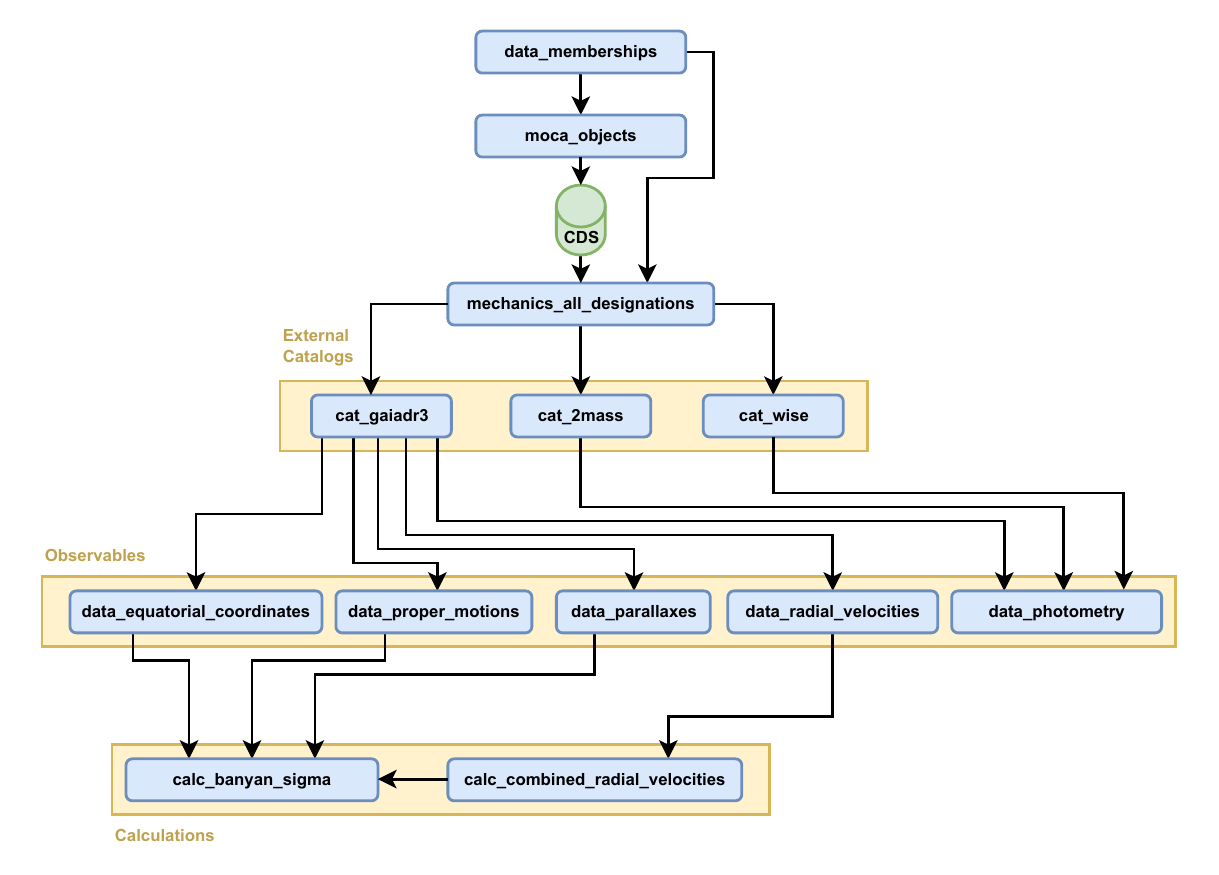}
 	\caption{Flow chart of update algorithm in MOCAdb. Black arrows represent dependencies for calculations or data transfer. Astrophysical objects are mostly inserted into MOCAdb starting from the \texttt{data\_memberships} table using the designations from the original paper. They are then cross-matched using CDS to compile all of their designations -- the main designation is added to the \texttt{moca\_objects} table with a new unique \texttt{moca\_oid} integer identifier assigned by the database. These \texttt{moca\_oid} identifiers are propagated back to the \texttt{data\_memberships} table, and all alternate designations are ported to the table \texttt{mechanics\_all\_designations}. All designations relevant to one of the known online databases (such as Gaia~DR3) are then used to download the relevant rows from the corresponding database into the corresponding \texttt{cat\_} table, and MySQL procedures then port several observable quantities (such as radial velocities and proper motions) in the corresponding \texttt{data\_} table in MOCAdb. These newly available observables trigger changes in the MD5 hashes used for calculations, and trigger a recalculation of the relevant properties for the \texttt{moca\_oid} objects which gained new data. See Section~\ref{sec:dbs} for more detail.}
 	\label{fig:db_schema}
\end{figure*}

A simplified flow chart that presents the mechanisms by which the MOCAdb tables are updated is shown in Figure~\ref{fig:db_schema}. Membership lists are gathered from literature papers, from which unique \texttt{moca\_oid} stars are imported, and cross-matched with a series of astronomical databases as described in Section~\ref{sec:catalog_data}. A set of various raw measurements are then derived from these astronomical databases (and from scientific papers directly), and form the base for various calculations performed in MOCAdb as described in Section~\ref{sec:discussion}.

Each table starting with the \texttt{data\_} prefix includes a column \texttt{object\_designation}, containing the identifier for the star listed in the original science paper or astronomical database where the data were obtained from. These unique identifiers are matched to a unique \texttt{moca\_oid} identifier in the \texttt{mechanics\allowbreak\_all\_designations} table, which contains a list of all known identifiers for each unique star. All known designations are downloaded from the CDS service \citep{2000AAS..143....9W} for stars with known CDS identifiers, and those additional designations are included in the \texttt{mechanics\_\allowbreak all\_designations} table. After all rows of the \texttt{data\_memberships} table have been connected to a \texttt{moca\_oid} identifier, those without a match in the \texttt{mechanics\_\allowbreak all\_designations} table are added to the database with Simbad or the relevant database matching their designation (e.g., Gaia~DR3) along all of their CDS identifiers.

\subsection{Cross-Matching and Catalog Data}\label{sec:catalog_data}

All stars listed in the \texttt{moca\_objects} table were automatically cross-matched to a set of \added{10} external astronomical databases using a set of custom libraries: \added{Simbad, Gaia DR1, DR2 and DR3, Hipparcos, the The GALactic Archaeology with HERMES (GALAH) DR3, the Apache Point Observatory Galactic Evolution Experiment (APOGEE) DR11, the Radial Velocity Experiment (RAVE) DR6, ROSAT, and GALEX as described below, unless a designation specific to the catalog was already listed in Simbad, in which case that specific match was assigned. Catalog data were also downloaded for any source with a 2MASS, WISE, AllWISE, or CatWISE designation in Simbad.}

\added{All sources without a Gaia DR3 entry (mostly substellar objects) and without a corresponding existing Simbad match were cross-matched with an additional set of 23 visible, red, or near-infrared catalogs: 2MASS, WISE, AllWISE, CatWISE, the Dark Energy Survey (DES) DR2, the Noirlab Souce Catalog (NSC) DR2, Pan-STARRS1 DR2, the Sloan Digital Sky Survey (SDSS) DR17, the UKIRT Hemisphere Survey (UHS) DR3, the VISTA Hemisphere Survey (VHS) DR6, the VISTA survey of the Magellanic Clouds system (VMC) DR5, VISTA Variables in The Via Lactea (VVV) DR5, the VVV Infrared Astrometric Catalogue (VIRAC) DR4, the SkyMapper Southern Survey (SMSS) DR4, the Southern Photometric Local Universe Survey (S-PLUS) DR4, DENIS DR3, the DECam Plane Survey (DECAPS) DR2, the VISTA Kilo-degree Infrared Galaxy Survey (VIKING) DR4, the VISTA Deep Extragalactic Observations (VIDEO) DR5, and four DR11PLUS UKIRT InfraRed Deep Sky Surveys (UKIDSS; Deep Extragalactic Survey, DXS; Large Area Survey, LAS; Galactic Clusters Survey, GCS; and Galactic Plane Survey, GPS). All catalogs are listed in Table~\ref{tab:missions} with their references.}

When a Gaia identifier is available, the ARI Gaia services\footnote{Available at \url{https://gaia.ari.uni-heidelberg.de/index.html}} are used to cross-match 2MASS \citep{2006AJ....131.1163S} or AllWISE \citep{2014ApJ...783..122K} designations with ARI's \texttt{gaiadr2.\allowbreak tmass\_best\_neighbour}, \texttt{gaiadr2.\allowbreak allwise\_best\_neighbour}, \texttt{gaiadr3.\allowbreak tmass\_psc\allowbreak\_xsc\_best\_neighbour}, \texttt{gaiaedr3.\allowbreak allwise\_best\_neighbour}, and \texttt{gaiaedr3.\allowbreak dr2\_neighbourhood} tables.

When no matches are found, cross-matches that account for the most accurate Gaia proper motion (when available) are made\footnote{The nearest entries within $\approx$\,100\,pc were vetted using the \texttt{finder\_charts} library available at \url{https://github.com/jgagneastro/finder\_charts}.}. Once all combinations of cross-matches have been exhausted, another SIMBAD query is performed with each of the available identifiers, and any resulting match is used to gather additional related identifiers. Data from the SIMBAD database, such as coordinates, spectral types, kinematics, photometry and object types are also automatically downloaded and included in the \texttt{cat\_simbad} database table.

Individual designations that were identified for the 2MASS, WISE, AllWISE, CatWISE, Gaia~DR1, Gaia~DR2, Gaia~DR3 \citep{2006AJ....131.1163S,2010AJ....140.1868W,2014ApJ...783..122K,2021ApJS..253....8M,2016AA...595A...2G,2018AA...616A...1G,2023AA...674A...1G} catalogs are used to automatically query the relevant entries of these catalogs, using the IRSA\footnote{Available at \url{https://irsa.ipac.caltech.edu}.} and ARI\footnote{Available at \url{https://gaia.ari.uni-heidelberg.de/tap.html}.} servers, and are included to the MOCAdb tables \texttt{cat\_2mass}, \texttt{cat\_wise}, \texttt{cat\_allwise}, \texttt{cat\_catwise}, \texttt{allwise}, \texttt{gaiadr1}, \texttt{gaiadr2}, and \texttt{gaiaedr3}, respectively. The Hipparcos table of \cite{2007AA...474..653V} was included in its entirety in the \texttt{cat\_hipparcos} table due to its modest size. Data from the ROSAT all-sky survey \citep{1999AA...349..389V} was included in the \texttt{cat\_rosat} table with automatic queries of the ViZieR tables \texttt{IX/29/rass\_fsc}, \texttt{IX/10A/1rxs}, and \texttt{IX/10A/1rxs\_cor}, with a \added{15$"$ cross-match radius to account for the low astrometric precision of ROSAT. The best-available proper motions were used to project the reference MOCAdb coordinate to the average ROSAT epoch (1995 $\pm$ 5\,yr) to perform this cross-match, and the resulting astrometric uncertainty in the expected coordinates were added in quadrature to the base 15$"$ cross-match radius.}

A manual cross-match was performed with the \texttt{GALEXDR6PlusDR7} table \citep{2014AdSpR..53..900B} hosted at the Mikulski Archive for Space Telescopes (MAST)\footnote{Available at \url{https://archive.stsci.edu/}.}, using the best available Gaia proper motion to the average GALEX epoch (2005.5) with a 30$"$ cross-match radius: this accounts for both the limited astrometric precision of GALEX (5$"$), and a further 25$"$ for the range of GALEX epochs (2003--2008) with proper motions as large as 8.3\,\asyr. Once all possible matches have been identified with a database source star, only those that match within the uncertainty of proper motions and GALEX epoch (2.5\,yr) were identified, and only the best-matching (nearest) database sources to a given GALEX entry were preserved in cases where a single GALEX entry was a potential match to more than one MOCA database entry.

\added{All sources without Gaia counterparts} were also automatically cross-matched with a set of infrared or red-optical catalogs, listed in Table~\ref{tab:missions}. Mis-matches can occasionally occur due to source confusion especially in crowded fields near the Galactic plane; these are identified based on color outliers and visual checks, and are manually corrected as they are encountered.

\subsection{Companions and Co-Moving Stars}\label{sec:companions}

A table \texttt{moca\_companions} is used to keep track of all \texttt{moca\_oid} identifiers that are related to each other gravitationally. This includes co-moving or companions stars, substellar objects, or exoplanets that have a distinct MOCAdb entry. Typically, only those objects with a resolved \added{spectroscopic or photometric spectral types, or separate entries in the Gaia, Pan-STARRS or WISE catalogs, have a} separate entry in the MOCAdb to simplify the treatment of unresolved objects. Unlike in the SIMBAD database, no unique \texttt{moca\_oid} identifiers are assigned to a system of objects. For example, the TW~Hya association member TWA~13~AB has three distinct entries in Simbad: one for TWA~13~AB as a system (CD--34~7390), one for CD--34~7390~A, and one for CD--34~7390~B. In MOCAdb, only two entries correspond to this system: TWA~13~A (\texttt{moca\_oid}=7254) and TWA~13~B (\texttt{moca\_oid}=7255), with a hierarchical link from TWA~13~B to TWA~13~A in the \texttt{moca\_companions} table. We assign the entry of any catalog (such as 2MASS) to the primary star when the companion is not resolved. This approach simplifies the treatment of binary stars, while still allowing to determine which stars are resolved or not based on the spatial resolution of individual catalog and projected angular separations between the components. More complex hierarchical cases are managed by establishing a link between any companion to the primary star of a given center of mass of a hierarchical link; for example, a hierarchical link points TWA~4~Bb to TWA~4~Ba, and another link points from TWA~4~Ba to TWA~4~Aa in \texttt{moca\_companions}. A further advantage of the simple approach adopted here is that a new entry in the \texttt{moca\_objects} table can be added for any newly identified companion, along with its relationship in the \texttt{moca\_companions} table, without having to change the entry of the parent object.

\subsection{Coordinates, Proper Motions and Parallaxes}\label{sec:crd_pm_plx}

All available coordinates were assembled from the various catalogs described in Section~\ref{sec:catalog_data} and ported to the \texttt{data\_equatorial\_coordinates} table along with epochs and measurement errors, when available. \added{The one row per \texttt{moca\_oid} with the most useful reference epoch, with a known \texttt{measurement\_epoch\_yr} when possible, and including the proper motion (\texttt{pm\_corrected}$=1$) and parallax motion (\texttt{plx\_corrected}$=1$) when possible, is flagged with \texttt{adopt\_as\_reference}$=1$. This is the row that should be preferred when calculating projected coordinates at a future epoch, for example. Galactic and ecliptic coordinates are also available in two MySQL views (\texttt{calc\_ecliptic\_coordinates} and \texttt{calc\_galactic\_coordinates}), and are computed on-the-fly only for the reference equatorial coordinates.}

\added{The proper motions} and parallaxes were compiled in the respective tables \texttt{data\_proper\_motions} and \texttt{data\_parallaxes} along with their measurement errors and epochs. The \texttt{adopted} and \texttt{ignored} flags allow the \db\ maintenance libraries to either choose the single best measurement for a source when desired (\texttt{adopted=1}), or choose all good-quality measurements (\texttt{ignored=0}) when multi-epochs measurements are required\footnote{\added{These tasks are recalculated every few minutes when required, using MySQL events.}}.

We have also included proper motions and parallaxes from a wide range of literature publications in the two respective tables, in addition to those obtained directly from the catalog tables.

\subsection{Radial Velocities}\label{sec:rv}

Individual radial velocity measurements were compiled in the \texttt{data\_radial\_velocities} table \added{from 916 distinct literature catalogs. 92\% of the total measurements came from 9 catalogs \citep{2023AA...674A...1G,2020AA...636A..74T,2018AA...616A...1G,2020AJ....160..120J,2018AA...616A...7S,2007AN....328..889K,2006AstL...32..759G,2018ApJS..235...42A,2020AJ....160...82S}.}

We avoided including any radial velocity measurements that are predictions based on kinematic models (e.g., \citealp{2002AA...381..446M,malo_banyan_2014,2007AstL...33..571B}) or combinations of other sources in the literature. We included a column \texttt{n\_measurements} to keep track of radial velocity measurements that arise from a combination of several raw measurements, such as is the case with Gaia~DR3 radial velocities.

There were a few catalogs that did not provide radial velocities on an absolute frame-of-reference, but only relative radial velocities with the explicit goal of identifying exoplanet systems. These relative radial velocities were included in the \texttt{data\_relative\_radial\_velocities} table for completeness, and were used to estimate the standard deviations of the radial velocity of individual sources in the \texttt{data\_relative\_radial\_velocities\_std} table.

\subsection{Projected Rotational Velocities}\label{sec:vsini}

Projected rotational velocities $v \sin i$ were collected in the MOCAdb table \texttt{data\_vsini} from spectroscopic measurements in the literature. The Gaia~DR3 release \citep{2023AA...674A...1G,2023AA...674A..26C} also provides two measurements that are related to projected rotational velocities, both of which have specific regimes where they are useful, and are subject to distinct biases. We describe in how we have included them in MOCAdb in the remainder of this section.

\subsubsection{Projected Rotational Velocity Biases in Gaia DR3}\label{sec:vsinibias}

\begin{figure*}
	\centering
	\subfigure[Gaia~DR3 \texttt{vbroad} versus literature $v\sin i$]{\includegraphics[width=0.49\textwidth]{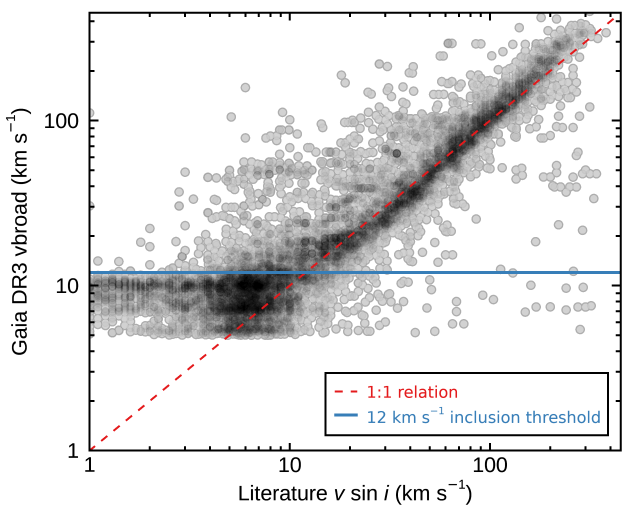}\label{fig:vbroad_lit}}
	\subfigure[Gaia~DR3 \texttt{astrophysical\_parameters.vsini\_esphs} versus literature $v\sin i$]{\includegraphics[width=0.49\textwidth]{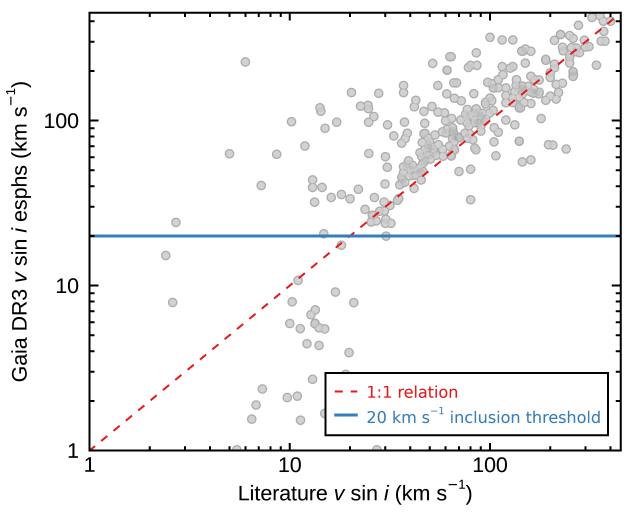}\label{fig:vsini_esphs_lit}}
	\caption{Gaia DR3 \texttt{vbroad} and $v\sin i$ measurements based on two distinct pipelines, compared with literature $v\sin i$ measurements. The Gaia~DR3 \texttt{vbroad} measurements are representative of literature values above 12\,\kms\, and the Gaia~DR3 $v\sin i$ measurements are representative of literature values above 20\,\kms. We have therefore chosen these respective thresholds for their inclusion in MOCAdb as a valid value for $v\sin i$. See Section~\ref{sec:vsinibias} for more details.}
	\label{fig:gaiadr3_vs_lit_vsini}
\end{figure*}

\begin{figure*}
	\centering
	\subfigure[Standardized \texttt{vbroad} residuals]{\includegraphics[width=0.49\textwidth]{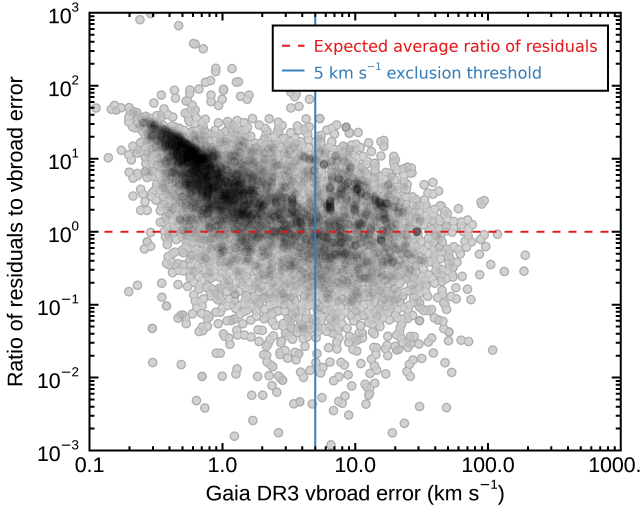}\label{fig:vbroad_res}}
    \subfigure[Standardized \texttt{vbroad} residuals after error inflation]{\includegraphics[width=0.49\textwidth]{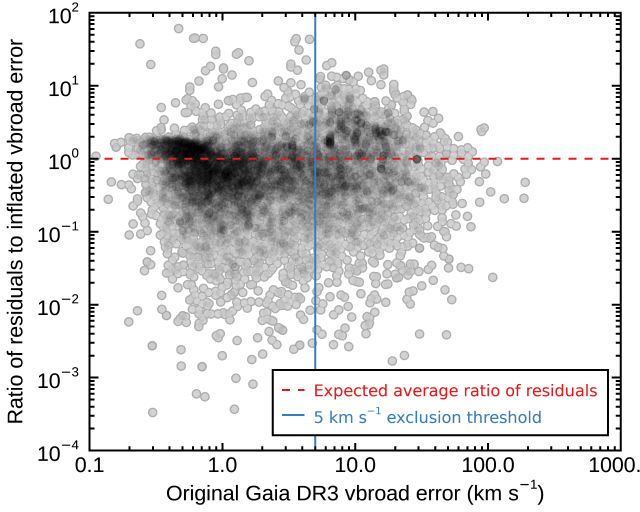}\label{fig:vbroad_res_corr}}
	\subfigure[standardized \texttt{vsini\_esphs} residuals]{\includegraphics[width=0.49\textwidth]{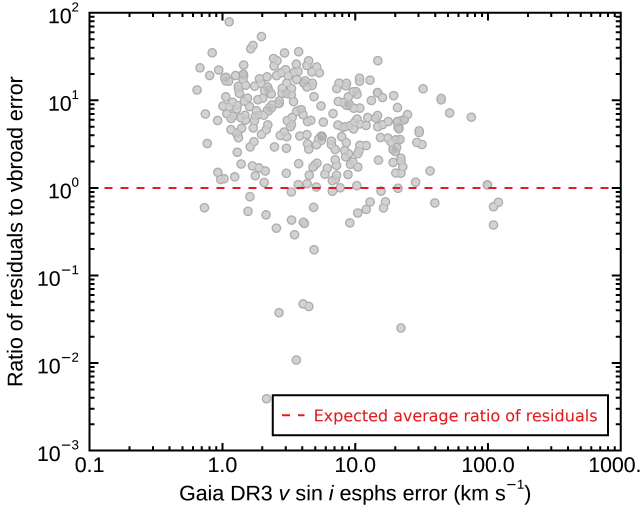}\label{fig:vsini_esphs_res}}
    \subfigure[standardized \texttt{vsini\_esphs} residuals after error inflation]{\includegraphics[width=0.49\textwidth]{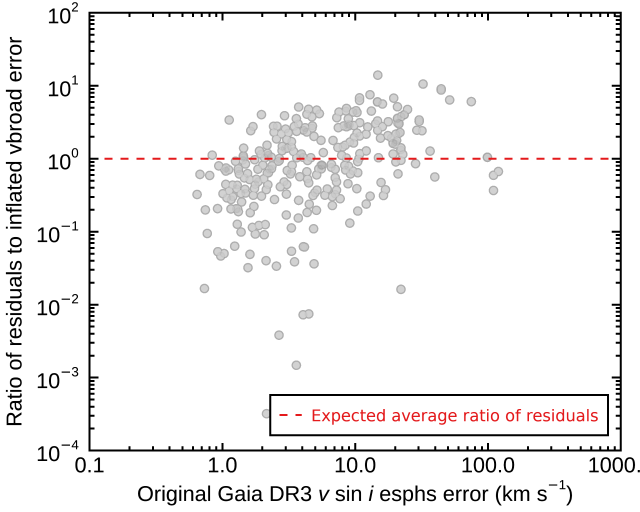}\label{fig:vsini_esphs_res_corr}}
	\caption{Upper panels: Gaia~DR3 \texttt{vbroad} errors compared with the error-normalized residuals for stars with a literature $v\sin i$ measurement (left). If the measurement errors of the Gaia~DR3 \texttt{vbroad} and the literature $v\sin i$ are realistic, normalized residuals should cluster around 1.0 (horizontal, red dashed line), which is not the case, indicating that an extra error term should be added in quadrature to the Gaia~DR3 \texttt{vbroad} errors. Once an extra error is added (5\,\kms\ for stars with \teff\ in the range 4,000--7,000\,K, or 14.5\,\kms\ for other stars), the normalized residuals (right) behave as expected. The selection threshold based on the raw Gaia~DR3 \texttt{vbroad} errors (5\,\kms) for MOCAdb inclusion is also displayed with a vertical blue line. Bottom panels: idem but for the Gaia~DR3 $v\sin i$ measurements determined by the Gaia~DR3 Apsis pipeline, which warranted the inclusion of a 26\,\kms\ error term in quadrature. See Section~\ref{sec:vsinibias} for more details.}
	\label{fig:gaiadr3_vsini_res}
\end{figure*}

 Stars observed by Gaia with $G_{\rm RVS}$ magnitudes brighter than 16.2 \citep{2023AA...674A...5K} have their 846--870\,nm Radial Velocity Spectrometer (RVS) spectra analyzed to determine their radial velocity. This is done by matching a Doppler-shifted synthetic spectrum \citep{2014AA...562A..97D,2018AA...616A...6S} broadened by the instrumental spectroscopic line spread function as well as an additional free line-broadening parameter  \texttt{vbroad} that is modelled with a traditional rotation kernel. This parameter \texttt{gaiadr3.vbroad}\footnote{See  \url{https://gea.esac.esa.int/archive/documentation/GDR3/Gaia_archive/chap_datamodel/sec_dm_main_source_catalogue/ssec_dm_gaia_source.html\#gaia_source-vbroad}.} is stored in the main Gaia~DR3 source catalog, and it is named $v_{\rm broad}$ rather than $v\sin i$ to outline the fact that it accounts not only for stellar rotation, but also other phenomena such as macro-turbulence or pulsations. It is worth noting, however, that this is also true of most other measurements and catalogs in the literature which are referred to as $v\sin i$, and hence Gaia~DR3's $v_{\rm broad}$ should be directly comparable to those, after consideration of systematics and measurement errors. Our comparison of Gaia~DR3's $v_{\rm broad}$ measurements to literature $v\sin i$ show reasonable agreement (see Figure~\ref{fig:vbroad_lit}). \citep{2023AA...674A...8F} performed a similar comparison with various $v\sin i$ catalogs and found that the \texttt{gaiadr3.vbroad} values tend to be over-estimated for $v\sin i < 10$\,\kms\ and that the measurement quality degraded for stars with \teff$ > 7500$\,K (spectral types earlier than approximately A8) and $G_{\rm RVS}$ magnitudes fainter than 10. We observe a similar trend, and find that applying a selection threshold of \texttt{gaiadr3.vbroad}$ \geq 12$\,\kms\ is sufficient to exclude these systematically over-estimated Gaia~DR3 measurements. We also observe a standard deviation larger than what is expected if we only consider the reported errors in the Gaia~DR3 main source catalog (\texttt{gaiadr3.vbroad\_error}). We find that an extra error term must be added in quadrature in order for the deviations to be comparable to the expected errors: for stars with effective temperatures (\texttt{gaiadr3.rv\_template\_teff}) in the range 4,000--7,000\,K, an additional 5\,\kms\ error is sufficient, but for stars both cooler and hotter, we find that an additional error term of 14.5\,\kms\ is required. Additionnally, we observe a noticeable increase of outliers above the expected error bars (even when inflated) for values of \texttt{gaiadr3.vbroad} with a measurement error larger than 5\,\kms\ (see Figure~\ref{fig:gaiadr3_vsini_res}). For these reasons, we chose to only include the measurements with \texttt{gaiadr3.vbroad}\ $\geq 12$\,\kms\ and \texttt{gaiadr3.vbroad\_error}\ $\leq 5$\,\kms\ in MOCAdb, and we added the extra error terms described above in quadrature.

In addition to the \texttt{gaiadr3.vbroad} measurement, another estimation of a star's $v\sin i$ is performed in the Gaia~DR3 Apsis pipeline \citep{2023AA...674A..26C}, and reported in \texttt{astrophysical\_parameters.vsini\_esphs} \citep{2023AA...674A..28F} \footnote{See  \url{https://gea.esac.esa.int/archive/documentation/GDR3/Gaia_archive/chap_datamodel/sec_dm_astrophysical_parameter_tables/ssec_dm_astrophysical_parameters.html\#astrophysical_parameters-vsini_esphs}.} for early-type stars with \teff$ > 7500$\,K. While these values are measured with a method similar to \texttt{gaiadr3.vbroad}, the measurements are performed by different pipeline which rely on a joint analysis of the BP/RP low-resolution spectra and the RVS spectra to determine the astrophysical parameters of early-type stars, therefore the resulting $v\sin i$ measurements should benefit from more carefully selected synthetic spectra that are specific to early-type stars and consistent with the low-resolution BP/RP spectra. A similar comparison of these \texttt{astrophysical\_parameters.vsini\_esphs} values with other projected rotational velocities cataloged in the literature show reasonable correlation for \texttt{astrophysical\_parameters.vsini\_esphs} values above 20\,\kms\ (see Figure~\ref{fig:vsini_esphs_lit}), which we select are our inclusion threshold. Figure~\ref{fig:gaiadr3_vsini_res} shows that an extra error term of 26\,\kms\ needs to be added in quadrature to the Gaia~DR3 \texttt{astrophysical\_parameters.vsini\_esphs} values to obtain reasonable residuals, which we included when importing the measurements in MOCAdb.

\subsection{Spectra}\label{sec:spectra}

The MOCAdb currently contains \nspectra\ non-duplicated spectra, mostly from the SpeX Prism Library\footnote{Available at \url{https://cass.ucsd.edu/~ajb/\allowbreak browndwarfs/spexprism}.}, the IRTF Spectral Library\footnote{Available at \url{https://irtfweb.ifa.hawaii.edu/~spex/IRTF_Spectral_Library/}.} \citep{2003PASP..115..362R}, the Montreal Spectral Library\footnote{Available at \url{https://jgagneastro.com/the-montreal-spectral-library/}.} and the Ultracool RIZzo spectral library\footnote{Available at \url{https://jgagneastro.com/the-ultracool-rizzo-spectral-library/}.}. Each unique spectrum is described in a row of the \texttt{moca\_spectra} table, with its header properties and a unique \texttt{moca\_specid} identifier. The data associated with these spectra (wavelengths, spectral flux densities and measurement errors) are stored in the \texttt{data\_spectra} table, which is linked through the \texttt{moca\_specid} identifier. The spectra can be visualized and downloaded at \url{https://dataviz.mocadb.ca/spectra}, and were used to calculate secondary quantities described in Section~\ref{sec:calcew}.

\subsection{Spectral Indices and Equivalent Widths}\label{sec:ew}

Measurements of spectral equivalent widths of the \li\ and \halpha\ lines are listed in the \texttt{data\_equivalent\_widths} table, and spectral indices (\logrhk\ and Mount Wilson $S$-index, mostly from \citealt{2020MNRAS.495.1252Z,2010ApJ...725..875I,2007AJ....133.2524W}) are listed in the \texttt{data\_spectral\_indices} table. Measurements currently listed in both tables are compilations from the literature.

\section{CALCULATION OF DERIVED QUANTITIES}\label{sec:calcs}

This section describes a number of parameters that are derived as part of the MOCAdb maintenance routines. In all MOCAdb calculations, the basic ingredients required for the calculation are combined into an MD5 hash that is stored in a column named \texttt{md5\_uid}, which allows to only refresh the calculations when the data have changed for a specific object and quantity. Custom IDL and Python routines are regularly started to refresh all of the calculated quantities, along with the associated time stamps (\texttt{created\_timestamp} or \texttt{modified\_timestamp}) associated with the calculations\footnote{\added{Specific calculation tasks are launched continuously on different hours and days of the week from an external computer, and simpler MySQL-only tasks, such as adopting the best calculation, are performed every few minutes with MySQL events}}. These operations are also cataloged in the \texttt{moca\_changelog} table.

\subsection{Distances}\label{sec:dist}

We used the method of \cite{2021AJ....161..147B} to estimate the distances of every source in MOCAdb with a parallax measurement available in the \texttt{data\_parallaxes} table. We elected to use the geometric prior of \cite{2021AJ....161..147B} in all cases, because a significant fraction of MOCAdb stars are young, which would complicate the use of photometry. A $10^4$-element Monte Carlo simulation is performed for every star to propagate the parallax measurement error onto the distance, and the 1- to 3-$\sigma$ measurement errors are listed along with the best estimate in the \texttt{data\_distances} table. A similar method was applied to all available parallaxes in MOCAdb, including the Hipparcos measurements and other literature sources.

The \texttt{data\_distances} table can also accept photometric distance estimates from the literature, indicated with the flag \texttt{photometric\_estimate}$=1$. For every \texttt{moca\_oid} object in the \texttt{data\_distances} table, the single best value is designated with the flag \texttt{adopted}$=1$. Trigonometric distances are systematically preferred over the photometric ones regardless of measurement errors; the Gaia~DR3 values are preferred over Gaia~DR2 distances, and otherwise distances with the smallest measurement errors are selected.

\added{A MySQL view \texttt{calc\_vtan} allows to query the tangential velocities (obtained by multiplying the distances with proper motions) of every object in MOCAdb, using their respective adopted values. These are computed on-the-fly when queried.}

\subsection{Galactic Coordinates}\label{sec:xyz}

\begin{figure*}
 	\centering
 	\subfigure[View from above the Galactic plane]{\includegraphics[width=0.49\textwidth]{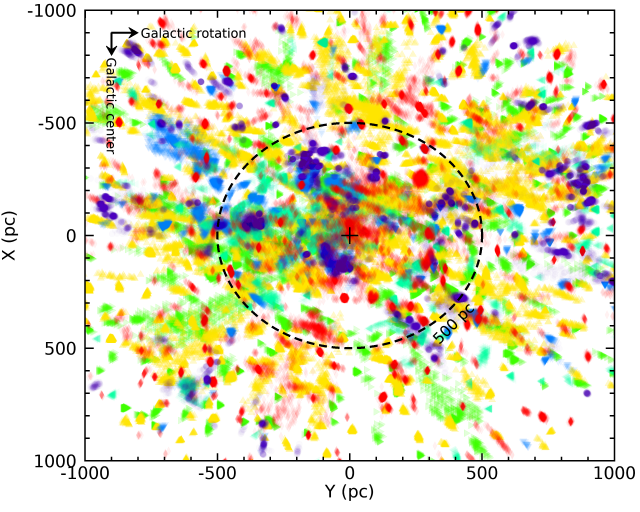}\label{fig:xy_stars_500}}
 	\subfigure[Cross-sectional view of the Galactic plane]{\includegraphics[width=0.49\textwidth]{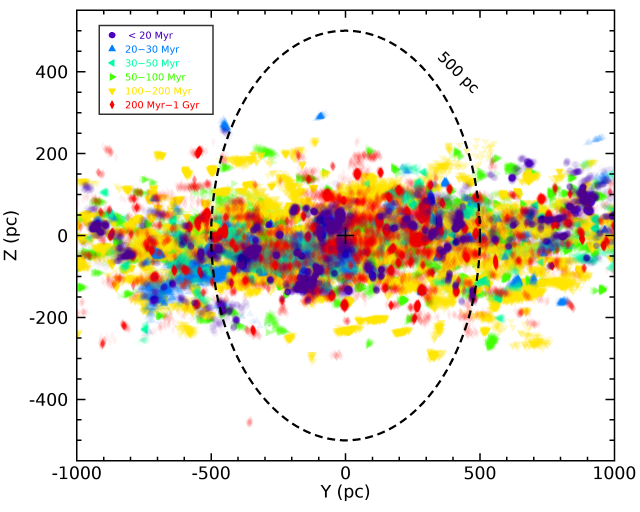}\label{fig:yz_stars_500}}
 	\caption{Galactic positions of individual stars in MOCAdb, color-coded by their association age. This figure shows how different age populations are not homogeneously distributed spatially, and how the nearby youngest stars are distributed along a plane that is slanted with respect to $Z = 0$. The MOCAdb census is mostly complete at distances up to 500\,pc, marked within the dashed line.}
 	\label{fig:xy_stars}
 \end{figure*}

The $XYZ$ Galactic coordinates of all objects with a distance in the \texttt{data\_distances} table were calculated and are listed in the table \texttt{calc\_xyz}. They are based on a $10^4$-element Monte Carlo simulation for error propagation, using the full probability distribution function of the distance estimates (which are in most cases based on the method of \cite{2021AJ....161..147B} as described in Section~\ref{sec:dist}). We adopt the definition of $XYZ$ axes used in the \texttt{IDL} \texttt{astrolib} package, which forms a right-handed rectangular coordinates system centered on the Sun, with $X$ pointing towards the Galactic Center. \added{The complete map of $XYZ$ coordinates for individual stars in MOCAdb is displayed in Figure~\ref{fig:xy_stars}.}

\subsection{Extinction}\label{sec:extinction}

The method described by \cite{2020ApJ...903...96G} was used to estimate the extinction towards each source in the database. A numerical integration of the 3D extinction map \emph{STructuring by Inversion of the Local InterStellar Medium} (STILISM; \citealp{2014AA...561A..91L,2017AA...606A..65C,2018AA...616A.132L})\footnote{Available at \url{https://stilism.obspm.fr}} towards each star is performed, based on the best-available coordinates and parallax measurement (inverted using the method of \cite{2021AJ....161..147B} with a geometric prior). A Monte Carlo simulation, where $10^4$ synthetic sets of equatorial coordinates and trigonometric distances are taken according to the best available measurements (the standard deviation of the Monte Carlo draws are set to the measurement errors). The corresponding $XYZ$ Galactic coordinates are then calculated and used to integrate the STILISM map to obtain Monte Carlo realizations of interstellar extinction, along with the asymmetrical uncertainties provided by the STILISM map for each realization. The standard deviation of the Monte Carlo realizations was added in quadrature to the average asymmetrical STILISM uncertainties to obtain a final measurement error on the extinction, and the numerical integration of the STILISM map toward the most likely Monte Carlo distance was taken as the center of the extinction measurement.

\subsection{Photometry}\label{sec:phot}

The individual photometric measurements of all surveys described in Section~\ref{sec:catalog_data} were compiled in the \texttt{data\_photometry} table, along with their epochs and measurement errors when available. The method of \cite{2020ApJ...903...96G} was used to correct the photometric reddening due to interstellar extinction on Gaia~DR3 photometry. This method consists of numerically integrating the instrumental bandpass of the corresponding photometric band with a \cite{1999PASP..111...63F} extinction curve anchored on the STILISM-based extinction measurements described in Section~\ref{sec:extinction}, weighed by the spectral energy distribution of the corresponding spectral type in the Pickles library \citep{1998PASP..110..863P} corresponding to the spectral type of the star\footnote{Consequently, extinction corrections were applied only for spectral types in the range O5--M6, and were not applied for giants, white dwarfs or substellar objects.}. Stars without known spectral types were corrected for extinction using the iterative method described by \cite{2020ApJ...903...96G}. Each iteration of this method consists in estimating a spectral type based on the Gaia colors, using it to estimate extinction, using the corrected photometry to estimate a new spectral type, and repeating until the estimated spectral type converges\footnote{Giants and white dwarfs were excluded from these calculations based on the position of the star in the Gaia~DR3 color-magnitude diagram.}. The \texttt{extinction\_corrected} flag is set to 1 in the \texttt{data\_photometry} table when the extinction was corrected successfully. \added{The Gaia~DR3 raw photometry that is not corrected for extinction is available in the \texttt{cat\_gaiadr3} table directly, and we avoided also storing it in \texttt{data\_photometry} table to improve the database efficiency.}

\subsection{Photometric Spectral Types}\label{sec:phot_spt}

\begin{figure}
 	\centering
 	\includegraphics[width=0.465\textwidth]{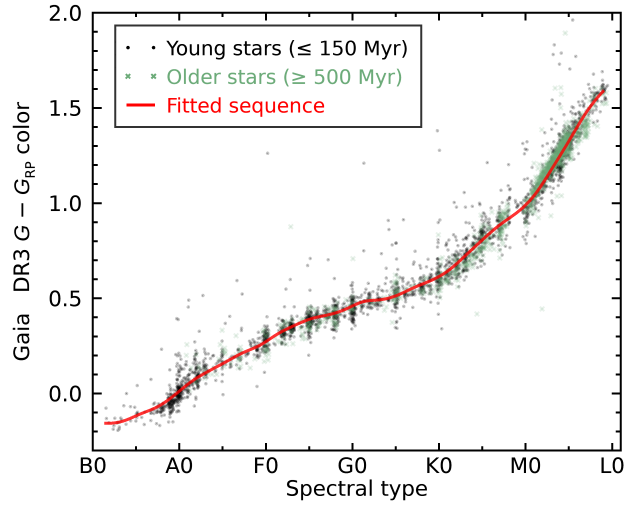}
 	\caption{The best-fitting spectral type to Gaia~DR3 sequence (red line) for stars younger than 150\,Myr (black circles) and older than 500\,Myr (green x symbols), as described in Section~\ref{sec:cmds}. We observe no significant systematic differences between younger and older stars in this particular parameter space, which allows us to assign color-based spectral type estimates independent of stellar ages.}
 	\label{fig:spt_gr}
\end{figure}

Photometric measurements and spectral types contained in the \texttt{data\_photometry} and \texttt{data\_spectral\_types} tables were used to construct empirical sequences of spectral type to Gaia~DR3 $G-G_{\rm RP}$ colors, which we found were not significantly affected by age (see Figure~\ref{fig:spt_gr}), similarly to the spectral type--color sequences based on Gaia~DR2 colors \cite{2020ApJ...903...96G}.

We developed a method for fitting astrophysical sequences that relies on a Lagrangian approach similar to the radial velocity fitting method of \citep{allers_radial_2016}. This approach is particularly useful for fitting spectral type--color or color--magnitude sequences, which typically contain many more objects at the red end due to the prevalence of M dwarfs, and simplifies determining the optimal number of fitting parameters. Standard polynomial fitting methods tend to over-fit the red end of such astrophysical sequences, and often require the fine-tuning of the polynomial order to obtain a reasonable fit.

In this Lagrangian approach, the free parameters are represented by anchor points \added{with horizontal locations equally spaced in the astrophysical diagram, and with vertical locations} adjusted to optimize the fit. A log-likelihood approach is used to determine the goodness-of-fit, assuming normally distributed measurement errors. The free model parameters that are optimized include \added{the} $N$ anchor points, plus one parameter to represent the intrinsic standard deviation of the observed data around the sequence, which added in quadrature to the measurement errors when computing the log-likelihood. Optional parameters allows to consider alternate hypotheses in the log-likelihood calculation, consisting of:

\begin{itemize}
    \item A penalty factor for high-order variations in the modelized sequence, which lowers the log-likelihood proportionally to the square of its second derivative;
    \item an outlier hypothesis, where the log-likelihood is computed as 10\% of the regular log-likelihood plus a fixed value corresponding to a 3$\sigma$ outlier; and
    \item a binary hypothesis (useful for color-magnitude diagrams), where the vertical location of the sequence is shifted by a 0.75\,mag with an adjustable prior probability.
\end{itemize}

We found that the use of these alternate hypotheses allowed us to model a wide number of astrophysical sequences without needing to adjust any parameter except for the number of anchor points, which is typically in the range 20--30. Crucially, the resulting sequences visually fit the data equally well independently of the chosen number of anchors, with the caveat that too few anchor points tend to smooth over physical variations, and too many anchor points tend to over-fit high-order variations in the data, especially on the blue end where fewer stars are usually sampled.

\begin{figure}
	\centering
	\includegraphics[width=0.49\textwidth]{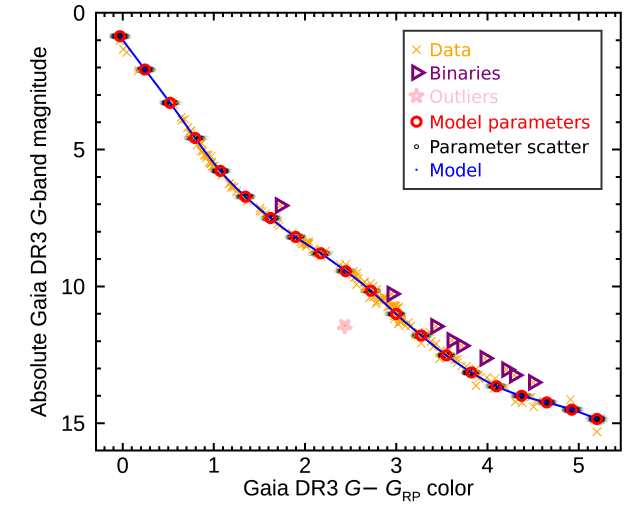}\label{fig:fitting_cmd}
	\caption{Example best-fitting sequence for members of the AB~Dor moving group in a Gaia~DR3 color-magnitude diagram, using the Lagrangian method described in Section~\ref{sec:cmds}. Model parameters are shown with red circles; their colors are fixed, and their absolute $G$-band magnitude are determined by the fitting algorithm, and are then used to construct the best-fitting sequence (blue line) with a spline. Individual stars that best matched the `young sequence' Bayesian hypothesis are shown as orange x symbols, whereas those that matched the `binary sequence' and `outlier' hypotheses are shown with purple triangles and pink stars, respectively.}
	\label{fig:mcmc_cmd_fitting}
\end{figure}

A Markov chain Monte Carlo approach was used to sample the log-likelihood of this Lagrangian approach, using $10^4$ Monte Carlo steps including a $3000$--element burn-in phase. This method was applied to derive an underlying sequence to the spectral type--Gaia~DR3 color data. The result is presented in Figure~\ref{fig:mcmc_cmd_fitting}.

\subsection{Fundamental Parameters}\label{sec:fundamental_parameters}

This section describes the determination of the fundamental parameters (ages, effective temperatures, masses, and radii) of stars and substellar objects in the MOCAdb database. The effective temperatures, masses, and radii are calculated with a focus on accuracy using empirical methods that are free of model assumptions when possible, at the cost of a significantly lower precision. We chose this approach in order to minimize biases and the impact of unresolved binaries on the determination of secondary quantities and allow reliable parameter estimates for a large number of objects with minimal human supervision. Robust but low-precision fundamental parameters are particularly useful to compute the initial mass function of a young association and to correct the impacts of gravitational redshift and convective blueshift on the heliocentric radial velocities of their members. We note that this approach is not optimal to constrain the properties of a star or its exoplanets with high precision. However, high-precision measurements calculated with other methods can still be included in MOCAdb for this purpose.

We did not attempt to determine the fundamental parameters of giant stars or white dwarfs as part of this work; however, literature measurements can be included in the relevant tables as needed in the future.

\subsubsection{Literature Ages}\label{sec:age}

A literature compilation of the ages of young associations is included in the \texttt{data\_association\_ages} table along with details of the age-dating method. We used the \texttt{association\_ages.adopted} column to designate the most reliable age estimate for each association. These association ages are taken as a reference to determine the age as well as the other fundamental parameters of their members, as described in the remainder of this section.

\subsubsection{Effective Temperatures}\label{sec:teff}

\begin{figure*}
 	\centering
 	\includegraphics[width=0.965\textwidth]{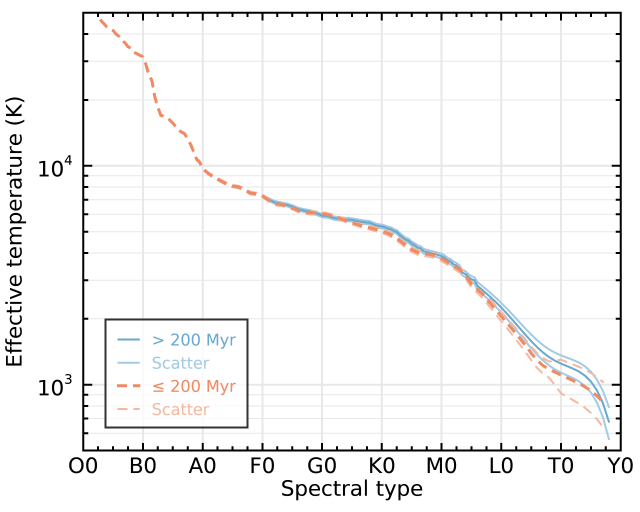}
 	\caption{Spectral type--\teff\ sequences of \cite{2013ApJS..208....9P} for main-sequence stars (blue line) and pre-main-sequence stars (orange dashed line). Pale lines around each sequence delineates their respective error bars, determined by the standard deviation of individual stars around the sequence. There are small but statistically significant differences in the effective temperature to spectral type sequence of mid-G stars and later, plausibly related to the lower surface gravity of younger stars affecting the classification based on main-sequence templates. The extension to the substellar regime are from \cite{2015ApJ...810..158F} and \cite{2025ApJ...979..145L}.}
 	\label{fig:tseq}
\end{figure*}

We used the spectral type--\teff\ sequences of \cite{2013ApJS..208....9P} extended with those of \cite{2015ApJ...810..158F}, both based on spectral energy distributions, to estimate the effective temperatures of all stars and substellar objects in MOCAdb. It is well known that pre-main sequence stars as well as brown dwarfs follow unique spectral type-to-effective temperature sequences. We have adopted the pre-main-sequence relations of \cite{2013ApJS..208....9P} and \cite{2015ApJ...810..158F} for stars and substellar objects that are likely members of associations 200\,Myr or younger, and the main-sequence or field brown dwarf relations for all other objects (as shown in Figure~\ref{fig:tseq}). As a consequence, stars that have plausible membership to more than one association have multiple effective temperature calculations (one per pair of unique \texttt{moca\_oid} and \texttt{moca\_aid} identifiers).

\begin{figure*}
 	\centering
 	\includegraphics[width=0.965\textwidth]{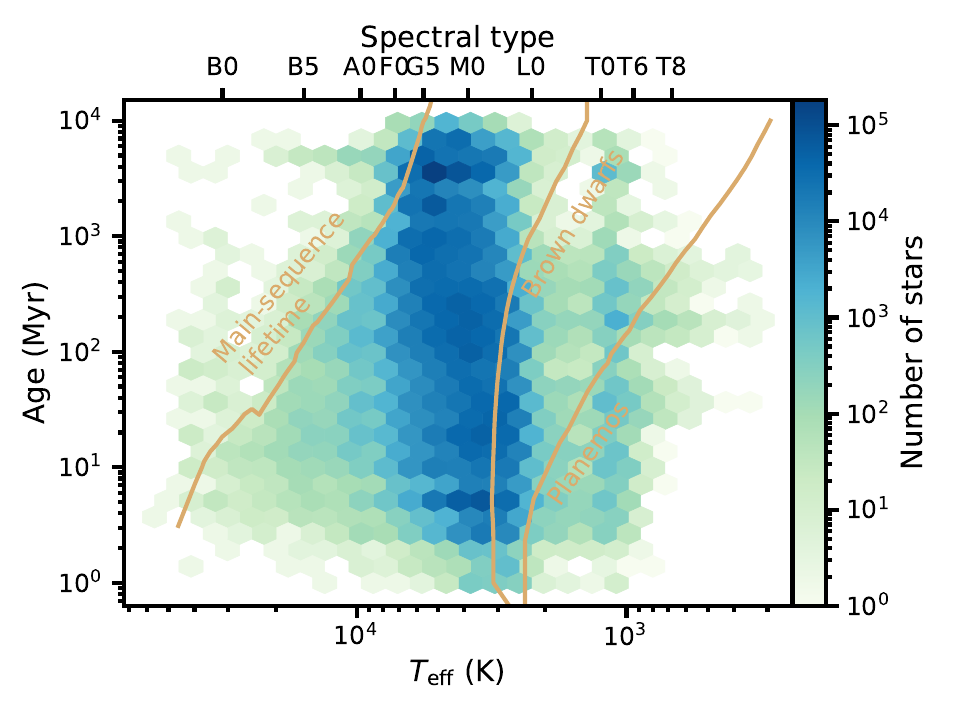}
 	\caption{Distribution of ages as a function of \teff\ and spectral types for individual stars in MOCAdb (hex bins). There is a relative dearth of hot stars at gradually older ages, which is a consequence of the limited lifetime of the most massive stars, indicated with the `Main-sequence lifetime' orange line. The mass regimes of brown dwarfs and planemos, based on the \cite{2001RvMP...73..719B} models, are also indicated.}
 	\label{fig:teffage_all}
\end{figure*}

The effective temperatures are calculated with a $10^4$--element Monte Carlo simulation. Synthetic values for spectral types are drawn randomly following the adopted spectral type measurement in the \texttt{data\_spectral\_types} table (the standard deviation is set to the measurement error, usually 0.5 subtypes). Each synthetic value is transformed to a \teff\ value using the sequence of the appropriate age, and the medial absolute deviation of the resulting \teff\ distribution in log space is taken as the measurement error that arises from the propagation of the spectral type error. Errors associated with the empirical spectral type--\teff\ sequences were taken as the standard deviation of reference stars used to build these sequences around the expected \teff\ values, from which we obtained uncertainties of 100\,K for stars M5 and earlier, and respectively 113\,K and 194\,K for the later-type objects at ages above or below 200\,Myr. These measurement errors were added in quadrature to the propagated spectral type measurement errors in order to obtain the final \teff\ measurement error in the \texttt{data\_teff} table of the \db. The \cite{2015ApJ...810..158F} sequence covers spectral types up to T8; The T9 data was taken from an average of currently known T9 brown dwarfs\footnote{\url{https://github.com/emamajek/SpectralType/blob/master/T9V.txt}}, and the temperatures of Y dwarfs were taken from \cite{2025ApJ...979..145L}. \added{The resulting effective temperatures of individual stars and substellar objects in MOCAdb are shown as a function of age as a 2D histogram in Figure~\ref{fig:teffage_all}.}

\subsubsection{Masses}\label{sec:masses}

\begin{figure*}
 	\centering
 	\includegraphics[width=0.965\textwidth]{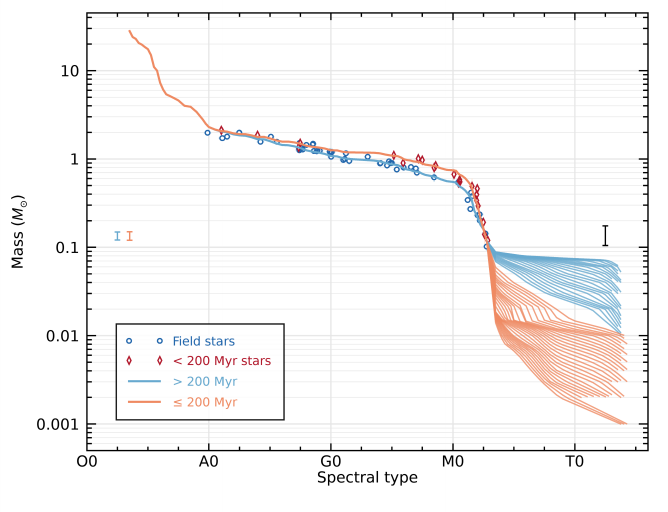}
 	\caption{Mass to spectral type sequences based on \cite{2013ApJS..208....9P} for main-sequence stars (blue lines) and extended to the substellar regime with \cite{1997ApJ...491..856B} evolutionary models. Dynamical mass measurements updated with Gaia~DR3 parallaxes are shown for main-sequence stars (blue circle) and young members of nearby moving groups (red diamonds) and were used to modify the \cite{2013ApJS..208....9P} sequence for young stars (orange lines). The discontinuity at 200\,Myr in the substellar tracks is due to our adoption of different spectral type to effective temperature sequences across this age boundary (see Figure~\ref{fig:tseq}). Typical log error bars based on the standard deviations of individual stars around their respective sequences are shown with blue and orange error bars. The black error bar represents the 25\% uncertainty that we adopted for the substellar regime to encompass possible systematic errors in model tracks.}
 	\label{fig:mseq}
\end{figure*}

We used a similar approach as described in Section~\ref{sec:teff} based on empirical spectral type sequences to estimate the masses of all stars and substellar objects in the database. We used the spectral type to mass sequences of \cite{2013ApJS..208....9P}\footnote{Available at \url{https://www.pas.rochester.edu/~emamajek/EEM_dwarf_UBVIJHK_colors_Teff.txt}.} for main-sequence stars, and we compared them with available dynamical masses for stars with spectral types in the range A2--M5 from the literature, for stars of various ages (2--150\,Myr and field stars)\footnote{See \citet{2004ApJ...604..741H}, \citet{2018AA...618A..23R}, \citet{2018AA...620A..33J}, \citet{2017AA...607A..10A}, \citet{2015ApJ...813L..11M}, \citet{2016AJ....152..175N}, \citet{2021ApJ...908...42P}, \citet{2021ApJ...908...46B} and \citet{2019ApJ...884...42S}}. In all cases where an updated Gaia~DR3 parallax was available, we used it to rescale the dynamical mass and its measurement errors. This allowed us to estimate an uncertainty around the \cite{2013ApJS..208....9P} sequence which did not need any modification to match these data from the standard deviation of measurements around the sequence in log space (0.11\,dex). For the young sequences, we used members of young associations with a known age to calculate a low-order polynomial deviation above the \cite{2013ApJS..208....9P} sequence in log mass, shown in Figure~\ref{fig:mseq}. We have then extrapolated these two sequences (both for younger and older ages) between spectral types M5.5 and the expected masses from the \cite{2001RvMP...73..719B} evolutionary sequences at spectral type M7 for each distinct age, and we have adopted the estimated masses from \cite{2001RvMP...73..719B} for later spectral types. As the \cite{2001RvMP...73..719B} evolutionary sequences provide effective temperatures rather than spectral types, we have used the empirical spectral type--\teff\ sequences described in Section~\ref{sec:teff} to estimate the spectral type at each point on the \cite{2001RvMP...73..719B} evolutionary tracks, using the sequence of the appropriate age. The resulting semi-empirical spectral type--mass sequences are shown in Figure~\ref{fig:mseq}. We have adopted conservative 20\% measurement errors for objects later than M5.

\subsubsection{Radii}\label{sec:radii}

\begin{figure*}
 	\centering
 	\includegraphics[width=0.965\textwidth]{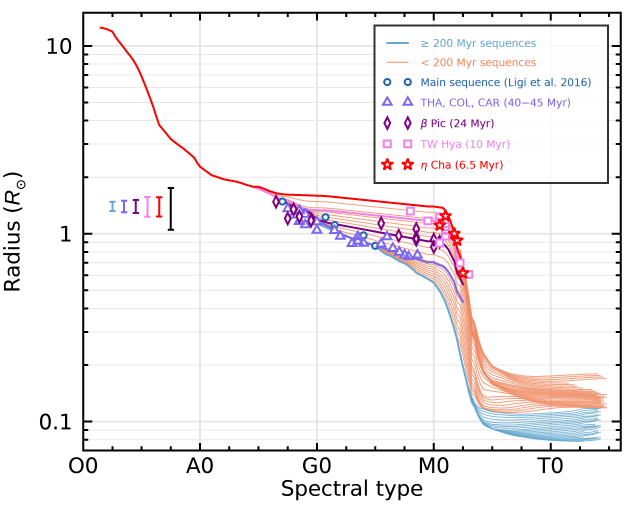}
 	\caption{Radius to spectral type sequences based on \cite{2013ApJS..208....9P} for main-sequence stars (blue line) extended to the substellar regime with \cite{1997ApJ...491..856B} evolutionary models. Measurements for individual stars obtained with the spectral energy distribution method of \cite{2006AA...450..735M} updated with Gaia~DR3 parallaxes are shown with different symbols depending on their membership in a nearby young association. These were used to adapt the \cite{2013ApJS..208....9P} sequence to different ages, and intermediate ages were interpolated in log age. Typical log error bars are shown with their corresponding sequence ages, and the black error bar represents the 25\% uncertainty that we adopted for the substellar regime to encompass possible systematic errors in model tracks.}
 	\label{fig:rseq}
\end{figure*}

We estimated the radii of each source and membership combination in the \db\ by constructing a spectral type--radii sequence with a similar method as described in Section~\ref{sec:masses}. We started from the spectral type to radius sequences of \cite{2013ApJS..208....9P}\footnote{Available at \url{http://www.pas.rochester.edu/~emamajek/EEM_dwarf_UBVIJHK_colors_Teff.txt}.} for main-sequence stars, and we compared them with available radius estimates using the spectral energy distribution method of \cite{2006AA...450..735M} with the current best available literature parallaxes (mostly from Gaia~DR3) for field stars and members of young associations with known ages (6--45\,Myr; \citealt{2013ApJS..208....9P}). We estimated an uncertainty around the \cite{2013ApJS..208....9P} sequence that is derived from the standard deviation of measurements around the sequence in log space (0.06\,dex). For young sequences, we used members of young associations with a known age to calculate a low-order polynomial deviation above the \cite{2013ApJS..208....9P} sequence in log radius, shown in Figure~\ref{fig:rseq}. We interpolated the resulting 5 sequences of distinct ages in logarithmic age to obtain sequences of intermediate ages. We then extrapolated these sequences between the M5.5 spectral types and the expected masses from the \cite{2001RvMP...73..719B} evolutionary sequences at the M7 spectral type for each distinct age, and we adopted the estimates of \cite{2001RvMP...73..719B} for later spectral types. We have again used the empirical spectral type--\teff\ sequences described in Section~\ref{sec:teff} to estimate the spectral type at each point on the \cite{2001RvMP...73..719B} evolutionary tracks, using the sequence of the appropriate age. The resulting semi-empirical spectral type--mass sequences are shown in Figure~\ref{fig:rseq}. We adopted conservative 20\% measurement errors for objects later than M5.

We used the same Monte Carlo method as described in Section~\ref{sec:masses} to propagate the spectral type and mass measurement errors to uncertainties on the radius estimates, which were added in quadrature to the uncertainties of the semi-empirical relations of Figure~\ref{fig:rseq}. The resulting radii and their measurement errors were included in the \texttt{data\_radii} \db\ table.

\subsection{Heliocentric Radial Velocities}\label{sec:hrv}

In this section, we discuss the corrections of convective blueshift (Section~\ref{sec:conv_blueshift}) and gravitational redshift (Section~\ref{sec:grav_redshift}) on radial velocity measurements, bias corrections for the Gaia~DR3 radial velocities (Section~\ref{sec:rv_bias}), and how multiple individual radial velocity measurements are used to estimate the center-of-mass heliocentric radial velocity of each system (Section~\ref{sec:rvc}).

\subsubsection{Estimation of Convective Blueshifts}\label{sec:conv_blueshift}

Convection cells at the surface of a star can bias the measurement of radial velocity obtained by measuring the Doppler shift because the centers of convection cells are both hotter (thus brighter) and occupy more surface area than the darker and narrower intergranular lanes. Because these convection cells contribute more light to the star's spectrum and they move towards the observer, they tend to shift the measured radial velocity by up to $\approx$\,0.4\,\kms. The exact bias depends on the detailed properties of the star, and on the resolving power and line spread function of the instrument used to measure the radial velocity, but the accuracy of radial velocity measurements can be improved by deriving average convective blueshifts as a function of stellar type for a range of available instruments.

We estimated convective blueshift with an approach similar to that of \cite{2023ApJ...946....6C}, based on a literature compilation of convective blueshifts as a function of spectral type \citep{1988AJ.....96..198G,meunier_variability_2017,2013AA...550A.103A,2019ApJ...871..119D,2021AA...654A.168L,2019AA...624A..57L,2019MNRAS.483.5026L,2020AA...641A..69B}. We fit a broken second-order polynomial to the data of \cite{meunier_variability_2017} in the range of spectral types F7--K4, which we extended with a linear extrapolation on either side, assuming a zero blueshift at spectral type F1 due to the expected radiative surface of hotter stars, and assuming a zero blueshift at spectral type M4 due to the constraint of \citep{2020AA...641A..69B}.

\begin{figure}
 	\centering
 	\includegraphics[width=0.465\textwidth]{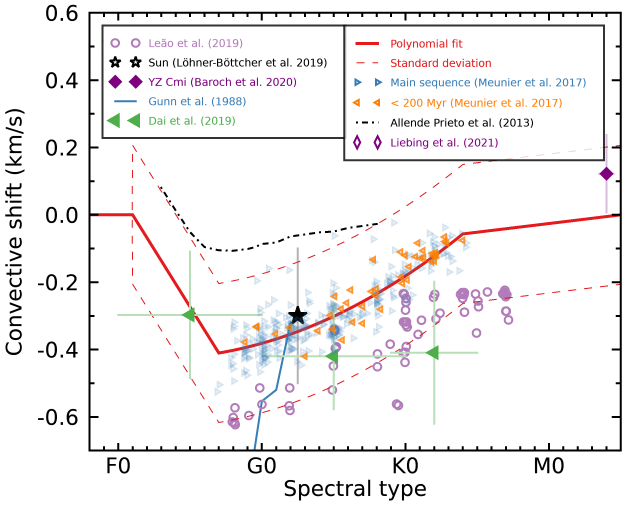}
 	\caption{Estimates of convective blueshift as a function of spectral type from various literature studies. Our adopted estimate is shown with a thick, red line and a conservative conservative envelope is shown with a dashed red lines, which encompasses most of the studies and instruments. We do not apply a convective blueshift correction to stars F0 and earlier due to their lack of a convective envelope. See Section~\ref{sec:conv_blueshift} for more details.}
 	\label{fig:conv_shift}
\end{figure}

We adopted a conservative error estimate on the convective blueshift that is taken as the standard deviation of all data sets around the sequence (0.21\,\kms), which was added in quadrature to the measurement errors of all corrected radial velocities. Achieving better precision in the spectral classes subject to convective blueshift will require an approach with a bias correction that is dedicated to every instrument design, configuration and radial velocity measurement algorithm. The measurement errors of the radial velocities of objects later than M5 were also inflated by the same amount given the lack of observational constraints on their convective blueshifts and the fact that they have a convective surface, but no error inflation was applied to the radial velocities of stars with spectral types F0 or earlier due to their lack of surface convection. The resulting sequence is compared to data from the literature in Figure~\ref{fig:conv_shift}. The convective blueshifts calculated for individual stars with a $10^4$-element Monte Carlo are listed in the table \texttt{calc\_convective\_blueshifts}.

\subsubsection{Estimation of Gravitational Redshifts}\label{sec:grav_redshift}

\begin{figure*}
	\centering
	\subfigure[Gravitational redshift]{\includegraphics[width=0.49\textwidth]{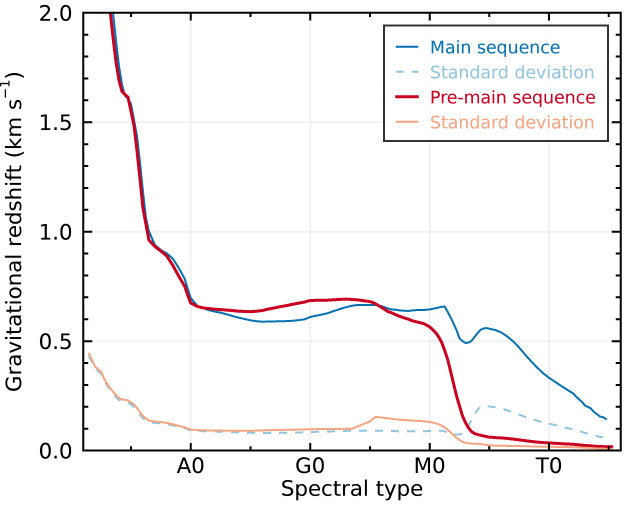}\label{fig:grav_red}}
	\subfigure[Total redshift]{\includegraphics[width=0.49\textwidth]{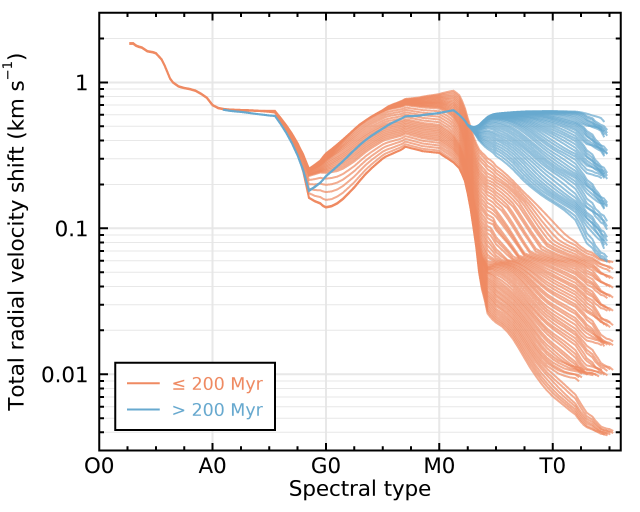}\label{fig:tot_red}}
	\caption{Left panel: The gravitational redshift adopted in this study for young (red) or main-sequence (blue) stars, derived from empirical spectral type to mass and radii relations. The associated uncertainties are shown with dashed lines. Right panel: The combined total velocity shift (convective blueshift + gravitational redshift) adopted in this work, shown in logarithmic shift for clarity. The total shift sequences split across ages for stars of spectral classes F or later because of the age dependence of their radii (shown in Figure~\ref{fig:rseq}.}
	\label{fig:red}
\end{figure*}

The wavelength of light leaving the surface of a star is elongated as it escapes the gravitational well, as expected by the theory of general relativity. This affects the measurements of stellar heliocentric radial velocities by the Dopper method by shifting the measured radial velocities to the red, typically by 0.5--0.7\,\kms\ for main-sequence stars of spectral classes AFGK (e.g., see \citealp{2011AA...526A.127P,2023ApJ...946....6C}), and more for earlier-type stars. We used the empirical mass and radius sequences as a function of spectral types (see Sections~\ref{sec:masses} and \ref{sec:radii}) to calculate the impact of gravitational redshift, and propagated the measurement errors using a $10^4$-element Monte Carlo. The resulting sequences are displayed in Figure~\ref{fig:grav_red}, and the combined impacts of gravitational redshift and convective blueshift are shown in Figure~\ref{fig:red}.

The gravitational redshifts calculated for individual stars with a $10^4$-element Monte Carlo are listed in the table \texttt{calc\_gravitational\_redshifts}. Because the mass and radius estimates of stars are age-dependent, so are the corrections to the impact of gravitational redshift on their radial velocities. For this reason, gravitational redshift corrections are calculated for every membership hypothesis discussed in MOCAdb, meaning that one measurement is listed for each star (identified with a \texttt{moca\_oid} identifier) and relevant young association (identified with a \texttt{moca\_aid} identifier) combination. Those that correspond to the current most likely membership are indicated \added{with the \texttt{adopted} flag.}

We note that the gravitational redshifts of white dwarfs are much more important than those of main-sequence stars, and those of giant stars are conversely much less important. We have not attempted to calculate them for either white dwarfs or giants in the MOCAdb. In the case of substellar objects, the gravitational redshift is strongly dependent on the object's age because these objects cool down over time. Consequently, a young object of a fixed spectral type (and thus temperature) has a much smaller mass, leading to a smaller surface gravity and gravitational redshift.

\subsubsection{Radial velocity biases in Gaia DR3}\label{sec:rv_bias}

This section describes radial velocity biases corrected at the step where MOCAdb pulls existing measurements from the table \texttt{cat\_gaiadr3} to insert them into \texttt{data\_radial\_velocities} (radial velocity values in the table \texttt{cat\_gaiadr3} are not corrected).

We corrected the hot-star radial Gaia~DR3 velocity biases described by \cite{2023AA...674A...7B}. This specific bias results in a systematic blueshift of the radial velocities measured for stars with \teff\ in the range 8500--14500\,K, and is caused by the proximity of the calcium infrared triplet\footnote{used by the Gaia~DR3 Radial Velocity Spectrometer (RVS) to calculate radial velocities \citep{2018AA...616A...5C}.}, to hydrogen Paschen lines in hot stars. The following quantity was therefore subtracted from the existing Gaia~DR3 radial velocities:

\begin{align}
    7.98\,\text{km\,s}^{-1} - \left(1.135\,\text{km\,s}^{-1}\,\text{mag}^{-1}\right)\cdot \texttt{grvs\_mag},
\end{align}

only for stars with the Gaia~DR3 column \texttt{rv\_template\_teff} in the range 8500--14500\,K and \texttt{grvs\_mag} in the range 6--12\,mag, as prescribed by \cite{2023AA...674A...7B}, where \texttt{grvs\_mag} is the Gaia~DR3 $G_{\rm RVS}$ magnitude measured by the Radial Velocity Spectrometer \citep{2018AA...616A...5C}. These biases range from $\approx$\,$+$1\kms\ to $\approx$ $-$8\kms.

We also applied the bias correction of \cite{2023AA...674A...5K} for cool and faint stars in Gaia~DR3 that is caused by RVS detector traps that delay the release of electrons as the science target moves across the CCD in the spectral direction. This causes a slight redshift in the radial velocities measured for faint stars, which can be corrected by subtracting the following quantity to the Gaia~DR3 radial velocities:

\begin{align}
    \left(0.0275\,\text{km\,s}^{-1}\,\text{mag}^{-2}\right)\cdot\texttt{grvs\_mag}^2 \\
    - \left(0.55863\,\text{km\,s}^{-1}\,\text{mag}^{-1}\right)\cdot\texttt{grvs\_mag}\\ +2.81129,
\end{align}.

This correction is applied to stars with \texttt{rv\_template\_teff} below 8500\,K and \texttt{grvs\_mag} fainter than 11\,mag, as prescribed by \cite{2023AA...674A...5K}. These biases range from zero to $\approx$\,$+$0.4\kms.

Curtis (priv. comm.) identified yet another bias in the Gaia~DR3 radial velocities of A-type stars \footnote{See \url{https://aas242-aas.ipostersessions.com/?s=0E-AC-3F-05-83-84-86-EA-54-A8-43-E1-57-13-A7-E1}.} by investigating the deviations in the Gaia~DR3 radial velocities of Pleiades members (corrected for the two biases mentioned above) with respect to the median space velocities $UVW$ of the cluster. This bias was observed to be independent of Gaia $G_{\rm BP} - G_{\rm RP}$ colors and is rather dependent on the \teff\ of the chosen radial velocity template in Gaia DR3. 

\begin{figure}
 	\centering
 	\includegraphics[width=0.495\textwidth]{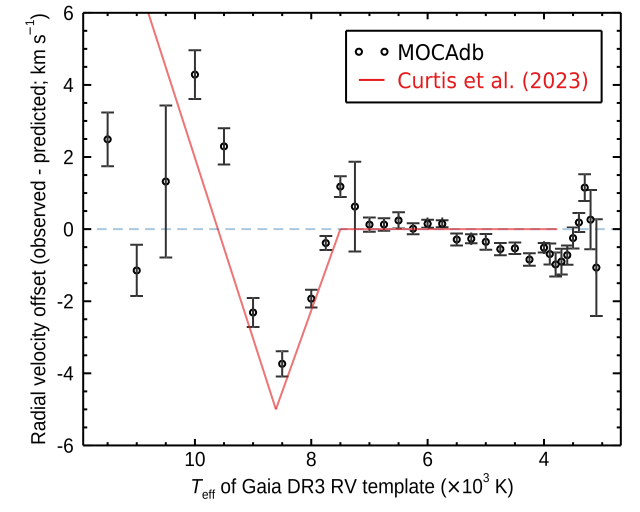}
 	\caption{Average radial velocity offset between Gaia DR3 measurements and predictions based on cluster membership using BANYAN~$\Sigma$, grouped by the effective temperature of the Gaia DR3 radial velocity template (black circles). The error bars represent measurement errors on the average; effective temperatures at both extremes have larger errors because of the smaller number of available cluster members. The blue dashed line represents zero offset, and the red line represents the Gaia DR3 radial velocity bias observed by Jason Curtis with the Pleaides members, consistent with our observed trend.}
 	\label{fig:curtis_rv_bias}
\end{figure}

Figure~\ref{fig:curtis_rv_bias} shows the resulting trend in the Gaia~DR3 radial velocities\footnote{Corrected for the two biases mentioned above, as well as gravitational redshift and convective blueshift.} minus the BANYAN~$\Sigma$ radial velocity prediction\footnote{We used the BANYAN~$\Sigma$ radial velocity prediction using all other observables but the radial velocity. This corresponds to the radial velocity that minimizes the distance between a star's Galactic space velocities $UVW$ and those of the corresponding BANYAN~$\Sigma$ model of the association.} using MOCAdb members of all nearby associations within 200\,pc of the Sun with ages 1\,Gyr or younger (excluding Theia groups, open cluster coronae or any group with a standard deviation above 1.5\,\kms\ in either dimension of its Galactic space velocity). We observe a trend similar to that reported in the Pleiades, where A-type stars of effective temperatures around 8\,500\,K appear artificially blueshifted and require an additive $\approx$\,3.7\,\kms\ correction to be applied. We also observe a more subtle blueshift of up to $\approx$\,1\,\kms\ for K dwarfs, which could be due to an insufficient correction of gravitational blueshift, specifically for the Gaia instrumental setup and this range of spectral types, rather than a systematic problem with the Gaia radial velocity templates. The underlying reason for this bias warrants future investigation and could be an age-dependent effect given that young low-mass stars tend to display enhanced magnetic activity.

The radial velocity bias shown in Figure~\ref{fig:curtis_rv_bias} does not disappear if the biases of \cite{2023AA...674A...7B} and/or \cite{2023AA...674A...5K} are left uncorrected; the three biases are independent of each other and must all be corrected, in the order described here.

\subsubsection{Combined Center-of-Mass Radial Velocities}\label{sec:rvc}

All radial velocity measurements from the \texttt{data\_radial\_velocities} table were combined in a \texttt{calc\_radial\_velocities\_combined} with the explicit goal of approximating the center-of-mass radial velocity of unresolved multiple systems. In order to achieve this, we have drawn $10^4$ synthetic radial velocity measurements for each individual literature radial velocity measurement (with a standard deviation set to the measurement error, for which we adopted a floor of 0.01\,\kms), scaled by the \texttt{n\_measurements} column to weigh combined radial velocities appropriately. Radial velocities taken within the same Julian day were combined with a weighted average before this Monte Carlo step, and were counted as a single measurement, to avoid over-representing single-epoch values from intensive high-accuracy radial velocity monitoring campaigns, which may otherwise \added{over-sample} a short phase of the radial velocity curve for multiple stars. We have adopted the weighted averages and weighted standard deviations (divided by the square root of the number of distinct epochs) of the resulting set of synthetic radial velocities for each given source as our estimated combined radial velocity and its measurement error. This method is similar to the approach adopted by \cite{miret-roig_dynamical_2020}, who also used a Monte Carlo simulation to estimate the center-of-mass radial velocities of unresolved multiple systems.

We have mutually excluded radial velocities from some radial velocity catalogs, listed in the MOCAdb table \texttt{mechanics\_combinatory\_logic}. One such example is \cite{2006AstL...32..759G} and \cite{2007AN....328..889K}: when individual measurements were available from both compilations, we only used the values from \cite{2007AN....328..889K} because both catalogs were constructed in part from the same sources. Another example is Gaia~DR3 radial velocities, which supersede those listed in Gaia~DR2.

Radial velocities corrected for the combined effects of gravitational redshift and convective blueshift are listed in the \texttt{calc\_radial\_velocities\_corrected} table.

We note that these combined radial velocities can be used to calculate approximate $UVW$ space velocities, but the combined effects of gravitational redshift and convective blueshift (described further in sections \ref{sec:grav_redshift}  and \ref{sec:conv_blueshift}) can have systematic impacts much larger than the measurement errors quoted in the \texttt{calc\_radial\_velocities\_combined} table. If these measurements were to be taken at face value for such calculations, we would recommend subtracting an average 0.6\,\kms\ from the combined radial velocities, and adding 0.6\,\kms\ in quadrature to the measurement errors (see e.g., \citealt{2023ApJ...946....6C}), to account for the two effects.

\newpage
\newpage
\newpage
\newpage
\newpage

\subsection{Galactic Space Velocities}\label{sec:uvw}

The $UVW$ Galactic space velocities are calculated for every MOCAdb object with available distance, proper motion, and radial velocity measurements in the respective MOCAdb tables \texttt{data\_distances}, \texttt{data\_proper\_motions} and \texttt{data\_radial\_velocities\_combined}, and are listed in the table \texttt{calc\_uvw\_raw}. They are based on a $10^4$-element Monte Carlo simulation for error propagation, using the full probability distribution function of the distance estimates (which are in most cases based on the method of \cite{2021AJ....161..147B} as described in Section~\ref{sec:dist}). We adopt the definition of $UVW$ axes used in the \texttt{gal\_uvw} routine of the \texttt{IDL} \texttt{astrolib} package, which forms a right-handed rectangular coordinates system centered on the Sun's velocty, with $U$ pointing towards the Galactic Center.

\begin{figure*}
 	\centering
 	\includegraphics[width=0.965\textwidth]{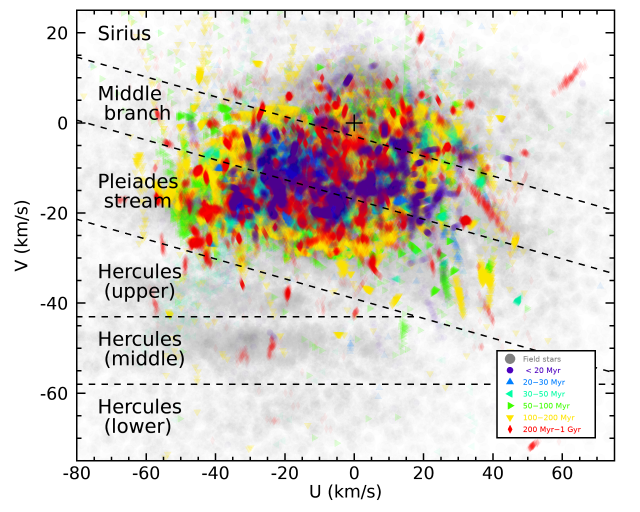}
 	\caption{Median $UV$ Galactic space velocities of individual stars in MOCAdb, color-coded by their association age, compared with nearby Gaia~DR3 stars (black circles) and the regions delimited by \cite{1958JBAA...78...21E} and \cite{1971PASP...83..251E}.}
 	\label{fig:uv_groups}
\end{figure*}

Radial velocities are subject to age-dependent biases caused by gravitational redshift and convective blueshift, as described in Section~\ref{sec:hrv}. As a consequence, so are the Galactic space velocities. We have thus calculated $UVW$ velocities using the corrected radial velocities from the \texttt{data\_radial\_velocities\_corrected} table, with a similar Monte Carlo method. Because these calculations are age dependent, a value is listed not only for every unique \texttt{moca\_oid} object, but also for every plausible \texttt{moca\_aid} association. The calculation which corresponds to the most likely host association is flagged with \texttt{adopted}$=1$. \added{The resulting Galactic space velocities of individual stars in MOCAdb are shown in the $UV$ plane in Figure~\ref{fig:uv_groups}.}

\subsection{Spectral Types}\label{sec:spt}

\added{All available spectral types are cataloged in the table \texttt{data\_spectral\_types}, along with a quality flag (mostly obtained from Simbad), and an associated \texttt{spectral\_type\_number} for convenience (0 for M0, 10 for L0, -10 for K0 and so on). The best-available spectral types (those with the highest quality flags and the smallest measurement errors) are flagged with \texttt{adopted}$=1$, and columns such as \texttt{luminosity\_class}, \texttt{gravity\_class} and \texttt{suffix} allow to store detailed peculiarities about different types of objects. The binary columns \texttt{wd\_like} and \texttt{giant\_like} allow to quickly identify white dwarfs and giants based on automated pattern matching, and the substellar-specific columns \texttt{lowg\_like}, \texttt{subdwarf\_like} and \texttt{field\_like} allow to quickly flag peculiar objects with low surface gravity or unusually red colors (typically young substellar objects); those with a high surface gravity or unusually blue colors (typically older, low-metallicity subdwarfs); or those that do not show any spectroscopic peculiarity, respectively.}

\subsection{Spectra}\label{sec:calcew}

Stellar and substellar spectra from the Montreal Spectral Library\footnote{Available at \url{https://jgagneastro.com/the-montreal-spectral-library/}.}, the SpeX Prism Library \citep{2014adap.prop..149B} and the IRTF spectral Library \citep{2009ApJS..185..289R}, and the SIMPLE archive \cite{zenodosimplearchive}\footnote{\added{A persistent and versioned copy of the SIMPLE archive is available on Zenodo at \href{https://doi.org/10.5281/zenodo.13937301}{doi:10.5281/zenodo.13937301}, and the current version is available at \url{https://simple-bd-archive.org}.}} were included in the MOCAdb. The table \texttt{moca\_spectra\_packages} lists individual packages of spectra that share similar properties; the table \texttt{moca\_spectra} lists individual spectra along with their header properties, and the table \texttt{data\_spectra} lists the wavelengths and spectral flux densities of all spectra stored in the database.

\subsection{Equatorial Rotational Velocity Predictions}\label{sec:predvsini}

A prediction of equatorial velocity can be produced for all objects in the MOCAdb that benefit from both a radius estimate in the \texttt{data\_radii} table and the adopted rotation period in the \texttt{data\_rotation\_periods} table. These estimations can be useful for example to exclude fast rotators from radial velocity-based exoplanet searches. The MOCAdb equatorial radial velocity predictions are calculated by generating a $10^4$-element Monte Carlo simulation, where the calculated radius corresponding to each membership hypothesis is randomly sampled in log space, and the rotation periods are also randomly sampled in log space. The corresponding equatorial rotational velocity prediction $v_{\rm pred}$ is then obtained for every Monte Carlo element with:

\begin{align}
    v_{\rm pred} = 50.607 \frac{\kms\ {\rm days}}{R_{\rm Sun}} \frac{R}{P_{\rm rot}}
\end{align}

A Cumulative Distribution Function (CDF) is then built for every Monte Carlo value. The CDF median is adopted as the best prediction, and asymmetrical error bars covering 68\% of the distribution based on the CDF are adopted.

We did not calculate equatorial rotational velocity predictions for objects that already have a measured projected rotational velocity in the \texttt{data\_vsini} table. These cases allow to pose constraints on the projected inclination instead, discussed in Section~\ref{sec:proji}.

\subsection{Projected Inclinations}\label{sec:proji}

It is possible to estimate a projected inclination for the objects that benefit from a radius estimate in the \texttt{data\_radii} table, a rotation period in the \texttt{data\_rotation\_periods} table and a projected rotational velocity $v\sin i$ measurement in the \texttt{data\_vsini} table. This is done with a $10^4$-element Monte Carlo similar to that described in Section~\ref{sec:predvsini}, where the inputs are sampled in log space, and the inclination is calculated with the following equation:

\begin{align}
    \sin i = \frac{v \sin i}{50.607\ \kms\ {\rm days}\ R_{\rm Sun}^{-1}} \frac{P_{\rm rot}}{R}
\end{align}

In this case, we have also included a $P(i) = \sin i$ prior appropriate for the projections of randomly distributed 3D inclinations.

\begin{figure*}
	\centering
	\includegraphics[width=0.98\textwidth]{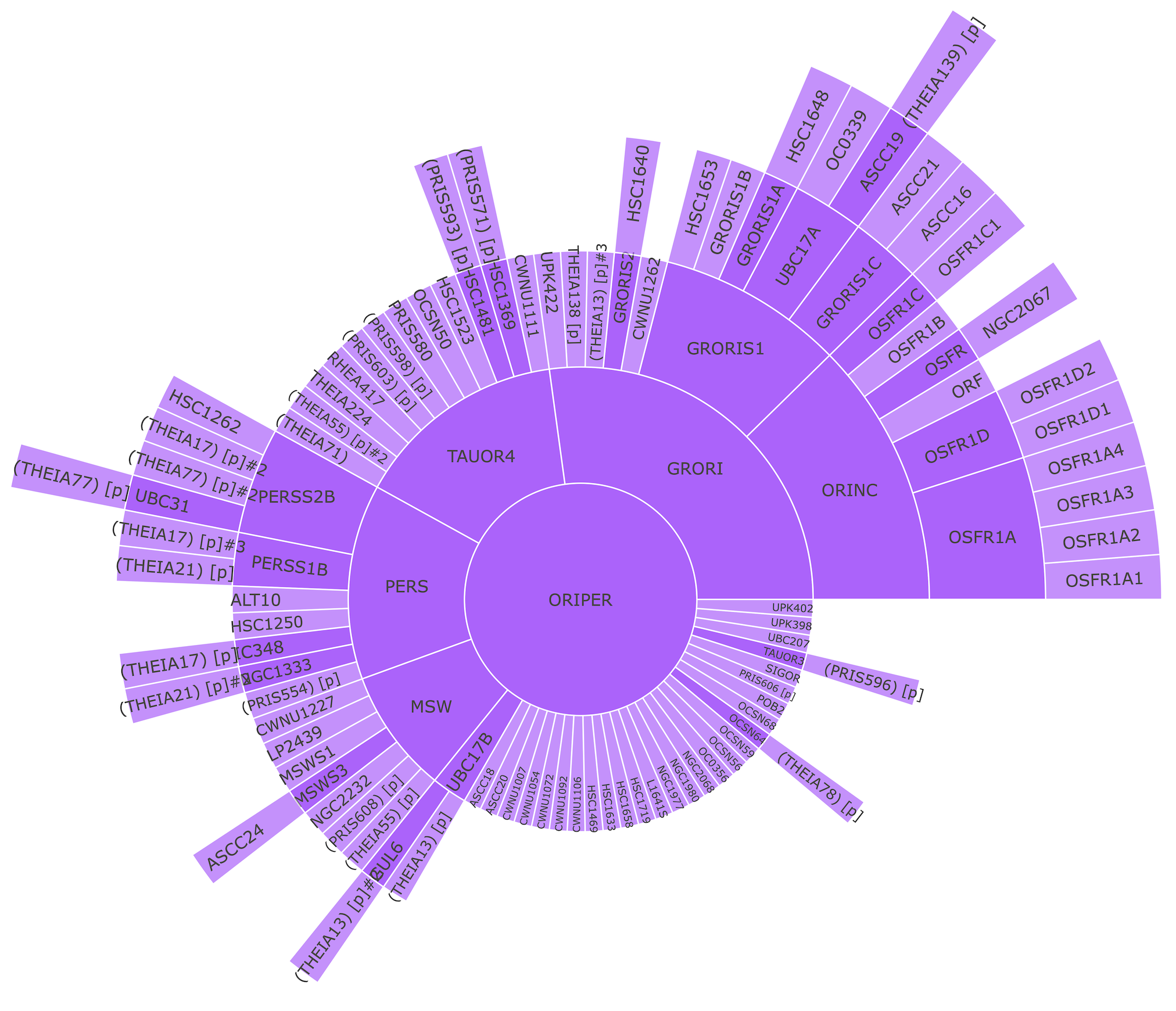}\label{fig:oriper_sunburst}
	\caption{A visualization of the Orion-Perseus branch in the hierarchical structure of young associations included in MOCAdb. The \texttt{moca\_aid} association codes with a [p] suffix indicate that only a part of that specific definition is related to the parent association. Associations with the flag \texttt{moca\_associations.suboptimal\_grouping} are indicated between parentheses, which means that they are not used in MOCAdb as an optimal way of delimiting stellar populations (i.e., either the group is split differently using another definition, or it is not recognized as real).}
	\label{fig:sunburst}
\end{figure*}

\section{DISCUSSION}\label{sec:discussion}

\subsection{Establishing a Hierarchical Structure of Nearby Associations}\label{sec:hier}

Some associations have been historically described using different names, which can sometimes refer to an ensemble or substructures (e.g., the Scorpius Centaurus OB association includes the smaller Lower Centaurus Crux, Upper Scorpius and Upper Centaurus Lupus associations among others), or refer to different parts of a larger group that had not yet been fully explored. For example, the Local Association Subgroup B4 of \cite{1999AA...341..427A}, with the database keyword \texttt{LAB4}, was later included with a spatially larger set of stars that now forms the AB~Doradus moving group (\citealp{2004ApJ...613L..65Z}; keyword \texttt{ABDMG}). As a consequence, the associations listed in the MOCAdb can overlap with each other and bear a hierarchical parent and child relationship. In the previous example, \texttt{LAB4} is registered as a child of \texttt{ABDMG} in the database. Cases where a young association is essentially an exact match to another, better characterized association (e.g., Group~1 of \citealp{2017AJ....153..257O} corresponds to the well-known $\alpha$~Per open cluster, see e.g. \citealt{1958ZA.....45..243H}) are flagged in the \texttt{data\_association\_relationships} table, with the flag \texttt{complete\_parent\_overlap}=1.

There are other cases where groups of stars defined in recent publications through automated clustering methods have merged several known structures that are not coeval. For example, Theia~55 of \cite{2019AJ....158..122K} includes a large number of members of the NGC~2232 open cluster ($\approx 18$\,Myr; CD20), the LP~2439 open cluster ($27 \pm 0.7$\,Myr; \citealp{2021ApJ...917...23K}), the Greater Taurus, Taurus-Orion and Monoceros Southwest regions of \cite{2021ApJ...917...23K}. Although this may indicate that these regions are related to each other in some way, there are many scenarios where grouping these stars together is not optimal. We have therefore preserved groupings such as \texttt{LAB4} or Theia~55 that are either obsolete, collections of non-coeval associations, or groupings with high contamination, but have assigned them a flag \texttt{suboptimal\_grouping}$=1$ in the database. This will make it possible to retrieve their lists of members, but they will not automatically be investigated further. The relationships between such associations are also listed in the table \texttt{data\_association\_relationships} with a flag \texttt{partial\_subgroup\allowbreak\_overlap} (e.g., \texttt{moca\_aid}=\texttt{THEIA55} is linked to \texttt{parent\_aid}=\texttt{NGC2232} with \texttt{partial\_subgroup\allowbreak\_overlap}=1).

Several known young associations also have a hierarchical structure, whether because star-forming regions contain multiple overdensities, open clusters, or because open clusters and their tidal tails or `coronae' can be grouped as a larger system \citep{2019AA...621L...3M,2021AA...645A..84M,2019AA...627A...4R,2019ApJ...877...12T}. These cases are also marked in the \texttt{data\_association\_relationships} table, with \texttt{complete\_parent\_overlap}=0 and \texttt{partial\_subgroup\allowbreak\_overlap}=0. Several of these structures were also identified directly in the MOCAdb database, based on 3-dimensional visualizations and overlaps in the lists of members.

In order to separate the lists of stars that belong to the core of an open cluster or its corona, we have created separate entries for such open clusters with known coronae, with the original name of the association (e.g., Hyades with the database keyword \texttt{HYA}) referring to the âcoreâ of the cluster, the corona being explicitly named (e.g., Hyades corona with a \texttt{C} preceding the database keyword: \texttt{CHYA}) and the âsystemâ referring to all stars from both the core and corona (e.g., Hyades System with an \texttt{S} preceding the database keyword: \texttt{SHYA}). In these cases, both the core and corona of the open cluster are registered as children of the full open cluster system.

The resulting hierarchical structure of a few complexes of young associations is shown in Figure~\ref{fig:sunburst} and can be dynamically explored in \url{https://mocadb.ca}.

\subsection{Compilation of Propagated Membership Lists}\label{sec:membership_lists}

Once all memberships were cataloged, we combined all membership claims while accounting for hierarchical relationships between associations into the table \texttt{mechanics\_memberships\allowbreak\_propagated}. Any claim of membership which refers to a stellar association (e.g., Upper Scorpius, or keyword \texttt{USCO}) with a non-null parent association (e.g., Scorpius-Centaurus or keyword \texttt{SCOCEN}) and without the \texttt{data\_association\_relationships.partial\_subgroup\allowbreak\_overlap}$=1$ flag are therefore repeated as a membership claim in both of the child (\texttt{USCO}) and parent (\texttt{SCOCEN}) association in the \texttt{mechanics\_memberships\allowbreak\_propagated} table. This allows a user to obtain the full set of membership claims in the literature at either hierarchical grouping level with a single filtering step of the \texttt{mechanics\_memberships\allowbreak\_propagated} table. All associations with a zero-valued \texttt{is\_real} flag are ignored in this step and excluded from the \texttt{mechanics\_memberships\allowbreak\_propagated} table.


Although the tables \texttt{data\_memberships} and \texttt{mechanics\allowbreak\_memberships\allowbreak\_propagated} contain an exhaustive list of membership claims in the literature, they will result in various levels of contamination, depending on the association, if they are used to construct a sample of members. A table that lists all currently accepted membership rejections is therefore included in the database table \texttt{mechanics\_automatically\allowbreak\_rejected\_memberships} to provide an easily managed mechanism for reducing the sample contamination. This table can link a specific star in the \texttt{moca\_objects} table to a young association that it is not a member of, or more generally specify a range of acceptable ages for the star with the \texttt{minimum\_age\_myr} and \texttt{maximum\_age\_myr} columns, and further details such as comments, a method and a literature reference.

We used external libraries to compile the lists of memberships from the \texttt{mechanics\_memberships\allowbreak\_propagated} table, and exclude those which constraints listed in \texttt{data\_rejected\allowbreak\_memberships} make the proposed membership claim unlikely. The membership claims are grouped by the unique star and association, and the membership confidence level and literature references are concatenated into a new table \texttt{mechanics\_memberships\allowbreak\_vetted}. Additionally, the table \texttt{data\_rejected\_memberships} allows to automatically reject members of a group based on fixed parameters such as a specific range in its $UVW$ velocity components; those are usually only useful to reject extreme outliers in a list of members. The resulting stars that are rejected based on these parameters are listed in \texttt{mechanics\_automatically\allowbreak\_rejected\_memberships}, and their status is also reflected in \texttt{mechanics\_memberships\allowbreak\_vetted}.

\subsection{Construction of Kinematic Models}\label{sec:bsigma_models}

\begin{figure*}
 	\centering
 	\includegraphics[width=0.965\textwidth]{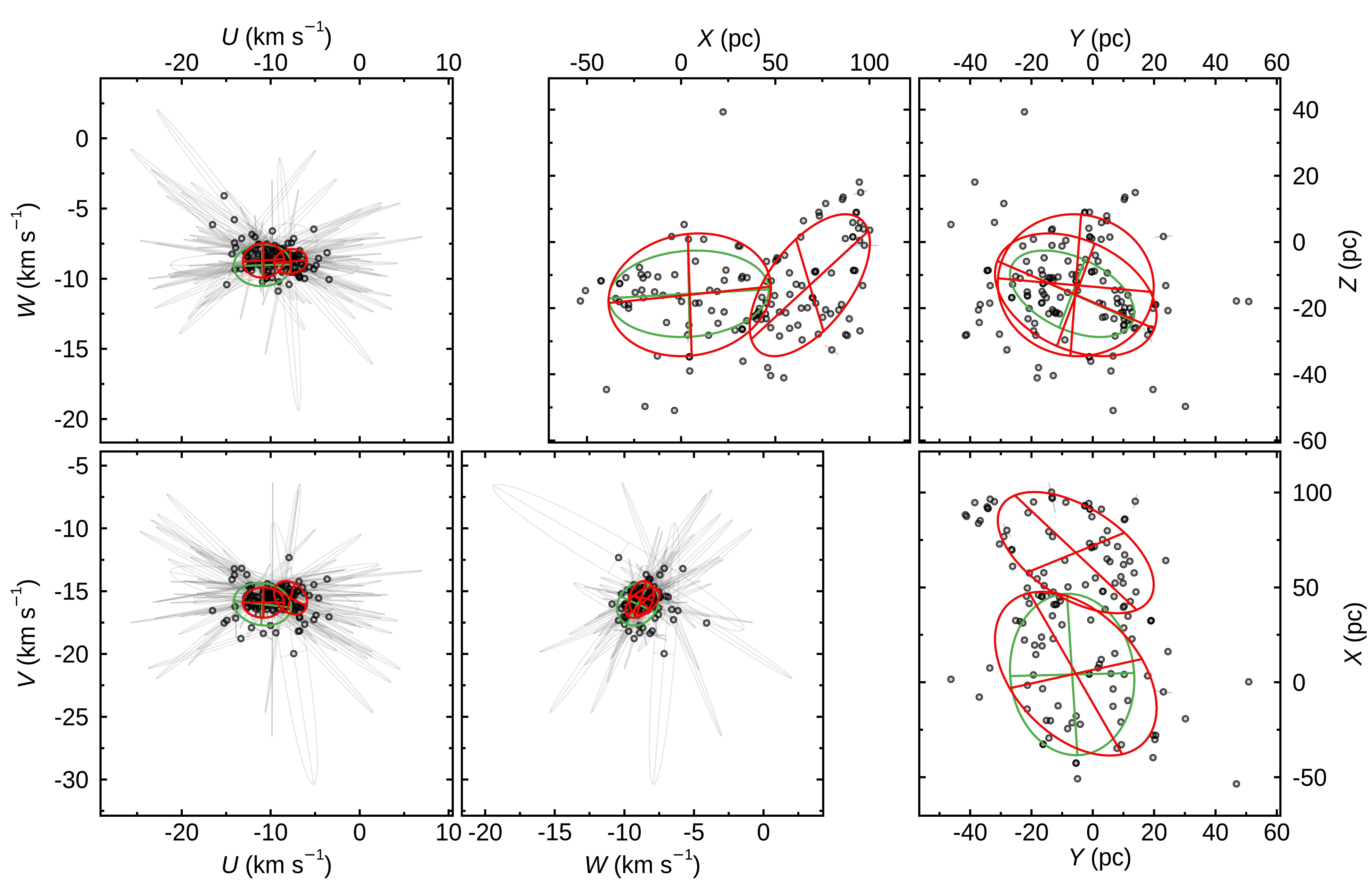}
 	\caption{Projections of the multivariate Gaussian models determined with the extreme deconvolution algorithm for members of the $\beta$~Pic moving group in $XYZ$ Galactic positions and $UVW$ space velocities (red ellipses). Projected semi-major axes are shown with red lines, and the previous model of \cite{2018ApJ...856...23G} based on Gaia~DR1 and a simpler model construction method are shown in green. Individual stars are shown with black circles, and their error bar ellipses are shown in gray, with the correct orientations accounting for the fact that the largest error contributions (radial velocities or parallaxes) are oriented differently in 6D space for stars in different directions in the sky. Stars that were automatically flagged as outliers and not included in our the model construction are displayed in pink \added{in the complete figure set (no such outliers were detected in the case of the $\beta$~Pic moving group, shown above)}. The extreme deconvolution algorithm properly accounts for the orientation of the error bar ellipses of the individual stars, which would otherwise artificially inflate the models and affect its ratios of semi-major axes. The complete figure set (491 images) is available in the online journal.}
 	\label{fig:banyan_models}
\end{figure*}

We used the lists of members compiled here to update the spatial and kinematic models of the BANYAN~$\Sigma$ tool \citep{2018ApJ...856...23G} for \bsigmaassociations\ associations that we consider potentially coeval and non duplicated (i.e., those with the flag \texttt{moca\_associations.in\_banyan}$=1$).

BANYAN~$\Sigma$ is a Bayesian model selection tool that allows users to determine whether a star is more likely to belong to a known population of stars based on 6D multivariate Gaussian models in $XYZ$ Galactic coordinates and $UVW$ space velocities for stars that belong in known associations and a 10-components multivariate Gaussian model of field stars in the Solar neighborhood. When required, BANYAN~$\Sigma$ uses analytical solutions to marginalization integrals over missing heliocentric radial velocities and/or parallaxes, allowing it to identify likely candidate members even if only sky coordinates and proper motions are available. With the advent of the latest Gaia data releases, complete $XYZUVW$ coordinates can be calculated for a larger number of stars, which makes it possible to use other clustering metrics such as HDBSCAN  \citep{2017arXiv170507321M} that are not model dependent and do not assume Gaussian distributions for young coeval associations, however, a number of white dwarfs, lower-mass stars or distant stars will still be missing heliocentric radial velocity measurements for the foreseeable future, and will still benefit from a Bayesian analysis that marginalizes over this dimension. Brown dwarfs and isolated planetary-mass objects will still benefit even more from such an approach, given that there is currently no future plan to measure their parallaxes or heliocentric radial velocities on a large scale.

The spatial and velocity distributions of the unique young associations compiled in MOCAdb were built using a Gaussian mixture model in 6D $XYZUVW$ space. For each association, a list of all members were first selected from the table \texttt{mechanics\_memberships\allowbreak\_propagated}, excluding the low-probability members (\texttt{mechanics\_memberships\allowbreak\_propagated.\allowbreak moca\_mtid}=`LM') or rejected members (\texttt{mechanics\_memberships\allowbreak\_propagated.\allowbreak moca\_mtid}=`R'), and known companions (using the \texttt{moca\_companions} table). The best-available kinematic observables (coordinates, proper motions, parallax, radial velocity) and spectral types were then compiled for each member. In the cases where a Gaia~DR3 parallax was adopted as the best-available parallax measurement, we automatically selected the proper motion and sky coordinates from the same catalog, and also compiled all the covariances between sky coordinates, proper motions, and parallaxes. In the cases where the only available kinematic measurements came from a range of different catalogs, we assumed zero covariances between the different observables. \added{The \cite{2021AJ....161..147B} geometric priors (described in Section~\ref{sec:dist}) were also included when converting the parallaxes of individual stars into distances during the spatial-kinematic models construction because the spatial-kinematic models of the most distant open clusters are subject to this bias in a non-negligible way. The information content of the prior is overwhelmed by the collective information of the parallax measurements of individual stars in most cases, however. Instead of relying on the average distance and asymmetric error bars stored in the \texttt{data\_distances} table, we used the complete reconstructed distance probability distribution function.}

The radial velocities obtained from a combination of all available literature measurements were used and corrected for the convective blueshift and gravitational redshift, using the best available spectral type and assuming the age of the specific association for which the model was being constructed.

The 6D $XYZUVW$ distribution of each member was then reconstructed using a $10^4$-element Monte Carlo using the appropriate covariances. In order to reflect the parallax covariances onto the distances based on the \cite{2021AJ....161..147B} prior, we first generated a Gaussian distance distribution from one slice of the 5-dimensional Gaussian based on all available covariances. We then determined the position of each synthetic star with respect to this Gaussian distance cumulative distribution function, and interpolated it on the desired cumulative distribution function which accounts for the proper parallax inversion and prior. Once we have obtained the appropriate set of Monte Carlo simulations, we used them to calculate the Galactic $XYZ$ coordinates and the $UVW$ space velocities with the method described in Sections~\ref{sec:xyz} and \ref{sec:uvw}. 

We determined the 6-dimensional median position of every star in a uniformized space with consistent units [$X$,$Y$,$Z$,$cU$,$cV$,$cW$] by \added{multiplying} all $UVW$ values by a factor of \added{$c = 12$\,\pckms}, a value chosen to penalize $UVW$ outliers significantly. We then computed the 6D Euclidian distance between each star and the median $XYZUVW$ position and rejected any star that is located at more than 5 median absolute deviations from the median from the model construction. This step is intended to remove only the most significant outliers that may otherwise affect automated outlier rejection algorithms.

In a second outlier rejection step, we used the Extended Isolation Forests (\texttt{eif}) Python package of \cite{2018arXiv181102141H} with the 6D spatial-position vectors of every star and rejected any star with an outlier score above 0.55, which we found was efficient in only rejecting stars which visually appear outside of the typical distributions.

Once a clean set of non-outlier stars was assembled, a covariance matrix was built for each star from the full Monte Carlo distribution of its $XYZUVW$ values, using medians and comedians \citep{comedians} instead of averages and covariances to avoid a disproportionate impact that wide-tailed distributions may otherwise have on the covariances. We have then applied the extreme deconvolution technique of \cite{bovy_extreme_2011} to estimate the underlying density distribution of members, iterating with one to 20 individual Gaussian components.
Stars without a radial velocity measurement were included in the model construction by assigning them optimal radial velocities that place them at the center of the median $UVW$ distribution of each group, with a very wide measurement error (30\,\kms), as prescribed by \cite{bovy_extreme_2011} for missing values. In rare cases where no member of a given association has a measured radial velocity, we forced the use of a single Gaussian model component and replicated each star twice, once with a radial velocity of 30\,\kms, and once with a value of $-30$\,\kms, to artificially extend the Gaussian model in the direction where the projected true velocity of the association is unknown.

We used an increasing number of Gaussian mixture components until the corrected Akaike Information Criterion (AICc; see \citealt{aicc}) started increasing, an indication that the addition of extra components is not justified by the data. We elected to use the AICc rather than the standard Akaike Information Criterion (AIC) or the Bayesian Information Criterion (BIC) because our goal is to use as many mixture components as required to improve the mapping of a specific young association, in a context where the computational cost of additional components on calculating membership probabilities is small (a consequence of BANYAN~$\Sigma$ using analytical solutions to the marginalization priors). Therefore, we required a metric that does not strongly penalize the addition of mixture components (such as the BIC), but we required a metric that does not allow for an arbitrarly large number of components in the situation where many stars are part of a unimodal distribution. The AIC metric would allow a large number of mixture components in this scenario because it includes no penalty for the number of stars. Because the AICc includes a corrective term involving the number of stars, we chose it as the preferred metric as it achieves our goal of using as many mixture components as justified in order to improve the mapping of stars (i.e., the average log-likelihood over all stars) without adding components that are not justified by the data with statistical significance.

There are scenarios where a relatively small number of stars presents a statistically complex spatial shape, but not necessarily in the spatial-kinematic or the purely kinematic subsets of the complete 6D space. In order to properly model these scenarios, we employed four categories of model complexities with a varying number of Gaussian mixture components (1--20 components for each category), to pick the single model with the smallest AICc: (1) The models with full 6D covariance matroces (27 free parameters per GMM component, plus $N-1$ mixture weights for $N$ Gaussian components); (2) without spatial-kinematic covariances (18 free parameters per component excluding the mixture weights); (3) without spatial-kinematic or purely kinematic covariances (15 free parameters per component excluding the mixture weights); and (4) without any covariance term (12 free parameters per component excluding the mixture weights).

We required that the effective number of fitted stars, multiplied by the number of dimensions for every individual star (up to 6 for complete measurements of the Galactic coordinates and space velocities), were at least as large as the number of parameters in the Gaussian Mixture Model. Any model category or number of Gaussian mixture components that did not satisfy this requirement were ignored in our final model selection.

In the cases of associations with 5 stars or less, a more stringent minimum eigenvalue of 15\,pc or 1.5\,\kms\ was imposed on the components of the Gaussian model to avoid artificially small distributions. Associations with a single cataloged member were modeled with a single spherically symmetric Gaussian component with a characteristic width of 15\,pc and 1.5\,\kms\ to represent the unknown true spatial extent of the association.

This overall process of model construction described above is similar in principle to the model construction used in the original BANYAN~$\Sigma$ tool \citep{2018ApJ...856...23G}, with the following improvements: (1) The outlier rejection steps are much more robust and we found did not need any form of human supervision (although every model was visually inspected); (2) multiple Gaussian components are allowed to fit the data to obtain a better representation of the associations with non-Gaussian distributions; and (3) the underlying measurement errors and covariances are treated more appropriately by using the extreme deconvolution method. This led to models with a smaller $UVW$ dispersion on average compared with the previous method, and to more spherical distributions in many cases where biases would otherwise be introduced because of large parallax or radial velocity errors (e.g., distant associations would otherwise be elongated toward the observer).

\begin{figure*}
 	\centering
 	\includegraphics[width=0.965\textwidth]{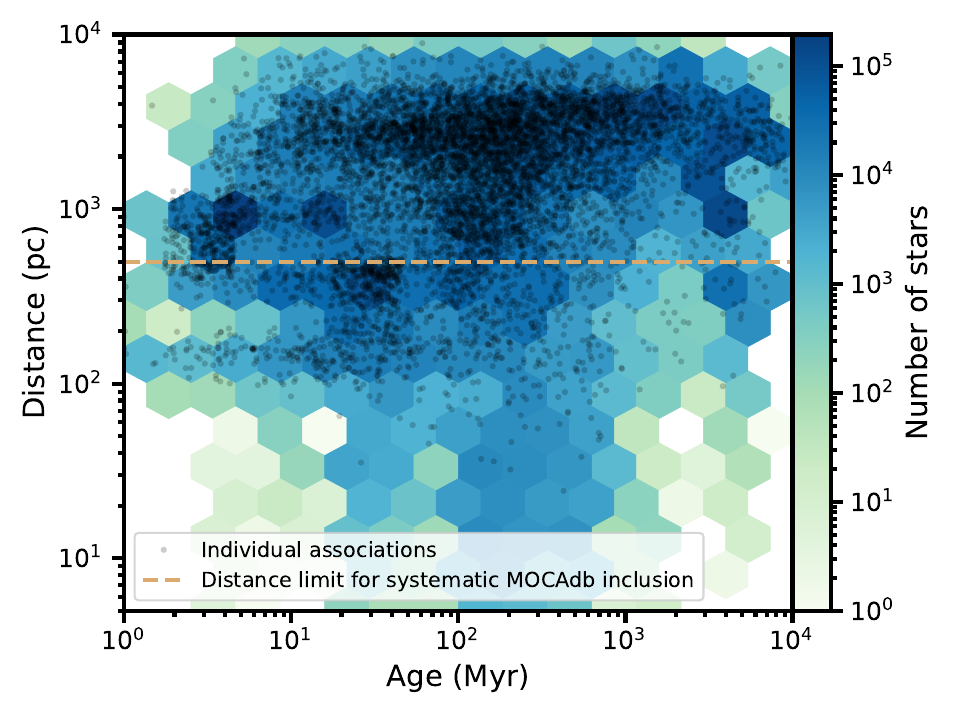}
 	\caption{Distribution of distances and ages for individual stars in MOCAdb (hex bins) compared with per-association averages (black circles). Although we included only associations with median distances within 500\,pc of the Sun (horizontal orange, dashed line), some or their members are located further away. There is a well-documented lack of stars younger than $\approx 10$\,Myr in the immediate Solar neighborhood ($\lesssim 100$\,pc).}
 	\label{fig:agedist_all}
\end{figure*}

The resulting fit to the members of the $\beta$ Pictoris moving group is shown in Figure~\ref{fig:banyan_models}, and the models of all other associations are included as an online-only figure set. Individual members are shown with black circles and their 6D measurement errors are represented by projected gray ellipsoids with their principal axes, and members that were automatically flagged as outliers are marked with purple circles instead. Projections of the single Gaussian cluster that was fitted by \texttt{xdeconv} are shown in red with its projected principal axes, and the previous multivariate Gaussian model of \cite{2018ApJ...856...23G}, which relies on a simpler algorithm that does not account for the proper orientations of measurement errors in $XYZUVW$ space, is shown with a green circle. In most cases where a previous BANYAN~$\Sigma$ model was available, the updated version is both much narrower in $UVW$ space, and the instances of models elongated along the line of sight in either $XYZ$ or $UVW$ space are drastically reduced. \added{The complete set of average association distances and ages are shown in Figure~\ref{fig:agedist_all}, along with a 2D histogram of the individual stars in MOCAdb.}

\added{The results from BANYAN~$\Sigma$ applied on every star in MOCAdb (such as the membership probability, and $XYZUVW$ separation from the core of the best model, the optimal distance and radial velocity) are available in the table \texttt{calc\_banyan\_sigma}. We provide calculations for every combination of available input observables to help diagnostics (the possibilities are: proper motion only, proper motion and radial velocity, proper motion and distance, or all of the above), and the one with the most available observables are set to \texttt{max\_observables}$=1$. The calculations can be performed on more than one set of BANYAN~$\Sigma$ models (\texttt{moca\_bsmdid}$=23$ for this work). The unique BANYAN~$\Sigma$ model versions are detailed in the table \texttt{moca\_banyan\_sigma\_models}; a single one is adopted with the \texttt{adopted}$=1$ flag. The view \texttt{calc\_banyan\_sigma\_best} allows to view only the BANYAN~$\Sigma$ results yielded by the adopted models and the full set of available observables, for a given star. The table \texttt{calc\_banyan\_sigma\_details} allows to view all detailed calculated quantities (such as the optimal distance and radial velocity) for every plausible association. These detailed outputs should be matched to the \texttt{calc\_banyan\_sigma} table using \texttt{calc\_banyan\_sigma\_details.cbds\_id}$=$\texttt{calc\_banyan\_sigma.id}.}

\subsection{Empirical Color-Magnitude Diagram Sequences}\label{sec:cmds}

\begin{figure*}
 	\centering
    \subfigure{\includegraphics[width=0.49\textwidth]{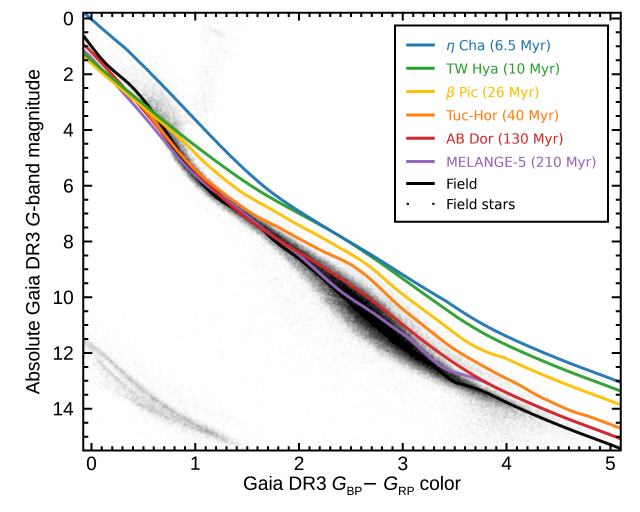}\label{fig:cmd_cal_br}}
    \subfigure{\includegraphics[width=0.49\textwidth]{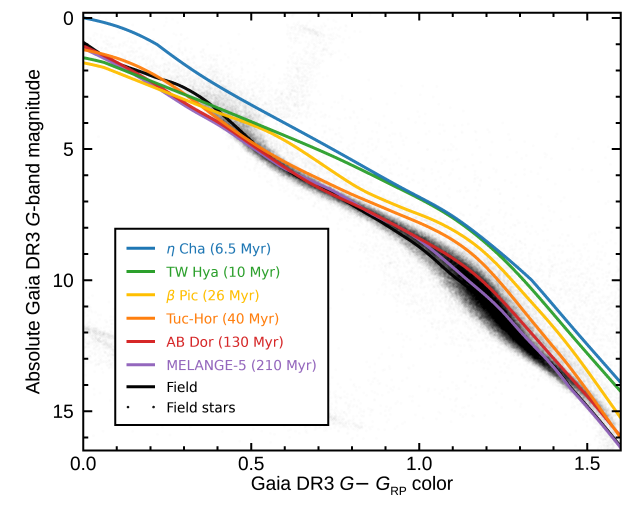}\label{fig:cmd_cal_gr}}
 	\caption{Gaia~DR3 color-magnitude diagram of nearby reference populations of distinct ages, compared with the best-fitting sequence constructed using the Lagrangian method described in Section~\ref{sec:cmds}. The reference populations are as follows: $\eta$~Cha association, 6.5 Myr \citep{2013MNRAS.435.1325M}; TW~Hya association, 10\,Myr \citep{2015MNRAS.454..593B}; $\beta$~Pic moving group, 26\,Myr \citep{2014ApJ...792...37M}; Tuc-Hor association, 40\,Myr \citep{2014AJ....147..146K}; AB~Dor moving group, 130\,Myr \citep{2018ApJ...861L..13G}; MELANGE--5 association, 210\,Myr \citep{2024AJ....168...41T}.}
 	\label{fig:cmd_seq}
\end{figure*}

We used the Lagrangian method described in Section~\ref{sec:phot_spt} to model the Gaia~DR3 color-magnitude diagram of nearby reference young associations with approximately solar metallicity, accounting for interstellar extinction. The resulting sequences, grouped by age, are shown in Figure~\ref{fig:cmd_seq}. These sequences provide empirical isochrones to compare the Gaia~DR3 photometry of solar-metallicity stellar populations, allowing users to determine approximate ages without model assumptions.

\subsection{Identification of New Candidate Members with Gaia DR3 and Hipparcos}\label{sec:new_members}

We used the updated BANYAN~$\Sigma$ models constructed in Section~\ref{sec:bsigma_models} to identify potentially missing candidate members of all likely coeval associations cataloged in the database using a subset of all Gaia~DR3 entries within 500\,pc of the Sun. In order to reduce the rate contamination that can be especially significant near the Galactic plane, we have only included Gaia~DR3 entries with a parallax measurement error below 0.4\,mas.

All stars with a membership Bayesian probability above 70\% as determined by BANYAN~$\Sigma$ that also have a most favorable $UVW$ position that is within 3\,\kms\ of the potentially related association are included in MOCAdb for a full compilation of their literature kinematics and future membership assessment, but only those with a membership probability above 90\% are considered as robust new candidate members. The constraint $UVW$ informs us about the best-case scenario separation between a star and the center of its potential host association in $UVW$ space, and is useful for excluding stars with unusual kinematics that do not fit any Bayesian hypothesis considered by BANYAN~$\Sigma$, including its model of nearby field stars.

Any potential candidate member with an existing entry in the \texttt{data\_memberships} database table has already been assessed in the existing literature, and was therefore ignored. The resulting candidate members were further vetted in a $M_G$ versus $G - G_{\rm RP}$ Gaia~DR3 color-magnitude diagram, and were rejected if they are at least 0.25\,mag fainter than the empirical sequence of the appropriate age (see Section~\ref{sec:cmds} and Figure~\ref{fig:cmd_seq}). Candidates that were flagged as likely interlopers based on this color-magnitude diagram cut were included in the \texttt{data\_memberships} table as rejected candidates (\texttt{moca\_mtid}$ = $`R'), to inform future searches for candidate members that may recover them.

Some of the brightest stars with parallax measurements in the Hipparcos mission \cite{1997AA...323L..49P} are either too bright for Gaia to measure their kinematics, or bright enough that the Gaia accuracy is decreased compared with the Hipparcos analysis of \cite{2007AA...474..653V}. Because of this, we have also used the complete Hipparcos compilation to identify some potentially missing brighter members of nearby young associations, using the same method and selection criteria.

\begin{figure*}
	\centering
	\subfigure[New Candidates Count]{\includegraphics[width=0.49\textwidth]{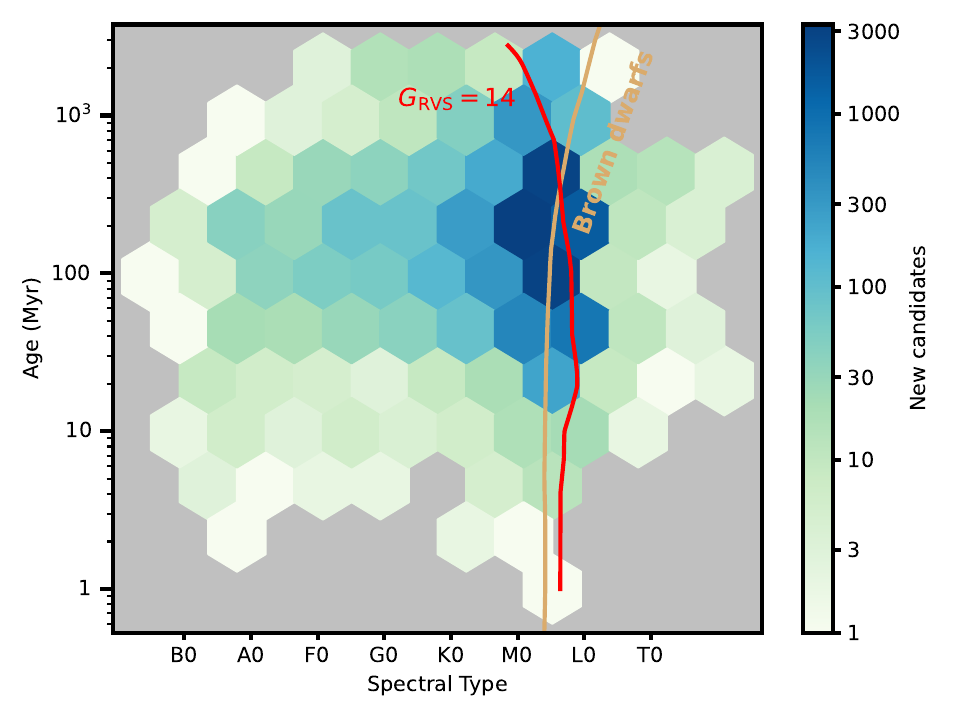}\label{fig:frac_increase}}
	\subfigure[Fractional Increase in Candidates]{\includegraphics[width=0.49\textwidth]{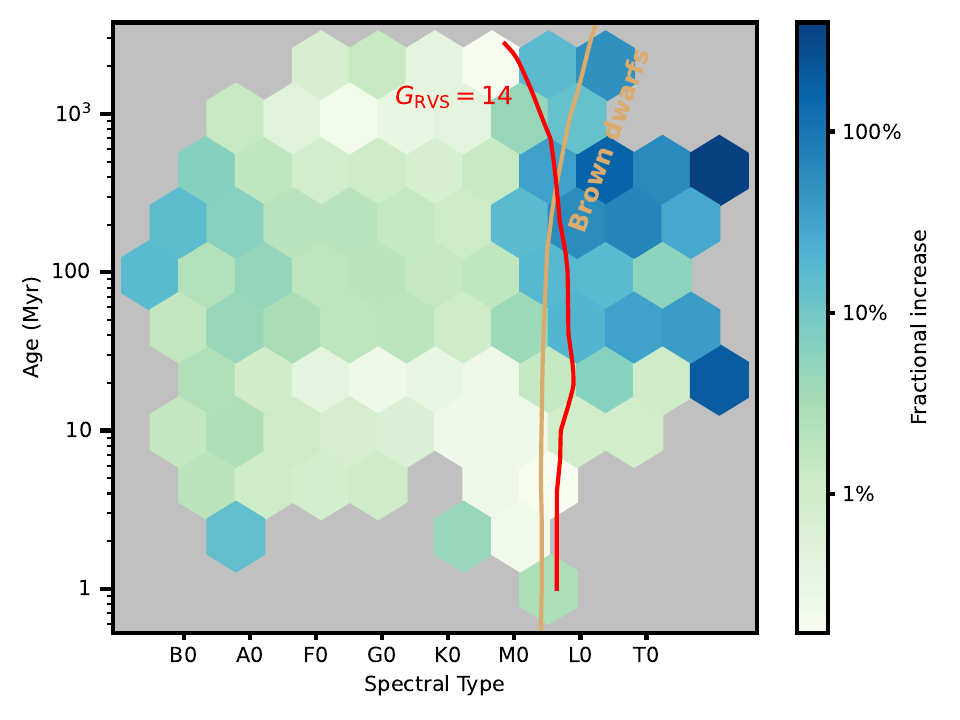}\label{fig:num_increase}}
	\caption{Spectral types and ages of the newly identified candidate members in this work. The left panel shows the absolute number of new candidates, and the right panel shows the fractional increase versus previous literature membership lists. Most of the newly identified candidates are early M dwarfs with ages in the 30--500\,Myr range, but the largest fractional increases are observed for later-type objects in the substellar regime or for BA stars, in part due to their lack of radial velocity measurements in Gaia~DR3, which BANYAN~$\Sigma$ does not require to make a probabilistic membership assessment. The red line outlines the spectral types past which the Gaia~DR3 radial velocities are unavailable, which correspond to an apparent $G_{\rm RVS}$--band magnitude of 14\,mag. We have assumed the distances of the nearest young association to the Sun at every age bin to calculate this delimiting apparent $G_{\rm RVS}$.}
	\label{fig:new_candidates_hist}
\end{figure*}

\begin{figure}
	\centering
	\includegraphics[width=0.49\textwidth]{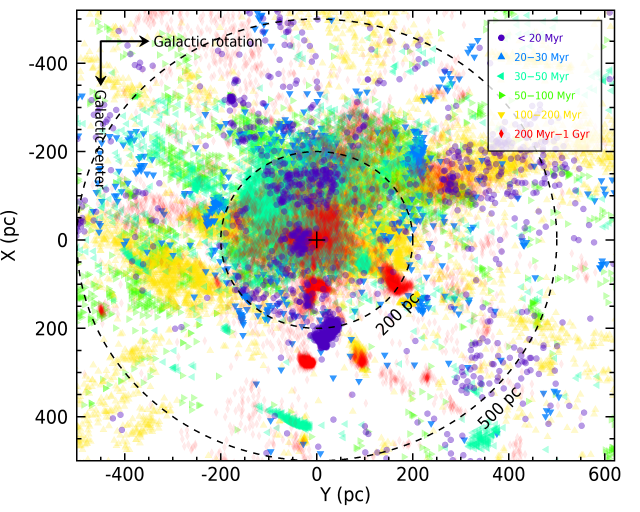}
	\caption{Galactic space coordinates of newly identified candidate members of young moving groups. Most newly identified candidates are located within 500\,pc of the Sun because they originate from the dedicated Gaia~DR3 + Hipparcos search within that radius. See Section~\ref{sec:new_members} for more details.}
	\label{fig:newmemxy}
\end{figure}

\begin{figure}
	\centering
	\includegraphics[width=0.49\textwidth]{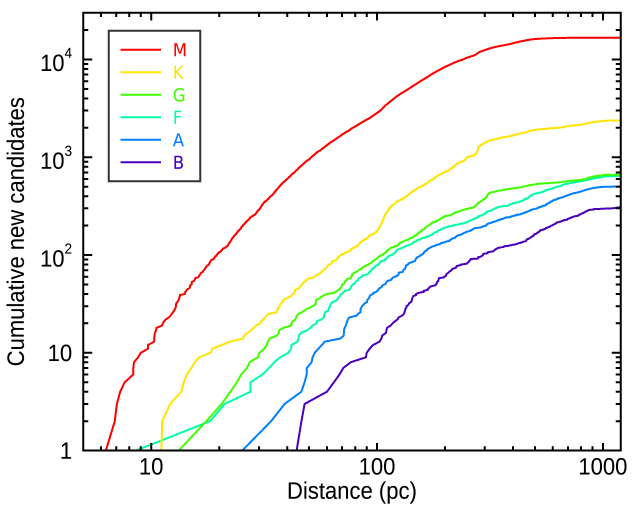}
	\caption{Distance cumulative distribution functions for all new candidate members reported here, by spectral class, and that most other types of stars were identified at distances further than 100\,pc. Follow-up work such as direct-imaging campaigns that target nearby young M dwarfs will therefore benefit from this refreshed sample of nearby, young M dwarfs. This figure illustrates that most of the newly identified members are M dwarfs. See Section~\ref{sec:new_members} for more details.}
	\label{fig:newmemdistcumul}
\end{figure}

In addition to the Gaia and Hipparcos samples, we included in our BANYAN~$\Sigma$ every object that were included in MOCAdb, which allowed us to identify a few other candidate members beyond 500\,pc of the Sun. Those mainly come from literature claims of non-memberships in a specific open cluster or association, which were included as non-members (\texttt{moca\_mtid}=`R') and which we categorized as a candidate member of another association. Additional sources come from catalog compilations of rotation periods or equivalent width measurements that were not previously identified as members of a young association. The candidate members resulting from this analysis are shown in Figures~\ref{fig:new_candidates_hist}, \ref{fig:newmemxy} and \ref{fig:newmemdistcumul}.

The \newcandidates\ resulting new candidate members are listed in Table~\ref{tab:new_candidates} and were included in the \texttt{data\_memberships} database table.

It is notable that the B2.5V star Nunki (Sig~Sgr) appears to be a candidate member in the $26 \pm 3$\,Myr-old $\beta$PMG \citep{2001ApJ...562L..87Z,2014ApJ...792...37M} with a 74\% membership probability in BANYAN~$\Sigma$ (at a $UVW$ separation of 4.3\,\kms from the center of the $\beta$PMG kinematic model). This massive star had not been previously identified as a candidate member in $\beta$PMG, possibly because it is somewhat of a kinematic outlier and for its lack of Gaia astrometry, which is a consequence of its brightness (its Hipparcos magnitude is 2.0\,mag). This would be the most massive member of the $\beta$PMG, to which $\beta$~Pictoris itself (an A6V star) is second. A B2.5V star such as Nunki is expected to leave the main sequence $\approx$\,60\,Myr after formation, and it is therefore plausible that it may be young enough to be a member of $\beta$PMG.

Nunki has long been known as a candidate equal-mass binary based on interferometry \citep{1974MNRAS.167..121H,1994AA...290..340B}, but it was more recently resolved as a near equal-mass $\approx$6.5\,\msol\ 0\farcs6 binary star with VLT/GRAVITY \citep{2009ASSP....9..361E}, along with an estimated system age of 30\,Myr \citep{2025RNAAS...9...71W}.

\subsection{Combined Age Diagnostic Sequences}\label{sec:age_prop}

\begin{figure*}
	\centering
	\subfigure[Lithium]{\includegraphics[width=0.49\textwidth]{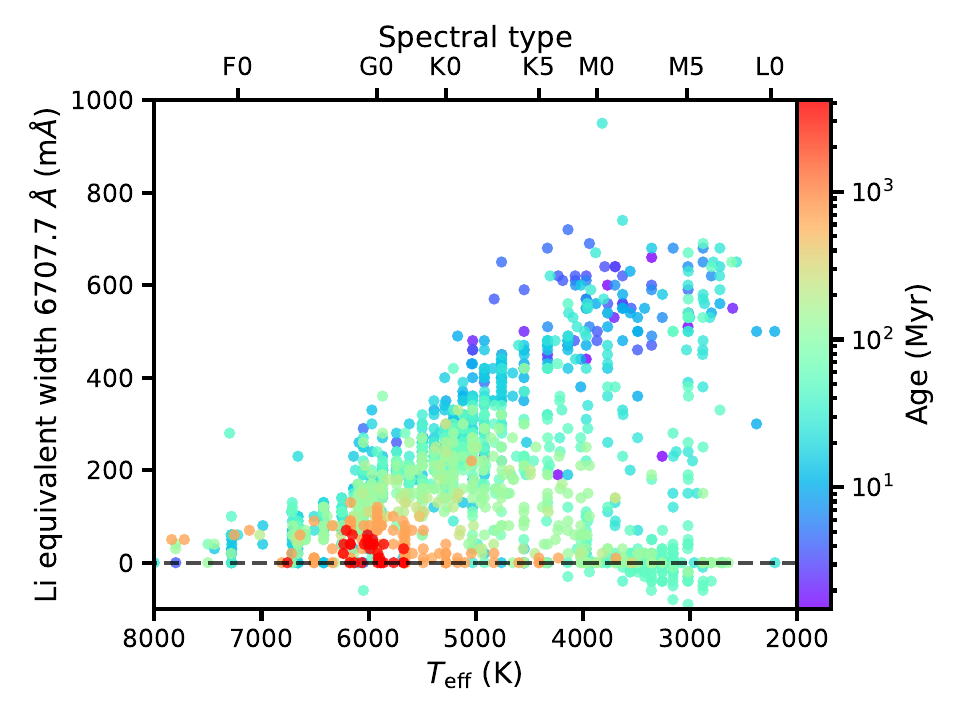}\label{fig:li_db}}
	\subfigure[H$\alpha$]{\includegraphics[width=0.49\textwidth]{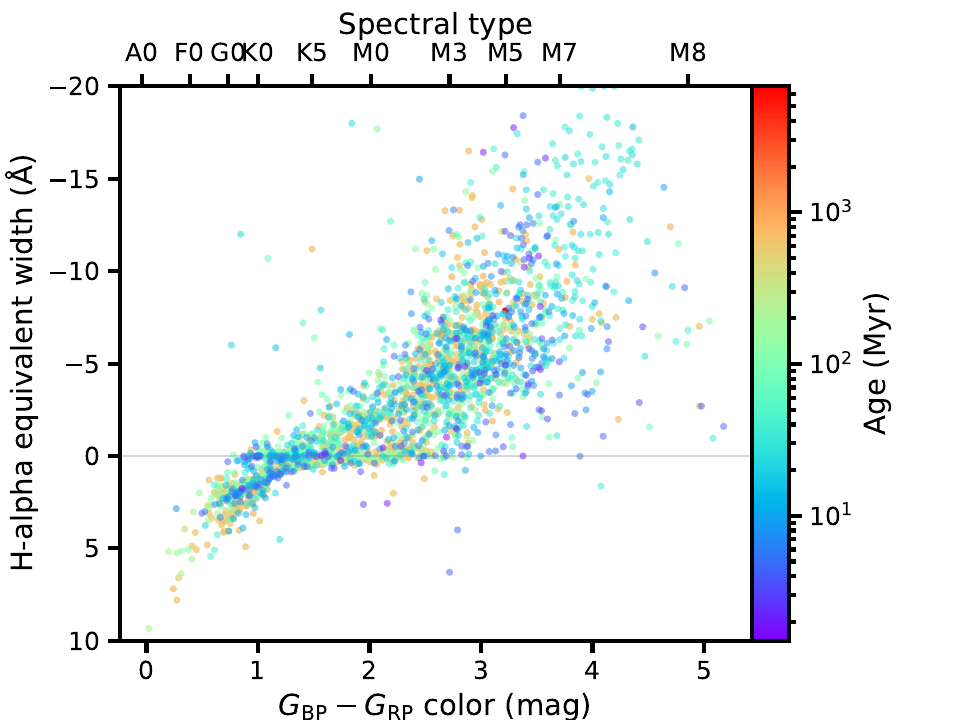}\label{fig:ha_db}}
	\subfigure[Mount Wilson $\log R^\prime_{\rm HK}$]{\includegraphics[width=0.49\textwidth]{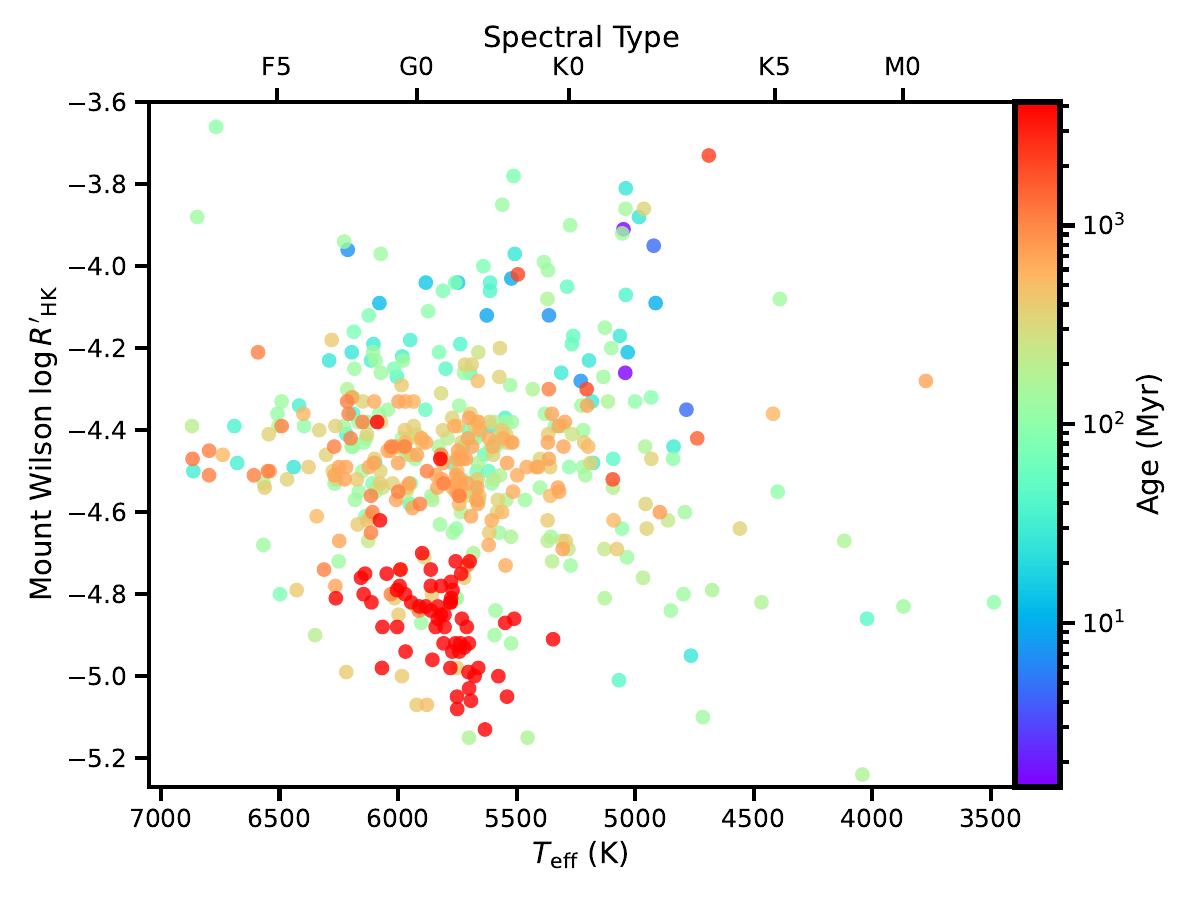}\label{fig:rhk_db}}
    \subfigure[Projected rotational velocities]{\includegraphics[width=0.49\textwidth]{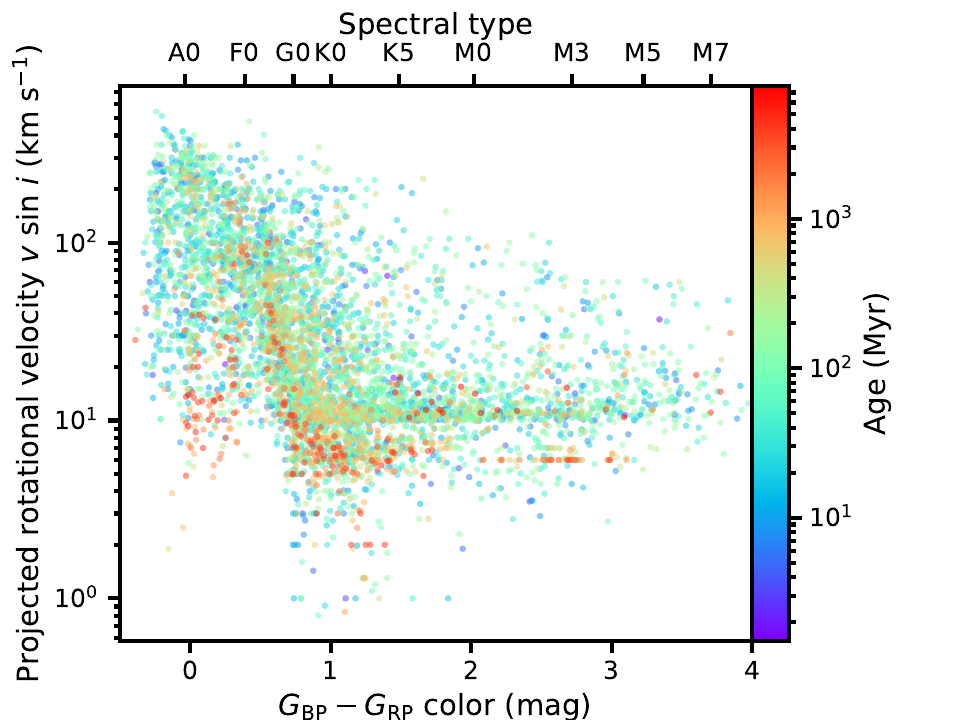}\label{fig:vsini}}
	\caption{Various age-dependent measurables available in MOCAdb as a function of stellar ages based on membership. These data sets will be useful as comparison sequences to determine the ages of non-members. See Section~\ref{sec:age_prop} for more details.}
	\label{fig:li_ha}
\end{figure*}

\begin{figure*}
	\centering
	\subfigure[Gaia activity index]{\includegraphics[width=0.49\textwidth]{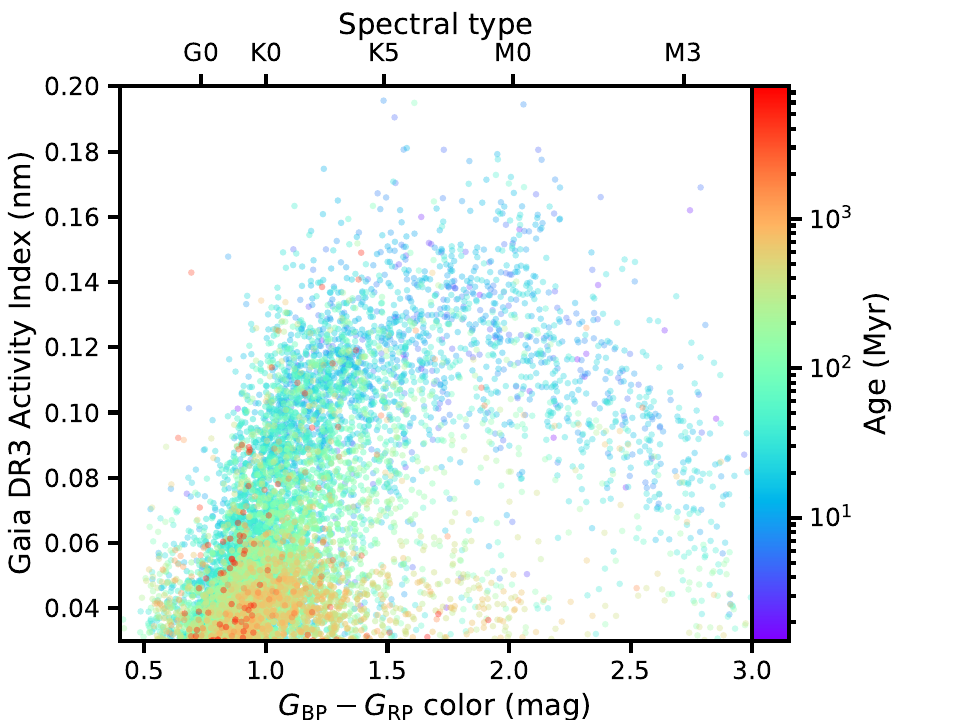}\label{fig:gaia_index}}
    \subfigure[Rotation periods]{\includegraphics[width=0.49\textwidth]{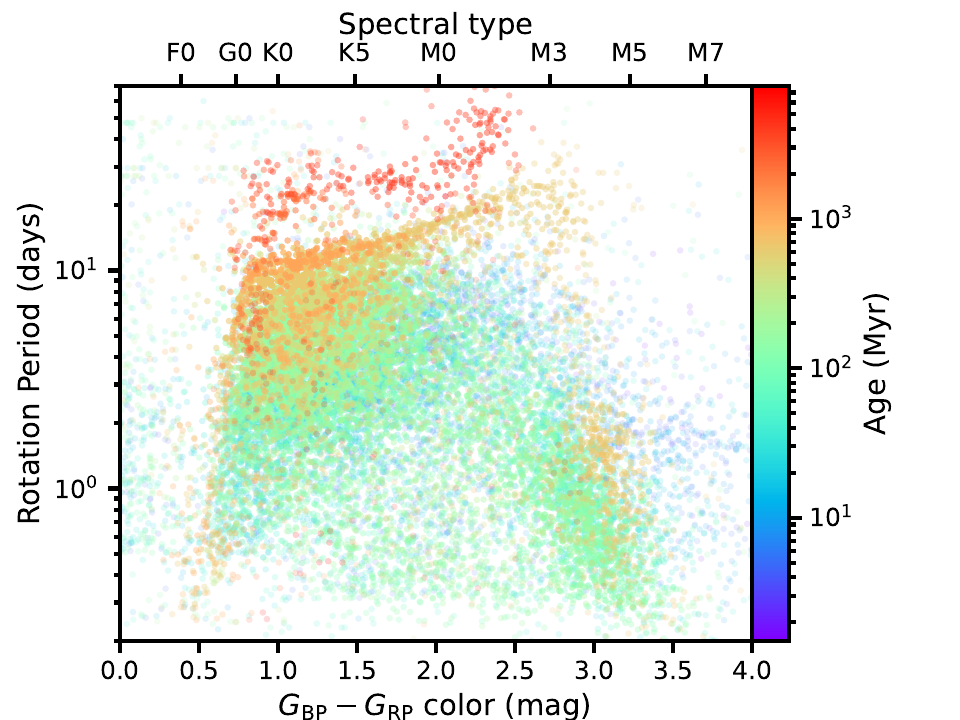}\label{fig:prot}}
    \subfigure[GALEX NUV]{\includegraphics[width=0.49\textwidth]{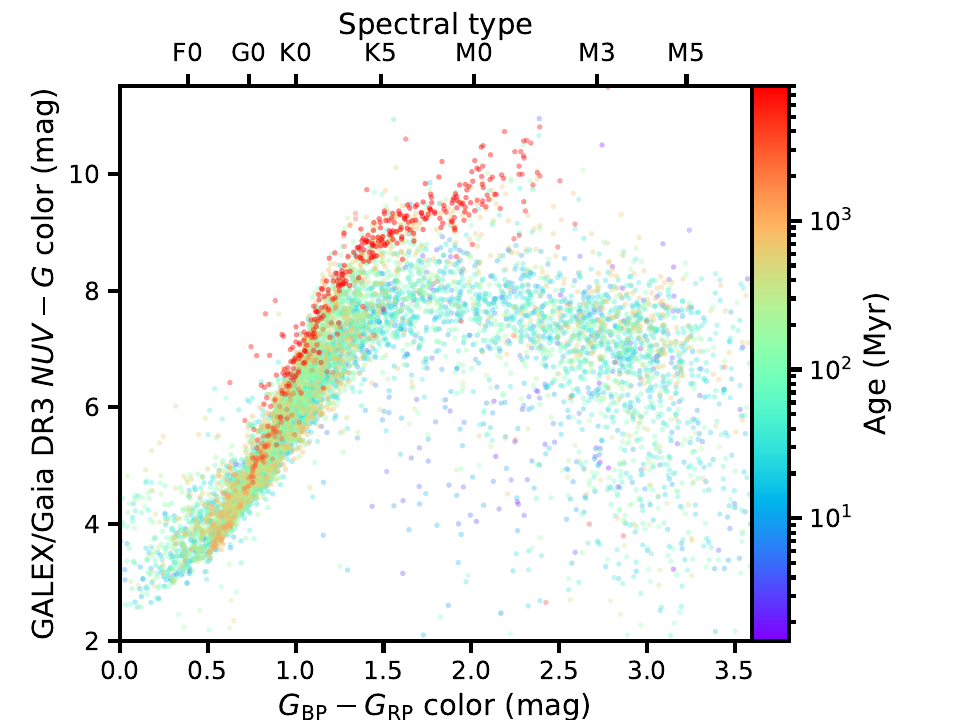}\label{fig:nuv_seq}}
    \subfigure[ROSAT $L_{\rm X}$]{\includegraphics[width=0.49\textwidth]{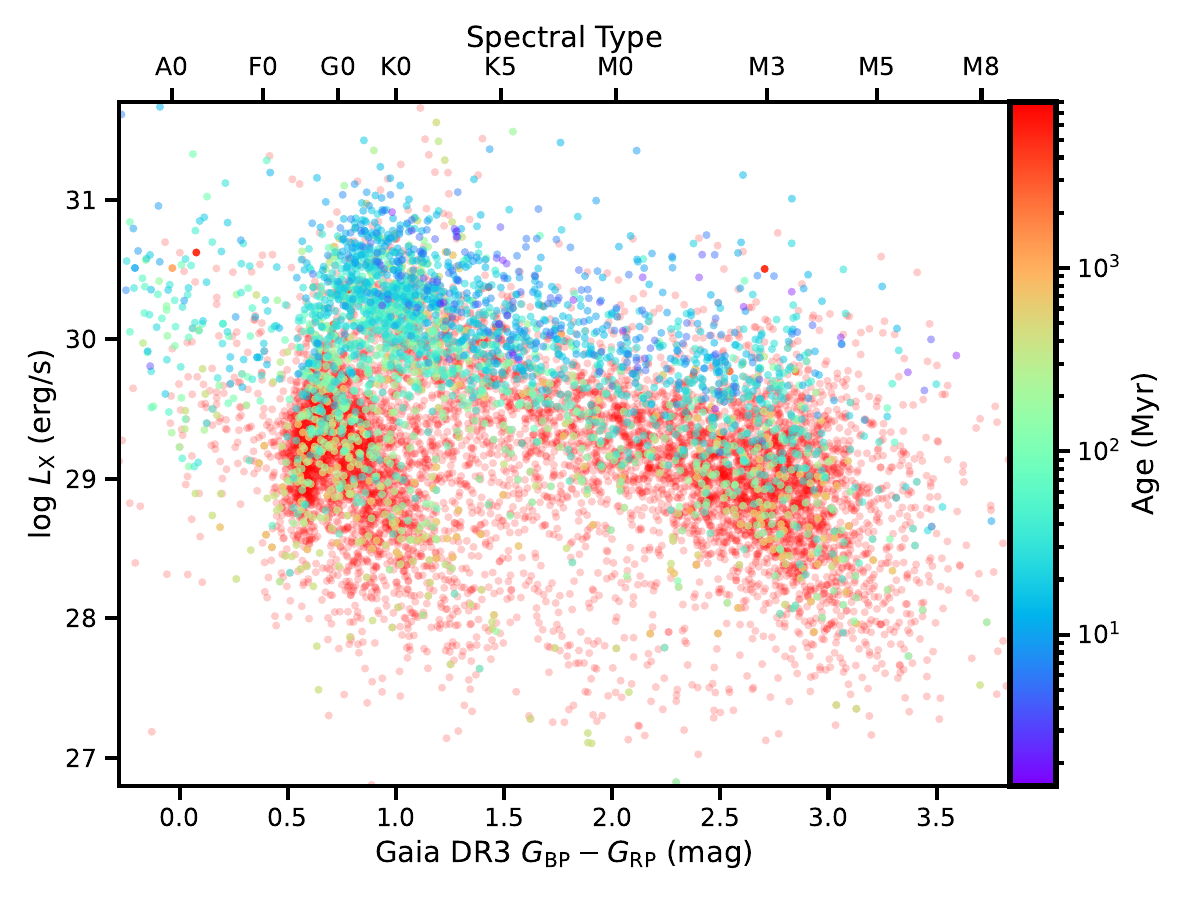}\label{fig:xray_db}}
	\caption{Various age-dependent measurables available in MOCAdb as a function of stellar ages based on membership. A transparency that favors high-density regions in ($x$,$y$,$\log {\rm age}$) space was used to improved visibility. These data sets will be useful as comparison sequences to determine the ages of non-members. See Section~\ref{sec:age_prop} for more details.}
	\label{fig:nuv_prot}
\end{figure*}

\begin{figure*}
	\centering
	\subfigure[VO$_z$]{\includegraphics[width=0.49\textwidth]{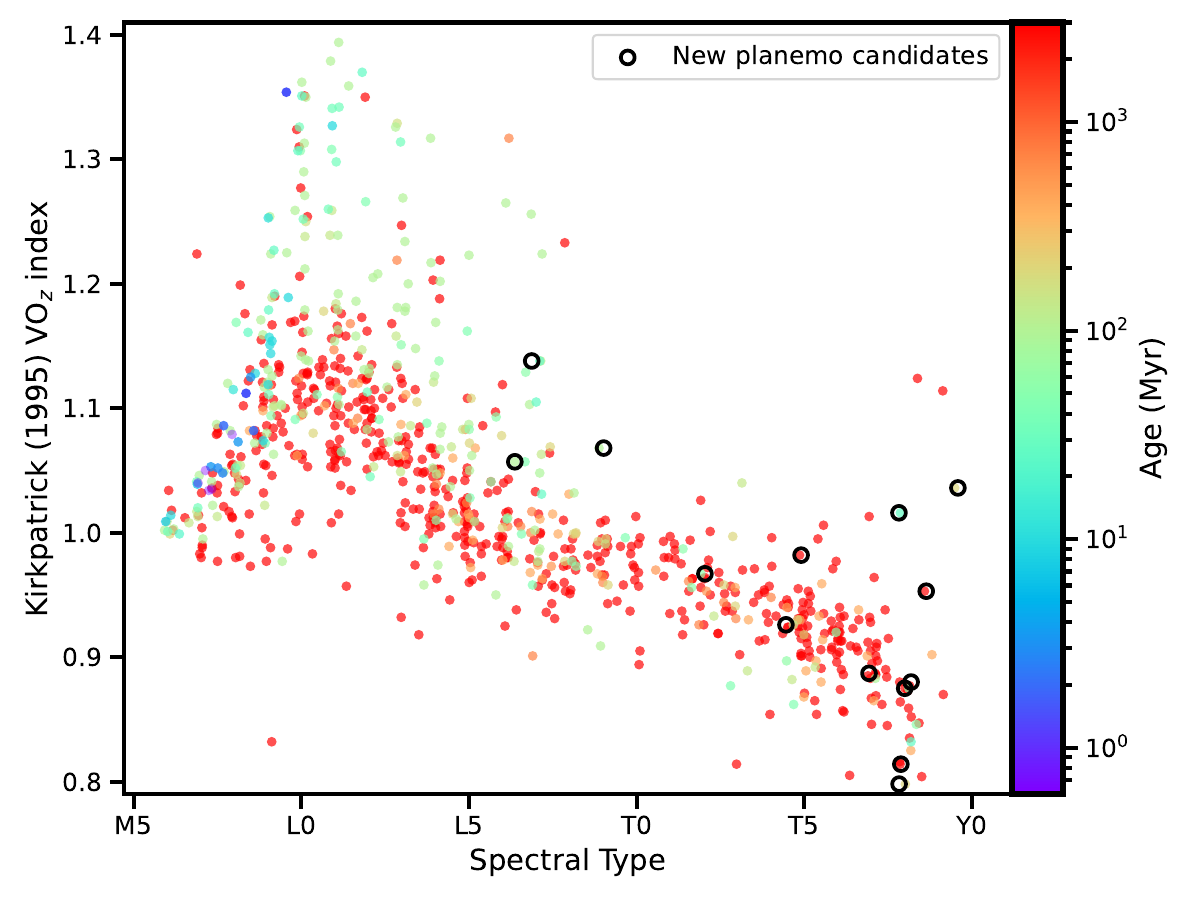}\label{fig:voz}}
    \subfigure[KI$_J$]{\includegraphics[width=0.49\textwidth]{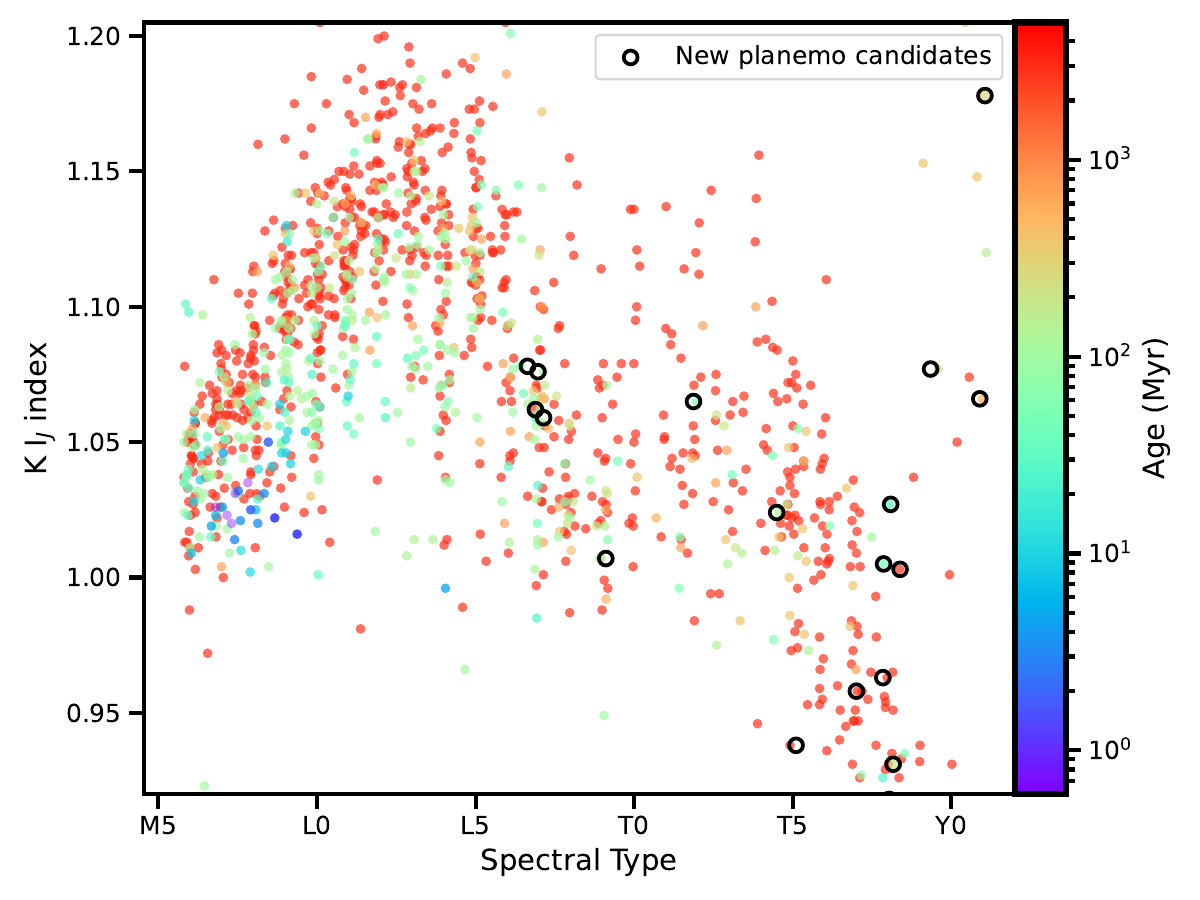}\label{fig:kij}}
    \subfigure[$H-$cont]{\includegraphics[width=0.49\textwidth]{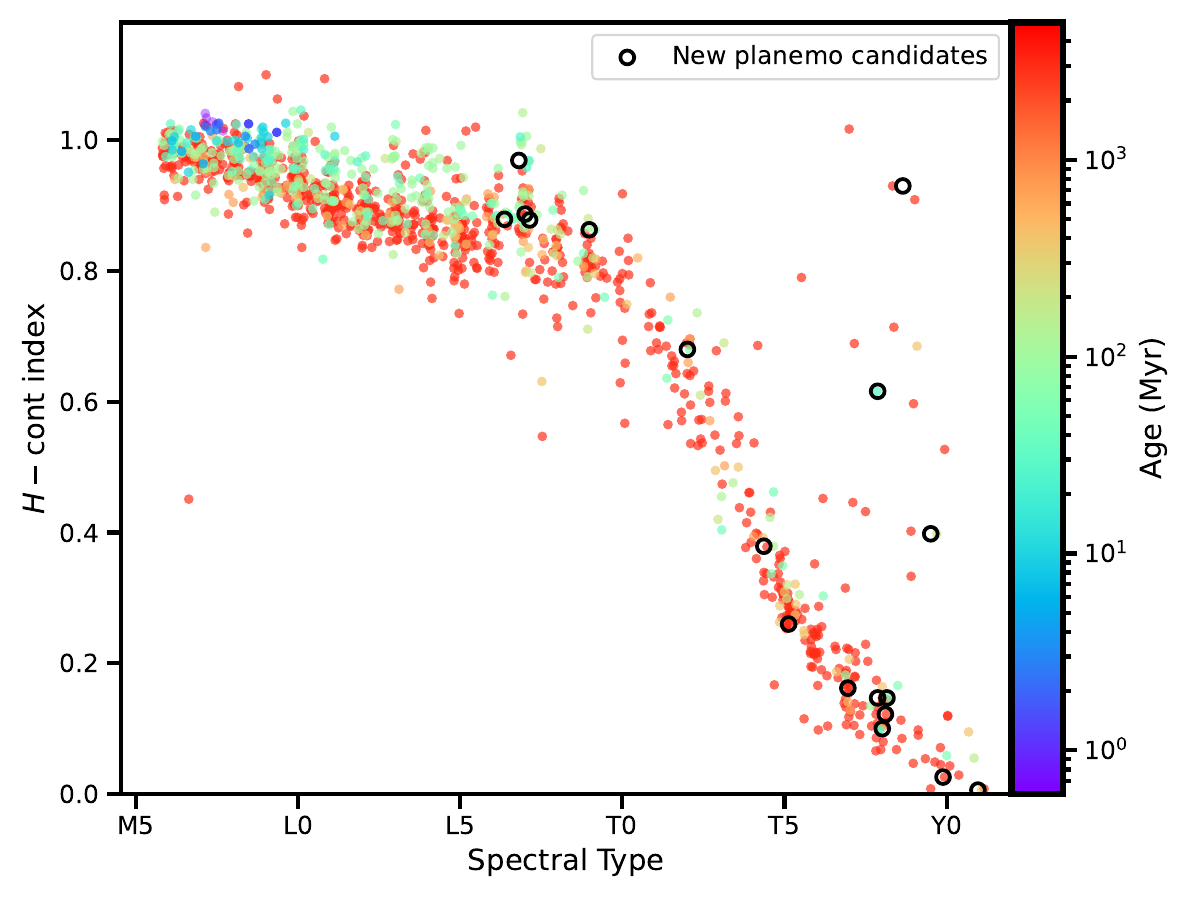}\label{fig:hcont}}
    \subfigure[FeH$_z$]{\includegraphics[width=0.49\textwidth]{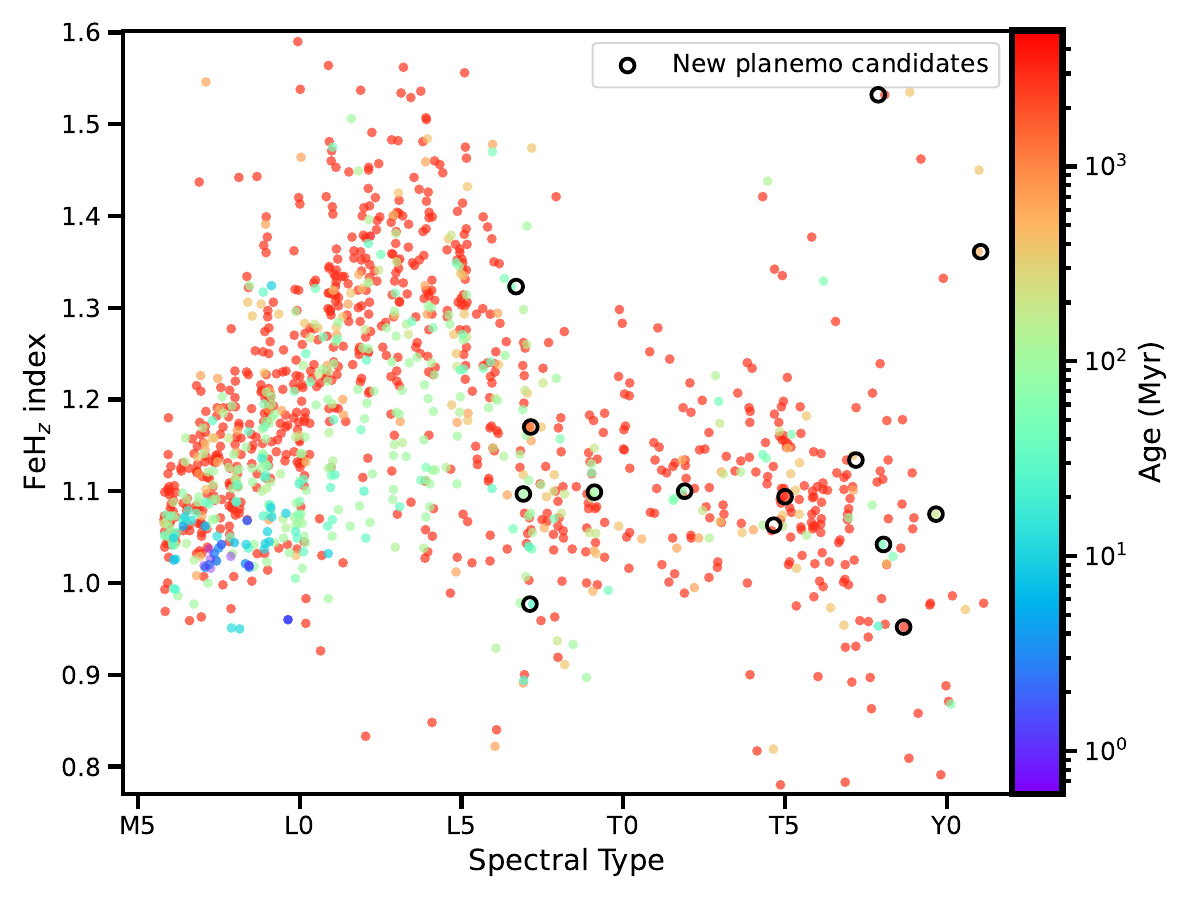}\label{fig:fehz}}
	\caption{Various surface gravity-dependent spectral indices (see e.g. \citealp{2013ApJ...772...79A}) computed in MOCAdb from the publicly available near-infrared spectra of substellar objects in MOCAdb as a function of stellar ages based on membership. For substellar objects without a known membership, we assumed an age of 3\,Gyr or 200\,Myr when a spectroscopic sign of low-gravity is noted in the literature, for display purposes. See Section~\ref{sec:age_prop} for more details.}
	\label{fig:spindex_bd}
\end{figure*}

In Figures~\ref{fig:li_ha}, \ref{fig:nuv_prot} and \ref{fig:spindex_bd}, we provide a preview of several age-dependent diagnostic measurements in the MOCA database. These data sets will provide valuable comparison sequences as well as training sets to build future age-dating tools that will allow inverting either a single age-dependent diagnostic measurement or a combination of them into an age estimation for a single star.

We also provide additional age probability density functions for several MOCAdb associations and their individual stellar members based on publicly available Bayesian age-dating Python algorithms:

\begin{itemize}
    \item \textbf{ChronoFlow} is an empirical gyrochronology model constructed in Gaia~DR3 color versus rotation period space based on normalizing flows \citep{2025ApJ...986...59V};
    \item \textbf{EAGLES v2} estimates ages using an artificial neural network trained on reference lithium equivalent width sequences \citep{2023eas..conf..193J,2024MNRAS.534.2014W};
    \item \textbf{BAFFLES} estimates ages based on empirical sequences in $R^\prime_{\rm HK}$ activity indices versus $B-V$ color for stars with spectral types as late as early M dwarfs (for those with available $B-V$ colors) \citep{2020ApJ...898...27S};
    \item The \cite{2023ApJ...954L..50E} relations similarly allow to determine Bayesian age posteriors based on empirical sequences of $R^\prime_{\rm HK}$ activity versus age, for spectral types M0--M2 and M2.5--M6.5, in cases where $B-V$ colors are not available;
    \item \textbf{EVA} uses an empirical relationship between the excess Gaia~DR3 photometric error and young association ages established by \cite{2023ApJ...953..127B}. The excess photometric error was established by \cite{2021ApJ...912..125G} as a proxy for variability, which explains its correlation with age.
\end{itemize}

In most cases, age probability density functions were constructed for each individual star and age-dating method. The only exception is EVA, where the age calibration is based on the 90th centile of an association's excess error distribution, meaning that only a single combined probability density function is generated for each young association.

The ages obtained from these methods are stored in the respective tables \texttt{data\_object\_ages}, \texttt{calc\_object\_age\_pdfs}, \texttt{data\_association\_ages} and \texttt{calc\_association\_age\_pdfs}. A more extensive characterization, including additional age-dating methods, will be presented in future work.

\subsection{Exoplanets}\label{sec:exoplanet}

\begin{figure*}
	\centering
	\subfigure[All exoplanet masses]{\includegraphics[width=0.49\textwidth]{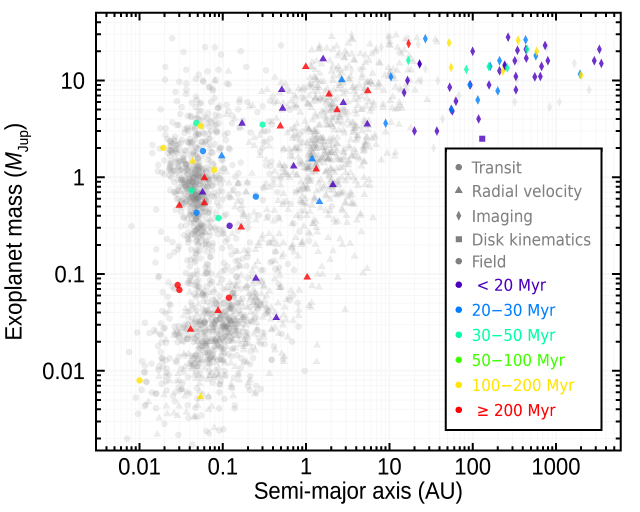}\label{fig:exo_mass}}
	\subfigure[Transiting exoplanet radii]{\includegraphics[width=0.49\textwidth]{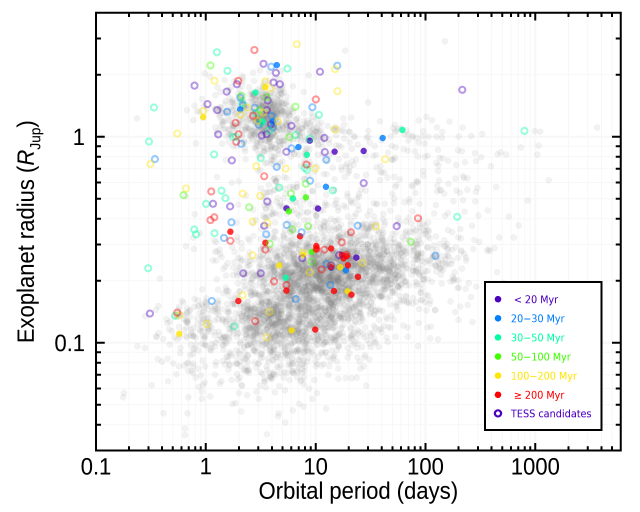}\label{fig:exo_rad}}
	\caption{Population properties of confirmed exoplanet systems listed in the NASA Exoplanet Archive (left panel and filled symbols in right panel) and the current TESS candidate exoplanets (empty circles in right panel). Systems that are members of a young association are indicated with a color and symbol that corresponds to the host association's age. See Section~\ref{sec:exoplanet} for more details.}
	\label{fig:exo}
\end{figure*}

The latest version of the NASA Exoplanet Archive was imported in the MOCAdb in order to identify exoplanet companions to members or candidate members of age-calibrated associations, shown in Figure~\ref{fig:exo}. This figure outlines the recently growing, albeit still sparse, set of known exoplanets with well-calibrated ages determined from membership in a young association.

We identify a current total of \nyoungexo\ confirmed exoplanet systems that may be members of age-calibrated associations, in addition to \ntessexo\ TESS \citep{2015JATIS...1a4003R} exoplanet candidates that are not yet confirmed. Among these, \nnewyoungexo\ confirmed exoplanets and \nnewtessexo\ TESS exoplanet candidates appear to be identified as candidate members of a young \added{association} for the first time here. The set of age-calibrated confirmed exoplanets and exoplanet candidates are listed in Tables~\ref{tab:exoplanets} and \ref{tab:exoplanet_candidates}, respectively. Focused searches for exoplanets around members of young associations (e.g., \citealp{2019ApJ...880L..17N,2023AJ....165...85W,2024AJ....168...41T,2025AJ....169..166D}) will enable population-level studies of how exoplanet properties evolve over time in the near future.

\subsection{Brown Dwarfs and Isolated Planetary-Mass Objects}\label{sec:planemos}

The MOCA database allows us to conveniently refresh the list of potential isolated planetary-mass objects that currently stand the membership test based on updated BANYAN~$\Sigma$ models.

In 19 cases where reliable substellar proper motions were not available, we used individual detections in the near-infrared and red-optical catalogs available in MOCAdb to calculate proper motions \citep{2006AJ....131.1163S,2010AJ....140.1868W,2021ApJS..253....8M,2021AJ....161..192N,2016arXiv161205560C,2018MNRAS.473.5113D,2013Msngr.154...35M,2005Msngr.119...56L,2023AJ....165...36M}. These cases are listed in Table~\ref{tab:substellar}.

\begin{figure*}
 	\centering
 	\includegraphics[width=0.965\textwidth]{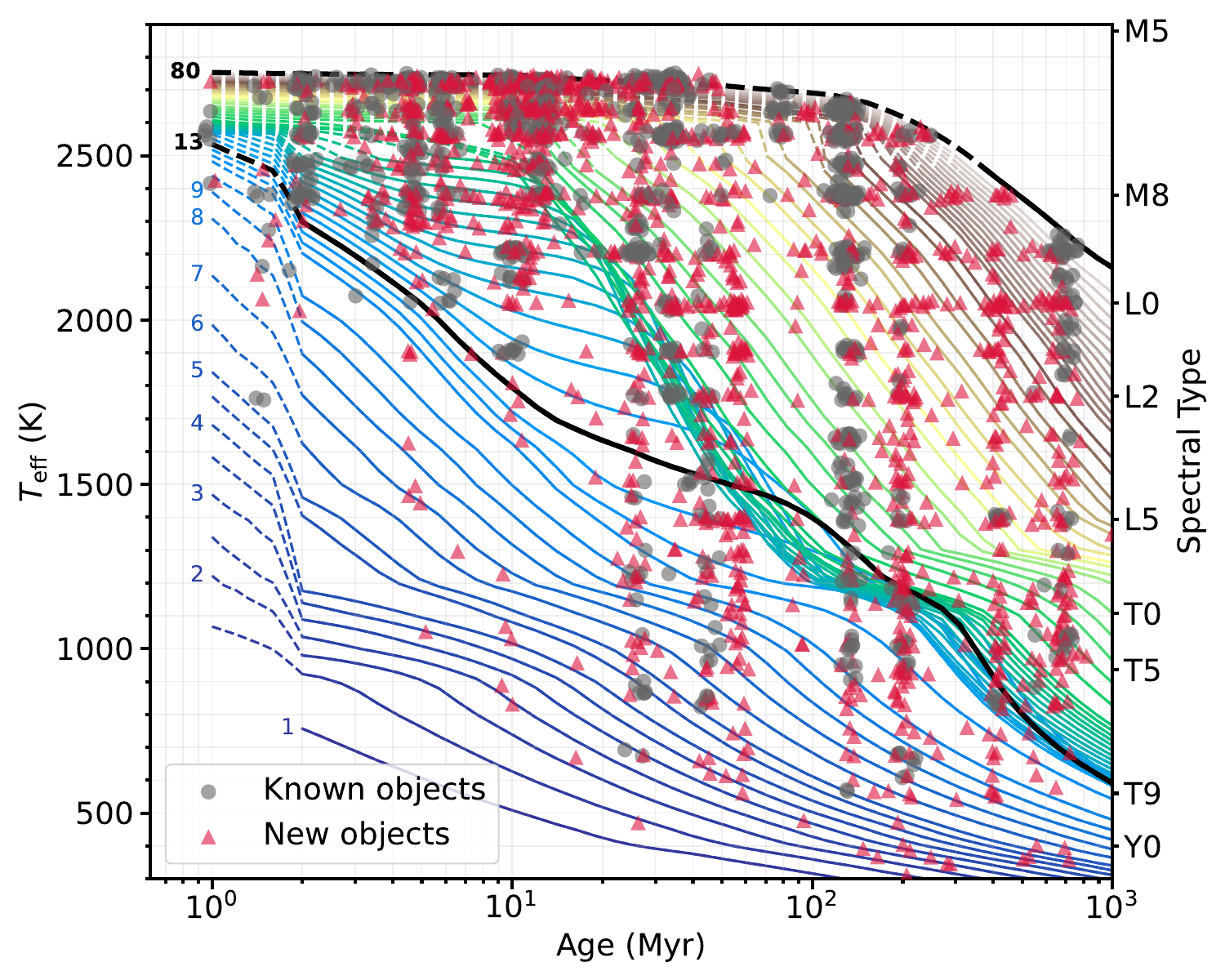}
 	\caption{Age and \teff\ of the current census of candidate age-calibrated substellar objects listed in Table~\ref{tab:substellar}, in cases where the host association currently benefits from an age estimate. Known candidate members of young association from the literature are shown as grey circles, and new candidate members are shown as red triangles. The \cite{2024ApJ...975...59M} isomass tracks are displayed in the background with thick lines, and their extensions onto the empirical mass tracks of Section~\ref{sec:masses} are shown as dashed lines. The 13\,\mjup\ and 80\,\mjup\ tracks highlighted in black, and model track masses are labeled in \mjup. A horizontal jitter of 0.02\,dex in age was added for clarity, as nearby associations have ages clustered around few distinct values. The isomass tracks overlap significantly around 13\,\mjup, significantly complicating reliable estimates around the planetary-mass regime boundary. This figure outlines that the age and \teff\ coverage of age-calibrated substellar objects has improved significantly in recent years, allowing for detailed studies of substellar atmospheres in a range of parameters. See Section~\ref{sec:planemos} for more details.}
 	\label{fig:substellar_tracks}
\end{figure*}

It was recently demonstrated that the impact of clouds is non-negligible on the cooling tracks of the coolest-mass substellar objects \citep{2024ApJ...975...59M}, causing a flattening, overlap and even crossings of isomasses near the planetary-mass boundary (13\,\mjup) in a temperature--age diagram (see Figure~\ref{fig:substellar_tracks}). Because of this, estimating the masses of isolated planetary-mass objects is not straightforward. While the \cite{2001RvMP...73..719B} evolutionary tracks used in Section~\ref{sec:masses} are appropriate for estimating the small impact of gravitational redshift on the heliocentric radial velocities of substellar objects, a more careful analysis is required to estimate which substellar objects have masses comparable to gas giant exoplanets.

We used a method based on the tracks of \citep{2024ApJ...975...59M} extended onto the empirical tracks of Section~\ref{sec:masses} by log-interpolation to achieve this. We converted the best-available spectral type of each object M6 or later in MOCAdb to an effective temperature with a $10^4$--element Monte Carlo and the \teff--spectral type sequence described in Section~\ref{sec:teff}\footnote{The probability density function representing the spectral type measurement is assumed to be Gaussian, but the resulting \teff\ is not assumed to be.}, and combined it with random draws from an asymmetrical log-normal distribution for the age (using asymmetrical age measurement errors when they are available). These random realizations in (\teff, age) space are then used to construct a two-dimensional kernel density estimate using Python's \texttt{scipy.stats.gaussian\_kde} \citep{1992mde..book.....S}, along which every isomass track is integrated to produce a probability for that specific mass. This was repeated for every isomass track to determine a mass probability density function, which is then translated to an estimate and asymmetric error bars by sampling the 15.9\%, 50\%, and 84.1\% centiles of the cumulative distribution function.

It is interesting to note that we revise the estimated mass of the Carina-Near member SIMP~J013656.5+093347.3 upwards at $17_{-5}^{+7}$\,\mjup\ (a $\approx$\,20\%\ increase), and revise the mass of the younger TW~Hya association member 2MASS~J12074836--3900043 (TWA~40) down to $14_{-2}^{+3}$\,\mjup\ (a 7\%\ decrease), illustrating the role that clouds play in pushing the masses of older planetary-mass candidates in particular to higher values. Furthermore, \cite{2015ApJS..219...33G} discussed an intriguing pile-up of objects with estimated masses around 13\,\mjup\ in the young association Tucana-Horologium, composed of a dozen objects with early-L spectral types. Although most of them still survive as likely members of this population, their masses are revised upward in the range 20--30,\mjup, which means that the previously observed piling-up at 13\,\mjup\ was likely an artifact of earlier cooling tracks that underestimated the masses of low-mass brown dwarfs at $\approx$40\,Myr.

We provide in Table~\ref{tab:substellar} a compilation of all currently known substellar objects that may belong to a nearby young association, based on updated BANYAN~$\Sigma$ membership probabilities, along with their estimated masses. Membership probabilities include radial velocities when they are available, trigonometric distances when available, and otherwise photometric distances (the absolute $K$-band magnitude versus $W1-W2$ color sequence is used when possible, otherwise the absolute $W2$-band magnitude versus $W1-W2$ sequence is used).

We catalog a total of \nyoungsubstellar\ potentially young substellar objects, \nyoungucd\ of those with spectral types L0 or later. About \nnewyoungucd\ of the L0 and later-type young substellar objects are newly identified, mostly in recently discovered or poorly studied nearby young associations. It is noteworthy that a number of young brown dwarfs appear to match the recently discovered tidal tails of Coma~Ber and the Hyades, and the corona of IC~2391. We find \ntotalplanemos\ objects whose central mass estimates fall in the planetary-mass regime, \nnewplanemos\ of which are newly recognized here.

We provide a list of \nrejectedplanemos\ previously suggested isolated planetary-mass candidates which memberships appear to be rejected by BANYAN~$\sigma$ in Table~\ref{tab:rej_planemos}. We caution that this could be subject to change if the measurement inputs are refined or if the BANYAN~$\Sigma$ models do not capture the full details of the spatial shapes of the respective associations.

It is important to note that the census of isolated planetary-mass objects presented here ignores some of the coldest Y dwarfs that do not benefit from a precise age estimation, as we focused here on age-calibrated objects. A number of Y dwarfs are so cold that, even at the age of the Universe, have estimated masses below 13\,\mjup\ (see e.g., \citealp{2014ApJ...786L..18L,2025ApJ...979..145L}).

Some of the newly identified candidate planetary-mass objects are of particular note. For example, the $\approx$\,T9.5 candidate brown dwarf CWISEP~J053644.82--305539.3 identified by \cite{2011ApJS..197...19K} benefits from a parallax and proper motion (but no spectrum), and could potentially be a $\approx$\,2\,\mjup\ isolated planetary-mass object in the $\beta$~Pic moving group, if corroborated with a spectral analysis and radial velocity measurement.

The Y0 CWISE~J201146.50--481259.8 from \cite{2020ApJ...889...74M} is similarly interesting, and may be a $\approx$\,4\,\mjup\ member of the $\approx$\,210\,Myr-old MELANGE--5 association of \cite{2024AJ....168...41T}. Our updated census of isolated planetary-mass objects includes a few more candidates with very low masses below $\approx$\,6\,\mjup\ that warrant further characterization, as very few isolated objects are currently known with such low masses.

The T1.5 PSO~J159.2399--26.3885 \citep{2015ApJ...814..118B} is also of particular interest as it is flagged here as a $\approx$\,5\,\mjup\ candidate member of the $\approx$\,10\,Myr TW~Hya association \citep{2015MNRAS.454..593B}. \citep{2015ApJ...814..118B} note that this object satisfies four of the six \cite{2010ApJ...710.1142B} spectroscopic binary classification criteria, making it a strong candidate substellar unresolved binary of unequal \teff, but they did not note anything else that may make it an immediately compelling low-gravity object. If confirmed, this would be one of the youngest known early-T objects, and it would make it a valuable benchmark to better understand the L/T transition at young ages.

\section{THE MOCAdb WEBSITE}\label{sec:website}

Most of the database content can be consulted at the \url{https://mocadb.ca} website with a simple user interface and a series of \texttt{Python} dash applications, allowing users to access distinct aspects of the database:
\begin{itemize}
    \item \textbf{Database Reports} on a single star or young association, listing a compilation of data relevant to this specific object;
    \item \textbf{The MOCAdb Explorer} allows users to display members of several young associations across different graphics (color-magnitude diagrams, spatial-kinematic positions, Lithium and rotation periods);
    \item \textbf{The Spatial-Kinematic Explorer} allows users to any three combinations of the $XYZUVW$ axes for members of specific associations with the relevant BANYAN~$\Sigma$ models;
    \item \textbf{The Substellar Photometry Explorer} allows users to view any combinations of spectral types, colors, absolute magnitudes, spectral indices or equivalent widths for substellar objects;
    \item \textbf{The Spectral Explorer} allows users to quickly view a spectrum stored in MOCAdb;
    \item \textbf{The Spectral Typing Tool} allows users to assign spectral types to substellar objects which spectra are stored in MOCAdb;
    \item \textbf{The Astrometric Explorer} allows users to view the astrometry of any object versus epoch and perform a quick fit of its proper motion and/or parallax;
    \item \textbf{Age PDF Explorers} allow users to view the individual age probability density functions for a given star or association, for each calculation method, with an optimal combined PDF across all methodste;
    \item \textbf{A Sunburst graph} of the hierarchical structure of the MOCAdb associations with a dynamic interface.
\end{itemize}

Users can also directly communicate with the database contents using \texttt{Python} \texttt{pandas} dataframes with the publicly available \texttt{mocapy} package\footnote{Available at \url{https://github.com/jgagneastro/mocapy}}.

For more advanced queries, users can directly communicate with the MOCAdb MySQL server using a MySQL client with the following credentials:

\begin{verbatim}
MySQL database  : mocadb
MySQL host      : 104.248.106.21
MySQL username  : public
MySQL password  : z@nUg_2h7_%?31y88
\end{verbatim}

\section{CONCLUSIONS}\label{sec:conclusion}

We presented the Montreal Open Clusters and Associations (MOCA) MySQL database, systematically compiling all known stellar associations and their members within 500\,pc, in addition to all known nearby substellar objects. A number of literature properties (e.g., proper motions, radial velocities, parallaxes, photometry, rotation periods) are compiled, combined into high-precision averages, or used as a basis for additional calculations (e.g. $UVW$ Galactic space velocities).

We used this updated census of nearby young associations with an updated Gaussian Mixture Model representation, based on automated outlier rejection and extreme deconvolution, to update the Bayesian membership classification tool BANYAN~$\Sigma$, allowing its models to capture much more complex shapes of nearby young associations, and updated its total number of modeled associations to \bsigmaassociations. Using this updated tool, we have identified \newcandidates\ new candidate members of nearby associations that were previously overlooked.

The MOCAdb is publicly accessible at \url{https://mocadb.ca}, through direct MySQL clients, or through the \texttt{mocapy} Python package, and allows users to directly interface with the data in a number of ways. The MOCAdb will enable the rapid construction of custom samples for a wide range of uses, such as sample construction for telescope follow-up or exoplanet searches, the construction of training sets for future age-dating tools, the population-level study of open cluster properties, or the search for isolated planetary-mass objects in nearby, young associations.

The MOCAdb will be updated continuously as new data become available, and as new features are needed\footnote{\added{The state of the MOCAdb at the moment of publication is available as a frozen version of CSV files on Zenodo at \url{https://zenodo.org/uploads/18166118} with DOI \texttt{10.5281/zenodo.18166118}.}}. The combination of MOCAdb with future data sets will enable the efficient identification of many more age-calibrated exoplanets. The fourth data release of Gaia will not only greatly improve the astrometric and radial velocity accuracy of stellar members especially at the low-mass end. Its enabling of large-scale searches for astrometric exoplanets will also greatly improve our census of exoplanets in age-calibrated associations cataloged in this work.

Combining future missions that enable large-scale discovery and characterization of substellar objects with MOCAdb will also enable the discovery of many new isolated planetary-mass objects. For example, the Vera Rubin Observatory will provide parallaxes for thousands of substellar objects that are too faint for Gaia astrometry \citep{2022ApJS..263...23G}, the Euclid space mission's Wide Survey \citep{2011arXiv1110.3193L} will allow the direct spectroscopic identification of many substellar objects, and its Deep Survey will even enable parallax measurements for additional substellar objects \citep{2024AA...686A.171Z} at further distances.

\begin{acknowledgements}

This project was developed in part at the Gaia FÃªte, hosted by the Flatiron Institute Center for Computational Astrophysics in 2022 June. This work was carried out partially with a Banting grant from the Natural Sciences and Engineering Research Council of Canada (NSERC). This work has benefited from The UltracoolSheet \citep{Best20US} at \url{http://bit.ly/UltracoolSheet}, maintained by Will Best, Trent Dupuy, Michael Liu, Rob Siverd, and Zhoujian Zhang, and developed from compilations by \cite{2012ApJS..201...19D}, \cite{2013Sci...341.1492D}, \cite{2016ApJ...833...96L}, \cite{2018ApJS..234....1B}, and \cite{2021AJ....161...42B}.

This research made use of: the SIMBAD database and VizieR catalog access tool, operated at the Centre de Donn\'ees astronomiques de Strasbourg, France \citep{2000AAS..143...23O}; data products from the Two Micron All Sky Survey (\emph{2MASS}; \citealp{2006AJ....131.1163S}), which is a joint project of the University of Massachusetts and the Infrared Processing and Analysis Center (IPAC)/California Institute of Technology (Caltech), funded by the National Aeronautics and Space Administration (NASA) and the National Science Foundation \citep{2006AJ....131.1163S}; data products from the \emph{Wide-field Infrared Survey Explorer} (\emph{WISE}; and \citealp{2010AJ....140.1868W}), which is a joint project of the University of California, Los Angeles, and the Jet Propulsion Laboratory (JPL)/Caltech, funded by NASA; the WEBDA database, operated at the Department of Theoretical Physics and Astrophysics of the Masaryk University; and the Unified Cluster Catalogue (UCC; \citealt{2023MNRAS.526.4107P}). This work has made use of the SIMPLE Archive of low-mass stars, brown dwarfs, and directly imaged exoplanets: \texttt{10.5281/zenodo.13937301}.

The Digitized Sky Surveys (DSS) were produced at the Space Telescope Science Institute under U.S. Government grant NAG W-2166. The images of these surveys are based on photographic data obtained using the Oschin Schmidt Telescope on Palomar Mountain and the UK Schmidt Telescope. The plates were processed into the present compressed digital form with the permission of these institutions. The Second Palomar Observatory Sky Survey (POSS-II) was made by the California Institute of Technology with funds from the National Science Foundation, the National Geographic Society, the Sloan Foundation, the Samuel Oschin Foundation, and the Eastman Kodak Corporation. The Oschin Schmidt Telescope is operated by the California Institute of Technology and Palomar Observatory.

This work presents results from the European Space Agency (ESA) space mission Gaia. Gaia data are being processed by the Gaia Data Processing and Analysis Consortium (DPAC). Funding for the DPAC is provided by national institutions, in particular the institutions participating in the Gaia MultiLateral Agreement (MLA). The Gaia mission website is https://www.cosmos.esa.int/gaia. The Gaia archive website is https://archives.esac.esa.int/gaia. The Digitized Sky Surveys were produced at the Space Telescope Science Institute under U.S. Government grant NAG W-2166. This research has benefitted from the SpeX Prism Spectral Libraries, maintained by Adam Burgasser at \url{https://cass.ucsd.edu/~ajb/browndwarfs/spexprism}. 
Part of this research was carried out at the Jet Propulsion Laboratory, California Institute of Technology, under a contract with the National Aeronautics and Space Administration (80NM0018D0004).
\end{acknowledgements}

\software{BANYAN~$\Sigma$ \citep{2018ApJ...856...23G}, EAGLES~v2 \citep{2023eas..conf..193J,2024MNRAS.534.2014W}, BAFFLES \citep{2020ApJ...898...27S}, ChronoFlow \citep{2025ApJ...986...59V}, EVA \citep{2023ApJ...953..127B}.}


\appendix

In this section, we provide a few examples of MySQL queries that can be used in the MOCAdb. Comments shown directly in the MySQL code blocks are preceded with the \texttt{--} symbols, which are ignored by MySQL upon execution.

In the first example, a user might aim to list all membership claims to the $\beta$~Pic moving group (\texttt{moca\_aid} = \texttt{BPMG}) for objects with spectral types T0 or later. This would be achieved by starting from the \texttt{data\_spectral\_types} table to filter on spectral types, and joining the resulting rows with the \texttt{data\_memberships} table, which contains an exhausting list of all association members, with the associated reference and type of membership. The potentially multiple membership claims related to a single object will be combined using a MySQL \texttt{GROUP BY} statement.

\lstset{basicstyle=\ttfamily, tabsize=4}
\begin{lstlisting}[language=SQL, frame=single]
SELECT
   -- Main object information
    mo.designation, mo.ra, mo.dec,
	-- Spectral type information
	spt.spectral_type, spt.moca_pid AS spt_ref,
	-- Membership information, stacked by moca_oid
	GROUP_CONCAT(CONCAT(dm.moca_mtid,' (',dm.moca_pid,')')) AS all_mem
-- Start from spectral types table to limit the spectral type range
FROM data_spectral_types spt
-- Join the main objects table to get the main designation
JOIN moca_objects mo USING(moca_oid)
-- Join the non-rejected memberships in BPMG
-- This will duplicate some rows for multiple membership claims
JOIN data_memberships dm ON(
    dm.moca_oid=spt.moca_oid AND
    dm.moca_mtid != 'R' AND
    dm.moca_aid='BPMG')
WHERE
	-- Only consider the best-available spectral type
	spt.adopted=1 AND
	-- Ignore spectral type estimates based on photometry
	spt.photometric_estimate=0 AND
	-- Limit the search to T0 or later (numeric spt = 20)
	spt.spectral_type_number >= 20
-- Combine the duplicated rows caused by multiple membership claims
GROUP BY spt.moca_oid;
\end{lstlisting}

\newpage
If a user would instead aim to list the objects which are currently categorized as high-probability candidate members of $\beta$PMG using the latest version of the BANYAN~$\Sigma$ models with all available observables, regardless of literature membership claims, the following query would achieve this:

\begin{lstlisting}[language=SQL, frame=single]
SELECT
    -- Main object information
    mo.designation, mo.ra, mo.dec,
	-- Spectral type information
	spt.spectral_type, spt.moca_pid AS spt_ref,
	-- Membership information
	cbs.ya_prob, cbs.observables, cbs.list_prob_yas
-- Start from spectral types table to limit the spectral type range
FROM data_spectral_types spt
-- Join the main objects table to get the main designation
JOIN moca_objects mo USING(moca_oid)
-- Join the BANYAN Sigma calculations that place an object in BPMG
--  with a probability above 90%
--  this table includes every combination of observables
--  and every available set of BANYAN models (moca_bsmdid)
JOIN calc_banyan_sigma cbs ON(
	-- Match the results with the moca_oid object
	cbs.moca_oid=spt.moca_oid AND
	-- Only consider BPMG
	cbs.moca_aid='BPMG' AND
	-- Only consider the rows using all available observables
	cbs.max_observables=1 AND
	-- Only consider objects with 90% membership or more
	cbs.ya_prob >= 90
	)
-- Join the list of available BANYAN models
--  to restrain results to the latest one (adopted=1)
JOIN moca_banyan_sigma_models mbsm ON(
    -- Match using the model version id
    mbsm.moca_bsmdid=cbs.moca_bsmdid AND
    -- Restrain to currently adopted BANYAN model version
    mbsm.adopted=1)
WHERE
	-- Only consider the best-available spectral type
	spt.adopted=1 AND
	-- Ignore spectral type estimates based on photometry
	spt.photometric_estimate=0 AND
	-- Limit the search to T0 or later (numeric spt = 20)
	spt.spectral_type_number >= 20;
\end{lstlisting}

\newpage
As a final example, let us consider a case where a user aims to gather all age-calibrated lithium equivalent width measurements as a function of the Gaia~DR3 $G-G_{\rm RP}$ color. This could be achieved by starting from the \texttt{calc\_equivalent\_widths\_combined} table, which already combines all available individual measurements per star, and joining the Gaia photometry, individual references, memberships, and the age associated to the host association. In this specific example, the ages are based on the membership classifications as directed by the best-available BANYAN~$\Sigma$ membership probabilities.

\begin{lstlisting}[language=SQL, frame=single]
SELECT
	-- Main object information
    mo.designation, mo.ra, mo.dec,
    -- Li EW information
    cewc.ew_angstrom*1000 AS ewli_milliangstrom,
    cewc.ew_angstrom_unc*1000 AS e_ewli_milliangstrom,
    -- Combine the EW Li references per object
    GROUP_CONCAT(ewref.bibcode) AS ew_refs,
    -- Gaia colors
    (gmag.magnitude - grpmag.magnitude) AS g_rp_color,
    -- Membership information
	cbs.moca_aid, cbs.ya_prob, cbs.observables,
	-- Age information
	daa.age_myr, ageref.bibcode AS age_ref
-- Start from the table that contains combined equivalent widths per-object
FROM calc_equivalent_widths_combined AS cewc
-- Join the main object table for generic information
JOIN moca_objects mo USING(moca_oid)
-- Join the individual EW measurements to gather the references
--  this will cause duplicated rows which will be re-combined
--  with a GROUP BY statement
JOIN data_equivalent_widths AS ew ON(ew.junct_setid=cewc.junct_setid)
-- Include the bibcodes for EW references
JOIN moca_publications AS ewref ON(ewref.moca_pid=ew.moca_pid)
-- Include Gaia DR3 G-band photometry (dereddened when necessary)
--  from the photometry table
JOIN data_photometry AS gmag ON(
	-- This constraint matches the EWs to the photmetry with the moca_oid
	gmag.moca_oid=cewc.moca_oid AND
	-- This selects the Gaia DR3 G-band photometry
	gmag.moca_psid='gaiadr3_gmag' AND
	-- This restrains the selection to the best-available measurement
	gmag.adopted=1)
-- Repeat with Gaia DR3 G_RP-band photometry
JOIN data_photometry AS grpmag ON(
	grpmag.moca_oid=cewc.moca_oid AND
	-- This selects the Gaia DR3 G_RP-band photometry
	grpmag.moca_psid='gaiadr3_rpmag' AND
	grpmag.adopted=1)
-- Join membership information based on BANYAN Sigma
JOIN calc_banyan_sigma cbs ON(
	-- Match the results with the moca_oid object
	cbs.moca_oid=cewc.moca_oid AND
	-- Only consider the rows using all available observables
	cbs.max_observables=1 AND
	-- Only consider objects with 90% membership or more
	cbs.ya_prob >= 90
	)
-- Join the list of available BANYAN models
--  to restrain results to the latest one (adopted=1)
JOIN moca_banyan_sigma_models mbsm ON(
    -- Match using the model version id
    mbsm.moca_bsmdid=cbs.moca_bsmdid AND
    -- Restrain to currently adopted BANYAN model version
    mbsm.adopted=1)
-- Join the best-available association age
JOIN data_association_ages daa ON(
	-- Age is selected based on BANYAN Sigma membership
	daa.moca_aid=cbs.moca_aid AND
	-- Restrain to the adopted age
	daa.adopted=1
	)
-- Include the bibcodes for age references
JOIN moca_publications AS ageref ON(ageref.moca_pid=daa.moca_pid)
-- This selects the Li equivalent widths
WHERE cewc.moca_spid='li'
-- Group by objects to combine the individual Li EW references
GROUP BY cewc.moca_oid;
\end{lstlisting}

\clearpage
\pagebreak
{
\ifpdf
\pdfpageattr{/Rotate 90}
\fi
\movetabledown=30mm
\begin{rotatetable*}
\begin{deluxetable*}{llcclcl}
\tablewidth{0.985\textwidth}
\tablecaption{List of Associations Considered in this Work.\label{tab:associations}}
\tablehead{
\noalign{\vskip 1.0ex}
\colhead{texttt{moca\_aid}} & \colhead{Designation} & \colhead{Real} & \colhead{\shortstack{Suboptimal\\ Definition}} & \colhead{\shortstack{Disc.\\ Ref.}} & \colhead{\shortstack{Age\\ (Myr)}} & \colhead{\shortstack{Age\\ Ref.}}
}
\startdata
ABDMG & AB Doradus moving group & 1 & 0 & 1 & 133$^{+15}_{-20}$ & 2 \\
BPMG & $\beta$ Pictoris moving group & 1 & 0 & 3 & 26 $\pm$ 3 & 4 \\
CAR & Carina association & 1 & 0 & 5 & 33.7$^{+2.0}_{-1.9}$ & 6 \\
CARN & Carina-Near moving group & 1 & 0 & 7 & 200 & 7 \\
CBER & Coma Berenices & 1 & 0 & -- & 675 $\pm$ 100 & 8 \\
COL & Columba association & 1 & 0 & 5 & 33.7$^{+2.0}_{-1.9}$ & 6 \\
EPSC & $\varepsilon$ Cha association & 1 & 0 & 9 & 3.7 & 10 \\
ETAC & $\eta$ Cha association & 1 & 0 & 11 & 6.5 & 10 \\
HYA & Hyades open cluster & 1 & 0 & -- & 695$^{+85}_{-67}$ & 12 \\
LCC & Lower Centaurus Crux OB association & 1 & 0 & 13 & 21 $\pm$ 4 & 14 \\
OCT & Octans association & 1 & 0 & 5 & 30-40 & 15 \\
PL8 & Platais 8 open cluster & 1 & 0 & 16 & 33.7$^{+2.0}_{-1.9}$ & 6 \\
PLE & Pleiades association & 1 & 0 & -- & 127.4 +6.3/-10 & 12 \\
THA & Tucana-Horologium association & 1 & 0 & 17 & 40 $\pm$ 3 & 18 \\
THOR & 32 Ori association & 1 & 0 & 19 & 21 $\pm$ 4 & 14 \\
TWA & TW Hya association & 1 & 0 & 20 & 10 & 21 \\
UCL & Upper Centaurus Lupus OB association & 1 & 0 & 13 & 21 $\pm$ 4 & 14 \\
UMA & Ursa Major association & 1 & 0 & -- & 414 & 22 \\
USCO & Upper Scorpius OB association & 1 & 0 & 13 & 10 & 23 \\
IC2602 & IC 2602 open cluster & 1 & 0 & -- & 40.0$^{+1.9}_{-1.4}$ & 6 \\
IC2391 & IC 2391 open cluster & 1 & 0 & -- & 51.0$^{+5.6}_{-3.9}$ & 6 \\
XFOR & $\chi$1 Fornacis cluster & 1 & 0 & 24 & 33.7$^{+2.0}_{-1.9}$ & 6 \\
ROPH & $\rho$ Ophiuchi star-forming region & 1 & 0 & -- & < 2 & 25 \\
CRA & Corona Australis star-forming region & 1 & 0 & -- & 4-5 & 26 \\
UCRA & Upper Corona Australis star-forming region & 1 & 0 & 27 & ~ 10 & 28 \\
\enddata
\tablecomments{Only the first 25 rows are shown here; the full table is available in the online version of this journal.}
\tablerefs{(1)~\cite{2004ApJ...613L..65Z}, (2)~\cite{2018ApJ...861L..13G}, (3)~\cite{2001ApJ...562L..87Z}, (4)~\cite{2014ApJ...792...37M}, (5)~\cite{2008hsf2.book..757T}, (6)~\cite{2024AJ....167...19L}, (7)~\cite{2006ApJ...649L.115Z}, (8)~\cite{2025arXiv250903461A}, (9)~\cite{2003ApJ...599.1207F}, (10)~\cite{2013MNRAS.435.1325M}, (11)~\cite{1999ApJ...516L..77M}, (12)~\cite{2022AA...664A..70G}, (13)~\cite{1964ARAA...2..213B}, (14)~\cite{2022AJ....164..151L}, (15)~\cite{2015MNRAS.447.1267M}, (16)~\cite{1998AJ....116.2423P}, (17)~\cite{2000ApJ...535..959Z}, (18)~\cite{2014AJ....147..146K}, (19)~\cite{2007IAUS..237..442M}, (20)~\cite{1997Sci...277...67K}, (21)~\cite{2015MNRAS.454..593B}, (22)~\cite{2015AAS...22511203J}, (23)~\cite{2016MNRAS.461..794P}, (24)~\cite{2002AA...389..871D}, (25)~\cite{2008hsf2.book..351W}, (26)~\cite{2012MNRAS.420..986G}, (27)~\cite{2000AAS..146..323N}, (28)~\cite{2018ApJ...856...23G}.}
\end{deluxetable*}
\end{rotatetable*}
}

\nocite{1914ApJ....40...43K, 1927pasr.book.....B, 1949PASP...61..183B, 1956ApJ...123..408B, 1958MNRAS.118..154E, 1959ApJ...130...69B, 1959PASP...71..145S, 1960PASP...72..205R, 1962ZA.....55..290H, 1964ARAA...2..213B, 1969AJ.....74.1021A, 1971PASP...83..251E, 1975AJ.....80...11V, 1975PASP...87...37E, 1977ATsir.966....7L, 1977ApJS...35..161S, 1983AJ.....88..642E, 1993ApJ...412..233S, 1994AJ....107.1796T, 1995ApJS..101..117K, 1997ESASP.402..571H, 1997Sci...277...67K, 1998AA...339..831B, 1998AJ....116.2423P, 1998ASPC..154.1793W, 1998PASP..110.1259G, 1999AA...341..427A, 1999AAS..135....5C, 1999AJ....117..354D, 1999ApJ...516L..77M, 2000AA...357..153F, 2000AAS..146..323N, 2000ApJ...535..959Z, 2000MNRAS.313L..23P, 2001ApJ...562L..87Z, 2002AA...383..631D, 2002AA...384..937Z, 2002AA...389..871D, 2003AA...410..565A, 2003ApJ...599.1207F, 2004AAS...204.0610T, 2004ARAA..42..685Z, 2004ApJ...613L..65Z, 2005AA...440..403K, 2006AJ....132.2198M, 2006ApJ...649L.115Z, 2007IAUS..237..442M, 2008hsf1.book..459B, 2008hsf2.book..351W, 2008hsf2.book..757T, 2009MNRAS.399..432N, 2009MNRAS.400L..20E, 2011AJ....141...92J, 2012MNRAS.419.1871D, 2012MNRAS.420..986G, 2013AA...558A..53K, 2013ApJ...778....5Z, 2013MNRAS.434..806B, 2013MNRAS.435.1325M, 2014AA...563A..94J, 2014AA...566A.132S, 2014AJ....147..146K, 2014ApJ...792...37M, 2015AAS...22511203J, 2015MNRAS.447.1267M, 2015MNRAS.452..173B, 2015MNRAS.454..593B, 2016AA...595A..22R, 2016MNRAS.458.3027M, 2016MNRAS.461..794P, 2017AJ....153..257O, 2018AA...615A.165B, 2018AA...618A..93C, 2018AA...619A.155S, 2018AJ....156..165C, 2018ApJ...856...23G, 2018ApJ...861L..13G, 2018ApJ...863...67G, 2018ApJ...865..136G, 2018MNRAS.473.2465B, 2019AA...621L...2R, 2019AA...622L..13M, 2019AA...623A.108B, 2019AA...624A.126C, 2019AA...627A...4R, 2019AJ....158...77C, 2019AJ....158..122K, 2019ApJ...870...27Z, 2019ApJ...877...12T, 2019JKAS...52..145S, 2019MNRAS.489.4418J, 2020AA...633A..99C, 2020AA...635A..45C, 2020AA...638A..85R, 2020AA...640A...1C, 2020AJ....159..105L, 2020AJ....160..279K, 2020ApJ...900L...4P, 2020ApJ...903...96G, 2020RNAAS...4...92G, 2021AA...645A..84M, 2021AA...647A..19T, 2021AJ....161..171T, 2021ApJ...912...16B, 2021ApJ...915L..29G, 2021ApJ...917...23K, 2021ApJS..254...20L, 2021MNRAS.504..356D, 2022AA...657L...3M, 2022AA...664A..70G, 2022AA...664A.175P, 2022AJ....164...88B, 2022AJ....164..115N, 2022AJ....164..151L, 2022ApJ...939...94M, 2022ApJ...941...49K, 2022ApJ...941..143K, 2022ApJS..262....7H, 2022arXiv220800070G, 2022arXiv221203266W, 2023AA...673A.114H, 2023AA...677A..59R, 2023AA...678A..71R, 2023AJ....165...37L, 2023ApJ...952...68S, 2023ApJ...954..134K, 2023ApJS..265...12Q, 2023ApJS..267...34H, 2023arXiv230503255S, 2024AJ....167...12C, 2024AJ....167...19L, 2024AJ....168...41T, 2024AJ....168..159L, 2024arXiv240605234T, 2025arXiv250605130O, 2025arXiv250903461A}

\clearpage
\pagebreak
{
\ifpdf
\pdfpageattr{/Rotate 90}
\fi
\movetabledown=40mm
\begin{rotatetable*}
\begin{deluxetable*}{lcccccccc}
\tablewidth{0.985\textwidth}
\tablecaption{New candidate members identified in this work.\label{tab:new_candidates}}
\tablehead{
\noalign{\vskip 1.0ex}
\colhead{Designation} & \colhead{\shortstack{Spectral\\ Type}} & \colhead{\texttt{moca\_aid}} & \colhead{\shortstack{Banyan~$\Sigma$\\ $P$ (\%)}} & \colhead{\shortstack{$\mu_\alpha\cos\delta$\\ (\masyr)}} & \colhead{\shortstack{$\mu_\delta$\\ (\masyr)}} & \colhead{\shortstack{Distance\\ (pc)}} & \colhead{\shortstack{RV\\(\kms)}} & \colhead{Refs\tablenotemark{a}}
}
\startdata
Gaia DR2 1576589693004879616 & A0 & UMA & 99.9 & $114 \pm 2$ & $2 \pm 2$ & $25.3 \pm 0.1$ & $\cdots$ & 1;2;--;3 \\
Gaia DR3 1563590579347125632 & A1.5Vs & UMA & 99.9 & $122.5 \pm 0.3$ & $-22.7 \pm 0.8$ & $24.9 \pm 0.3$ & $-5.6 \pm 0.9$ & 4;5;--;5 \\
Gaia DR3 3966634844566024960 & M2.5 & UMA & 99.9 & $85.16 \pm 0.04$ & $-48.24 \pm 0.02$ & $24.83 \pm 0.02$ & $-10.7 \pm 0.9$ & 6;5;--;5 \\
Gaia DR3 843042092397033472 & M3 & UMA & 99.9 & $100.9 \pm 0.4$ & $23.5 \pm 0.5$ & $24.1 \pm 0.3$ & $-11.3 \pm 0.8$ & 7;5;--;5 \\
2MASS J09161018+0153088 & M4Ve & UMA & 99.9 & $54.82 \pm 0.03$ & $-101.31 \pm 0.02$ & $15.652 \pm 0.007$ & $-12 \pm 1$ & 8;5;--;5 \\
2MASS J16352740+3500577 & M4Ve & UMA & 99.9 & $139.4 \pm 0.1$ & $-143.6 \pm 0.1$ & $17.24 \pm 0.03$ & $3.0 \pm 0.3$ & 8;5;--;5 \\
PM J07472+5020 & M4Ve & UMA & 99.9 & $-16.79 \pm 0.02$ & $79.62 \pm 0.02$ & $14.109 \pm 0.005$ & $-15.0 \pm 0.8$ & 8;5;--;5 \\
2MASS J11430780+6220365 & (M6.0) & UMA & 99.9 & $65.61 \pm 0.05$ & $18.76 \pm 0.06$ & $36.90 \pm 0.08$ & $\cdots$ & --;5;--;5 \\
2MASS J10390493+4352266 & (M6.0) & UMA & 99.9 & $61.57 \pm 0.03$ & $11.54 \pm 0.03$ & $28.71 \pm 0.03$ & $-11 \pm 2$ & --;5;--;5 \\
Gaia DR3 1713327707813549696 & M7e & UMA & 99.9 & $73.8 \pm 0.2$ & $-3.1 \pm 0.2$ & $41.8 \pm 0.3$ & $\cdots$ & 9;5;--;5 \\
2MASS J13445832+7715513 & (M7.1) & UMA & 99.9 & $73.8 \pm 0.2$ & $-3.1 \pm 0.2$ & $41.8 \pm 0.3$ & $\cdots$ & --;5;--;5 \\
2MASS J11240487+3808054 & M8.5 & UMA & 99.9 & $125.7 \pm 0.1$ & $-9.2 \pm 0.2$ & $18.41 \pm 0.05$ & $-14 \pm 3$ & 10;5;--;5 \\
2MASS J01474488+6351090 & K0V & CARN & 97.5 & $581.68 \pm 0.03$ & $-246.46 \pm 0.04$ & $10.041 \pm 0.004$ & $2.74 \pm 0.06$ & 11;5;--;5 \\
HD 99279 & K5-V & CARN & 98.1 & $-525 \pm 2$ & $82 \pm 2$ & $12.1 \pm 0.2$ & $8 \pm 5$ & 12;13;--;13 \\
TYC 8576-789-1 & K6Ve & CARN & 91.8 & $-48.11 \pm 0.01$ & $66.59 \pm 0.01$ & $70.08 \pm 0.05$ & $17 \pm 1$ & 14;5;--;5 \\
Gaia DR3 5448465052078103040 & (K7.0) & CARN & 93.7 & $-101.2 \pm 0.1$ & $34.5 \pm 0.1$ & $55.4 \pm 0.6$ & $\cdots$ & --;5;--;5 \\
Gaia DR3 2489349339521604864 & M1.5 & CARN & 94.6 & $193.06 \pm 0.03$ & $7.68 \pm 0.02$ & $30.71 \pm 0.02$ & $15.9 \pm 0.3$ & 15;5;--;5 \\
Gaia DR3 3409167094776800384 & M1.5V & CARN & 96.4 & $102.36 \pm 0.02$ & $-135.32 \pm 0.02$ & $20.608 \pm 0.010$ & $\cdots$ & 16;5;--;5 \\
Gaia DR3 5584660801498379264 & (M1.6) & CARN & 98.2 & $-37.77 \pm 0.01$ & $87.92 \pm 0.02$ & $47.71 \pm 0.03$ & $24.3 \pm 0.7$ & --;5;--;5 \\
LP 794-87 & M3.5 & CARN & 98.0 & $-453 \pm 4$ & $-33 \pm 4$ & $\cdots$ & $7 \pm 8$ & 17;18;--;-- \\
\enddata
\tablenotetext{a}{References for spectral types, proper motions, radial velocities, and distances, respectively.}
\tablecomments{Only the first 20 rows are shown here; the full table, which also contains the right ascension, and declination columns, is available in the online version of this journal. Distances between parentheses are photometric rather than trigonometric. Spectral types between parentheses are estimated based on colors.}
\tablerefs{(1)~\cite{2003AJ....125.1980K}, (2)~\cite{2025AJ....170...86S}, (3)~\cite{2007AA...474..653V_DUP}, (4)~\cite{2003AJ....126.2048G}, (5)~\cite{2023AA...674A...1G}, (6)~\cite{2012RAA....12..443B}, (7)~\cite{2023ApJS..264...17Z}, (8)~\cite{2013AJ....145..102L}, (9)~\cite{2000AJ....120.1085G}, (10)~\cite{2003AJ....126.2421C}, (11)~\cite{2007AJ....133.2524W}, (12)~\cite{2006AJ....132..161G}, (13)~\cite{2018AA...616A...1G}, (14)~\cite{2006AA...460..695T}, (15)~\cite{2004AJ....128..463R}, (16)~\cite{2015AA...577A.128A}, (17)~\cite{2007AJ....133.2825R}, (18)~\cite{2021ApJS..253....8M}.}
\end{deluxetable*}
\end{rotatetable*}
}

\nocite{1935bsdn.book.....S, 1955AnCap..18....0J, 1955ApJ...121..337S, 1957ArA.....2...55O, 1960AJ.....65...60S, 1963ApJ...138..832W, 1966AnCap..21....0S, 1966ApJ...146..142S, 1969AJ.....74..916H, 1969PASP...81..643C, 1969PhDT.........5M, 1970PASP...82.1360F, 1971PW&SO...1a...1S, 1973AA....22..203C, 1973ARAA..11...29M, 1974PhDT........10S, 1975ApJ...197..137S, 1975mcts.book.....H, 1976AJ.....81..245E, 1977AJ.....82..598G, 1978AAS...32...25R, 1978mcts.book.....H, 1979PhDT.........8F, 1980AJ.....85.1341S, 1981AnTok..18..125Y, 1981ApJS...45..437A, 1982mcts.book.....H, 1983ApJ...272..182A, 1984AJ.....89..702L, 1984ApJS...55..657C, 1985ApJS...59...95A, 1985ApJS...59..197B, 1986AJ.....91..144S, 1987ApJS...65..581G, 1988ApJ...331..958F, 1988mcts.book.....H, 1989ApJS...71..245K, 1990AAS...85.1069S, 1993ApJS...87..197G, 1993yCat.3135....0C, 1994AJ....107.1556G, 1994AJ....108.1924T, 1995AAS..110..367N, 1995AJ....109..797K, 1995AJ....110.1838R, 1995ApJS...99..135A, 1995PASP..107..846D, 1997AJ....113.1733H, 1997PASP..109..849G, 1998AA...338..563H, 1999MSS...C05....0H, 2000AJ....120..447K, 2000AJ....120.1085G, 2000AcApS..20...25L, 2000ApJ...538..363B, 2000ApJ...543..299M, 2001AA...373..625P, 2001AJ....121.2148G, 2001KFNT...17..409K, 2002AA...384..180F, 2002AJ....123.2002H, 2002AJ....123.3409H, 2002ApJS..143..513G, 2003AJ....125..984M, 2003AJ....125.1980K, 2003AJ....126.2048G, 2003AJ....126.2421C, 2004AJ....128..463R, 2005AA...442..211S, 2005AJ....130.1871B, 2005PASP..117..676R, 2006AA...446..515P, 2006AA...460..695T, 2006AJ....132..161G, 2006AJ....132..866R, 2006ApJ...645..676L, 2006ApJ...649..894L, 2007AA...474..653V, 2007AA...474..653V_DUP, 2007AJ....133..439C, 2007AJ....133.2524W, 2007AJ....133.2825R, 2007ApJS..173..104L, 2007MNRAS.376.1109J, 2007MNRAS.379.1599L, 2008AJ....135..785W, 2008AJ....136.1290R, 2008ApJS..178..339C, 2009AA...494..373M, 2009AJ....137....1F, 2009ApJ...699..649S, 2010AJ....139.1808S, 2010AJ....139.2566D, 2010ApJS..190..100K, 2010MNRAS.403.1949K, 2010MNRAS.404.1817Z, 2010MNRAS.406.1885B, 2011AA...527A..24L, 2011AJ....141...97W, 2011ApJ...743..138G, 2012AA...542A.105L, 2012AA...548A..53L, 2012AJ....144...14R, 2012ApJ...752...56F, 2012ApJ...752...59H, 2012RAA....12..443B, 2012yCat.1322....0Z, 2013AA...556A..15R, 2013AJ....145..102L, 2013AJ....145..134C, 2013AJ....146...85H, 2013ApJ...762...88M, 2013ApJ...772..129B, 2013ApJ...776..128K, 2013MNRAS.428.3104E, 2013MNRAS.430.1171D, 2013MNRAS.431..240O, 2013MNRAS.434.1422M, 2014AJ....147...20N, 2014AJ....147...34S, 2014AJ....148...36A, 2014ApJ...794..143B, 2014ApJS..211...10S, 2014MNRAS.443.2561G, 2014yCat....1.2023S, 2015AA...575A.117S, 2015AA...577A.128A, 2015AA...577A.148B, 2015AJ....150...42Z, 2015ApJ...801....4W, 2015ApJ...806...62B, 2015ApJ...812....3W, 2015ApJ...814..118B, 2015ApJS..219...33G, 2015MNRAS.446.3878M, 2015MNRAS.447.1267M, 2015MNRAS.449.3651M, 2015MNRAS.453L.103S, 2016AA...593A.119M, 2016AA...594A..39F, 2016AA...595A...2G, 2016AJ....152..190A, 2016ApJ...817..112S, 2016ApJS..224...36K, 2016ApJS..225...10F, 2016PhDT.......189A, 2016arXiv161205560C, 2017AJ....153...14W, 2017AJ....153..188F, 2017AJ....153..196S, 2017AJ....154...69S, 2017AJ....154..147D, 2017ApJ...836..167D, 2018AA...612A..96F, 2018AA...616A...1G, 2018AJ....155..265H, 2018AJ....156...76L, 2018ApJ...854..101B, 2018ApJ...867..109M, 2018MNRAS.480.5447Z, 2018MNRAS.481.3244G, 2019AJ....157..234S, 2019AJ....158...75H, 2019AJ....158...87D, 2019AJ....158..182G, 2019ApJ...877...60B, 2019MNRAS.485.2167G, 2020AA...637A..43K, 2020AA...639A..81B, 2020AJ....159..257B, 2020ApJ...889...74M, 2020ApJ...892...31B, 2020ApJ...899..123M, 2020ApJ...904..146F, 2020ApJS..251....6M, 2020MNRAS.491.2280S, 2020MNRAS.495.1252Z, 2020MNRAS.497..130T, 2020NatAs...4.1102M, 2020PASP..132j4401A, 2021AA...650A.190G, 2021AJ....161....5W, 2021AJ....161..192N, 2021ApJS..253....7K, 2021ApJS..253....8M, 2022AA...660A..38W, 2022AJ....163...64E, 2022AJ....163..242S, 2022ApJS..259...63S, 2022arXiv220800070G, 2023AA...672A..94D, 2023AA...674A...1G, 2023AA...674A.212D, 2023AA...675A..54V, 2023ApJ...959...63S, 2023ApJS..264...17Z, 2023ApJS..266...14T, 2023ApJS..267....7Z, 2023MNRAS.518.5106J, 2023RNAAS...7..184H, 2023Sci...380..198C, 2024AA...682A...5V, 2024ApJS..271...55K, 2025AJ....170...86S, Schn23b, SimNoPub, bywunpub}

\begin{deluxetable*}{lcc}
\tablewidth{0.985\textwidth}
\tablecaption{List of Tables in MOCAdb.\label{tab:mocadb_tables}}
\tablehead{
\noalign{\vskip 1.0ex}
\colhead{Table Name} & \colhead{Number of Columns} & \colhead{Number of Rows}
}
\startdata
\texttt{moca\_associations} & 18 & 10261 \\
\texttt{moca\_changelog} & 11 & 61408 \\
\texttt{moca\_companions} & 17 & 36887 \\
\texttt{moca\_spectral\_indices} & 15 & 78 \\
\texttt{moca\_instruments} & 6 & 73 \\
\texttt{moca\_missions} & 21 & 42 \\
\texttt{moca\_banyan\_sigma\_models} & 16 & 1 \\
\texttt{moca\_objects} & 18 & 3234191 \\
\texttt{moca\_spectra\_packages} & 30 & 28 \\
\texttt{moca\_object\_properties} & 10 & 22 \\
\texttt{moca\_publications} & 19 & 5507 \\
\texttt{moca\_extinction\_by\_spectral\_reference} & 12 & 2093 \\
\texttt{moca\_spectral\_type\_references} & 16 & 161 \\
\texttt{moca\_sequences} & 35 & 7 \\
\texttt{moca\_chemical\_species} & 14 & 10 \\
\texttt{moca\_spectra} & 69 & 4064 \\
\texttt{moca\_photometry\_systems} & 18 & 109 \\
\texttt{moca\_object\_types} & 5 & 13 \\
\texttt{moca\_membership\_types} & 7 & 6 \\
\texttt{cat\_2mass} & 73 & 2152213 \\
\texttt{cat\_apogeedr16\_allstars} & 176 & 439475 \\
\texttt{cat\_apogeedr16\_allvisits} & 70 & 1679797 \\
\texttt{cat\_allwise} & 305 & 1371102 \\
\texttt{cat\_catwise} & 200 & 1769974 \\
\texttt{cat\_decapsdr2} & 223 & 2004 \\
\texttt{cat\_denisdr3} & 87 & 23366 \\
\texttt{cat\_desdr1} & 184 & 22169 \\
\texttt{cat\_desdr2} & 238 & 41963 \\
\texttt{cat\_erassdr1} & 256 & 833615 \\
\texttt{cat\_euclidq1} & 664 & 589 \\
\texttt{cat\_gaiadr1} & 66 & 145784 \\
\texttt{cat\_gaiadr2} & 103 & 3081410 \\
\texttt{cat\_gaiadr3} & 164 & 3064316 \\
\texttt{cat\_gaiaedr3} & 109 & 609388 \\
\texttt{cat\_galahdr3} & 191 & 6583 \\
\texttt{cat\_galex\_gr67} & 103 & 43920 \\
\texttt{cat\_hipparcos} & 33 & 118182 \\
\texttt{cat\_gaiadr3\_binary\_masses} & 19 & 10549 \\
\texttt{cat\_gaiadr3\_vari\_rotation\_modulation} & 72 & 16433 \\
\texttt{cat\_exoplanets\_nasa} & 295 & 5154 \\
\enddata
\tablecomments{Only the first 40 rows are shown here; the full table is available in the online version of this journal.}
\end{deluxetable*}

\begin{deluxetable*}{lcl}
\tablewidth{0.985\textwidth}
\tablecaption{List of Columns in MOCAdb.\label{tab:mocadb_columns}}
\tablehead{
\noalign{\vskip 1.0ex}
\colhead{Table Name} & \colhead{Column Name} & \colhead{Data Type}
}
\startdata
\texttt{moca\_associations} & \texttt{id} & int(11) unsigned \\
\texttt{moca\_associations} & \texttt{moca\_aid} & varchar(12) \\
\texttt{moca\_associations} & \texttt{moca\_pid} & varchar(20) \\
\texttt{moca\_associations} & \texttt{name} & varchar(255) \\
\texttt{moca\_associations} & \texttt{simbad\_id} & varchar(255) \\
\texttt{moca\_associations} & \texttt{is\_real} & varchar(1) \\
\texttt{moca\_associations} & \texttt{coeval} & varchar(1) \\
\texttt{moca\_associations} & \texttt{suboptimal\_grouping} & varchar(1) \\
\texttt{moca\_associations} & \texttt{in\_banyan} & tinyint(1) \\
\texttt{moca\_associations} & \texttt{in\_banyan\_sigma\_2020} & tinyint(1) \\
\texttt{moca\_associations} & \texttt{in\_partiview} & tinyint(1) \\
\texttt{moca\_associations} & \texttt{physical\_nature} & varchar(50) \\
\texttt{moca\_associations} & \texttt{comments} & text \\
\texttt{moca\_associations} & \texttt{created\_timestamp} & timestamp \\
\texttt{moca\_associations} & \texttt{modified\_timestamp} & timestamp \\
\texttt{moca\_associations} & \texttt{bibcode} & varchar(30) \\
\texttt{moca\_changelog} & \texttt{id} & int(11) unsigned \\
\texttt{moca\_changelog} & \texttt{created\_timestamp} & datetime \\
\texttt{moca\_changelog} & \texttt{user} & varchar(50) \\
\texttt{moca\_changelog} & \texttt{machine\_name} & varchar(50) \\
\texttt{moca\_changelog} & \texttt{modified\_tables} & text \\
\texttt{moca\_changelog} & \texttt{nrows\_modified} & int(11) \\
\texttt{moca\_changelog} & \texttt{user\_description} & text \\
\texttt{moca\_companions} & \texttt{moca\_cid} & int(11) unsigned \\
\texttt{moca\_companions} & \texttt{moca\_oid\_parent} & int(11) unsigned \\
\texttt{moca\_companions} & \texttt{moca\_oid\_child} & int(11) unsigned \\
\texttt{moca\_companions} & \texttt{moca\_pid} & varchar(20) \\
\texttt{moca\_companions} & \texttt{origin} & varchar(255) \\
\texttt{moca\_companions} & \texttt{object\_designation\_parent} & varchar(60) \\
\texttt{moca\_companions} & \texttt{object\_designation\_type\_parent} & varchar(20) \\
\texttt{moca\_companions} & \texttt{object\_designation\_child} & varchar(60) \\
\texttt{moca\_companions} & \texttt{object\_designation\_type\_child} & varchar(20) \\
\texttt{moca\_companions} & \texttt{publication\_comments} & varchar(255) \\
\texttt{moca\_companions} & \texttt{comments} & text \\
\texttt{moca\_companions} & \texttt{created\_timestamp} & timestamp \\
\texttt{moca\_companions} & \texttt{modified\_timestamp} & timestamp \\
\texttt{moca\_companions} & \texttt{ignored} & tinyint(1) \\
\texttt{moca\_companions} & \texttt{bibcode} & varchar(30) \\
\texttt{moca\_spectral\_indices} & \texttt{id} & int(11) unsigned \\
\texttt{moca\_spectral\_indices} & \texttt{moca\_siid} & varchar(20) \\
\enddata
\tablecomments{Only the first 40 rows are shown here; the full table is available in the online version of this journal.}
\end{deluxetable*}

\clearpage
\pagebreak
\begin{deluxetable*}{llc}
\tablewidth{0.985\textwidth}
\tablecaption{External Catalogs Cross-Matched with MOCAdb Entries.\label{tab:missions}}
\tablehead{
\noalign{\vskip 1.0ex}
\colhead{\shortstack{MOCAdb\\ Table}} & \colhead{\shortstack{Mission\\ Name}} & \colhead{Reference}
}
\startdata
\texttt{cat\_2mass} & 2MASS PSC & 1 \\
\texttt{cat\_apogeedr16\_allstars} & APOGEE DR16 & 2 \\
\texttt{cat\_decapsdr2} & DECAPS DR2 & 3 \\
\texttt{cat\_denisdr3} & DENIS DR3 & 4 \\
\texttt{cat\_desdr2} & DES DR2 & 5 \\
\texttt{cat\_gaiadr1} & Gaia DR1 & 6 \\
\texttt{cat\_gaiadr2} & Gaia DR2 & 7 \\
\texttt{cat\_gaiaedr3} & Gaia EDR3 & 8 \\
\texttt{cat\_gaiadr3} & Gaia DR3 & 9 \\
\texttt{cat\_galahdr3} & GALAH DR3 & 10 \\
\texttt{cat\_galex\_gr67} & Galex GR67 & 11 \\
\texttt{cat\_hipparcos} & Hipparcos 2007 reduction & 12 \\
\texttt{cat\_nscdr2} & NSC DR2 & 13 \\
\texttt{cat\_ps1dr2} & PS1 DR2 & 14 \\
\texttt{cat\_ravedr6} & RAVE DR6 & 15 \\
\texttt{cat\_rosat} & ROSAT All-Sky & 16 \\
\texttt{cat\_splusdr4} & S-PLUS DR4 & 17 \\
\texttt{cat\_sdssdr17} & SDSS DR17 & 18 \\
\texttt{cat\_smssdr4} & SMSS DR4 & 19 \\
\texttt{cat\_uhsdr3} & UHS DR3 & 20 \\
\texttt{cat\_ukidssdxsdr11plus} & UKIDSSDXS DR11PLUS & 21 \\
\texttt{cat\_ukidssgcsdr11plus} & UKIDSSGCS DR11PLUS & 21 \\
\texttt{cat\_ukidssgpsdr11plus} & UKIDSSGPS DR11PLUS & 21 \\
\texttt{cat\_ukidsslasdr11plus} & UKIDSSLAS DR11PLUS & 21 \\
\texttt{cat\_vhsdr6} & VHS DR6 & 22 \\
\texttt{cat\_videodr5} & VIDEO DR5 & 23 \\
\texttt{cat\_vikingdr4} & VIKING DR4 & 24 \\
\texttt{cat\_viracdr4\_plx} & VIRAC DR4 & 25 \\
\texttt{cat\_vmcdr5} & VMC DR5 & 26 \\
\texttt{cat\_vvvdr5} & VVV DR5 & 27 \\
\texttt{cat\_wise} & WISE All-Sky SC & 28 \\
\texttt{cat\_allwise} & WISE AllWISE & 29 \\
\texttt{cat\_catwise} & WISE CatWISE & 30 \\
\enddata
\tablerefs{(1)~\cite{2006AJ....131.1163S}, (2)~\cite{2020ApJS..249....3A}, (3)~\cite{2023ApJS..264...28S}, (4)~\cite{1994ApSS.217....3E}, (5)~\cite{2021ApJS..255...20A}, (6)~\cite{2016AA...595A...2G}, (7)~\cite{2018AA...616A...1G}, (8)~\cite{2021AA...649A...1G}, (9)~\cite{2023AA...674A...1G}, (10)~\cite{2025PASA...42...51B}, (11)~\cite{2017ApJS..230...24B}, (12)~\cite{2007AA...474..653V_DUP}, (13)~\cite{2021AJ....161..192N}, (14)~\cite{2016arXiv161205560C}, (15)~\cite{2020AJ....160...82S}, (16)~\cite{1999AA...349..389V}, (17)~\cite{2024AA...689A.249H}, (18)~\cite{2022ApJS..259...35A}, (19)~\cite{2024PASA...41...61O}, (20)~\cite{2025AJ....170...86S}, (21)~\cite{2007MNRAS.379.1599L}, (22)~\cite{2013Msngr.154...35M}, (23)~\cite{2022RNAAS...6..109H}, (24)~\cite{2013Msngr.154...32E}, (25)~\cite{2018MNRAS.474.1826S}, (26)~\cite{2011AA...527A.116C}, (27)~\cite{2012AA...537A.107S}, (28)~\cite{2010AJ....140.1868W}, (29)~\cite{2014ApJ...783..122K}, (30)~\cite{2021ApJS..253....8M}.}
\end{deluxetable*}

\clearpage
\pagebreak
\begin{deluxetable*}{lcclcccc}
\tablewidth{0.985\textwidth}
\tablecaption{Age-calibrated exoplanets currently known.\label{tab:exoplanets}}
\tablehead{
\noalign{\vskip 1.0ex}
\colhead{\shortstack{Host Star\\ Designation}} & \colhead{$N_{\rm exo}$} & \colhead{Method} & \colhead{\texttt{moca\_aid}} & \colhead{\shortstack{Banyan~$\Sigma$\\ Prob. (\%)}} & \colhead{\shortstack{Mem.\\Ref.}} & \colhead{\shortstack{Age\\ (Myr)}} & \colhead{\shortstack{Age\\Ref.}}
}
\startdata
GJ 393 & 1 & RV & ABDMG & 99.3 & 1 & 133$^{+15}_{-20}$ & 2 \\
GJ 849 & 2 & RV & CRIUS197 & 94.8 & 3 & $\cdots$ & -- \\
HD 147513 & 1 & RV & CRIUS135 & 95.2 & 3 & 100-700 & 4 \\
GJ 163 & 3 & RV & HSC1711 & 85.1 & 3 & 260 (-190,+670) & 5 \\
GJ 3021 & 1 & RV & CRIUS203 & 97.3 & 3 & 670.0 +376.7/-241.1 & -- \\
LSPM J2116+0234 & 1 & RV & CARN & 92.7 & 3 & 200 & 6 \\
bet Pic & 2 & RV & BPMG & 99.0 & 7 & 26 $\pm$ 3 & 8 \\
GJ 9714 & 1 & RV & VELA & 91.7 & 3 & 2-9960 & -- \\
HD 114783 & 2 & RV & OCTN & 89.8 & 3 & 30-100 & 9 \\
HIP 12961 & 1 & RV & CRIUS197 & 98.7 & 3 & $\cdots$ & -- \\
HD 179949 & 1 & RV & VELA & 85.2 & 3 & 2-9960 & -- \\
HD 142415 & 1 & RV & CRIUS209 & 81.0 & 3 & $\cdots$ & -- \\
HS Psc & 1 & RV & ABDMG & 99.4 & 10 & 133$^{+15}_{-20}$ & 2 \\
30 Ari B & 1 & RV & CRIUS203 & 83.2 & 3 & 670.0 +376.7/-241.1 & -- \\
eps Tau & 1 & RV & HYA & 99.9 & 11 & 695$^{+85}_{-67}$ & 12 \\
HD 285507 & 1 & RV & HYA & 99.9 & 11 & 695$^{+85}_{-67}$ & 12 \\
HD 212301 & 1 & RV & MEL5 & 93.6 & 13 & 210 $\pm$ 27 & 14 \\
HD 135625 & 1 & RV & CRIUS209 & 97.4 & 3 & $\cdots$ & -- \\
HD 114082 & 1 & RV & GRSCOS27C & 99.5 & 3 & 14.6 & 15 \\
TAP 26 & 1 & RV & OH29 & 99.9 & 16 & 29.3 $\pm$ 1 & 17 \\
V830 Tau & 1 & RV & L1529 & 99.9 & 18 & 3.4 & 15 \\
CI Tau & 1 & RV & TAUMGLIU4 & 99.9 & 16 & 2.0 & 16 \\
Pr0201 & 1 & RV & PRA & 99.9 & 19 & 617 & 20 \\
Pr0211 & 2 & RV & CPRA & 99.9 & 21 & 617.0 & 20 \\
NGC 2682 Sand 364 & 1 & RV & NGC2682 & 99.7 & 22 & ~4000 & 22 \\
NGC 2682 YBP 401 & 1 & RV & NGC2682 & 99.9 & 22 & ~4000 & 22 \\
NGC 2682 Sand 978 & 1 & RV & NGC2682 & 99.9 & 22 & ~4000 & 22 \\
NGC 2682 Sand 1429 & 1 & RV & NGC2682 & 99.9 & 22 & ~4000 & 22 \\
NGC 2682 YBP 1194 & 1 & RV & NGC2682 & 99.8 & 22 & ~4000 & 22 \\
NGC 2682 YBP 1514 & 1 & RV & NGC2682 & 99.9 & 22 & ~4000 & 22 \\
\enddata
\tablecomments{Only the first 30 rows are shown here; the full table is available in the online version of this journal.}
\tablerefs{(1)~\cite{2009AA...508..833D}, (2)~\cite{2018ApJ...861L..13G}, (3)~This Paper, (4)~\cite{2022ApJ...939...94M}, (5)~\cite{2023AA...673A.114H}, (6)~\cite{2006ApJ...649L.115Z}, (7)~\cite{2001ApJ...562L..87Z}, (8)~\cite{2014ApJ...792...37M}, (9)~\cite{2013ApJ...778....5Z}, (10)~\cite{2010AJ....140..119S}, (11)~\cite{1998AA...331...81P}, (12)~\cite{2022AA...664A..70G}, (13)~\cite{2024arXiv240605234T}, (14)~\cite{2024AJ....168...41T}, (15)~\cite{2021ApJ...917...23K}, (16)~\cite{2021ApJS..254...20L}, (17)~\cite{2024AJ....167...19L}, (18)~\cite{2023AJ....165...37L}, (19)~\cite{2014ApJ...795..161D}, (20)~\cite{2018ApJ...863...67G}, (21)~\cite{2019AA...627A...4R}, (22)~\cite{2021AA...647A..19T}.}
\end{deluxetable*}

\nocite{1998AA...331...81P, 2001ApJ...562L..87Z, 2003AA...404..913S, 2003ApJ...599..342S, 2004ApJ...602..816L, 2006ApJ...649L.115Z, 2007AJ....134.2340K, 2008hsf2.book..351W, 2009AA...508..833D, 2010AA...520A..15M, 2010AJ....140..119S, 2011AJ....141...92J, 2011ApJ...732...61Z, 2012AJ....143...80S, 2013ApJ...778....5Z, 2013MNRAS.435.1325M, 2014ApJ...788...81M, 2014ApJ...792...37M, 2014ApJ...795..161D, 2015ApJ...813...83C, 2015MNRAS.454..593B, 2017AJ....153..257O, 2018AJ....156..165C, 2018ApJ...861L..13G, 2018ApJ...862..138G, 2018ApJ...863...67G, 2019AA...621L...3M, 2019AA...623A.112D, 2019AA...626A..80C, 2019AA...627A...4R, 2019AJ....158..122K, 2019ApJ...870...27Z, 2019ApJ...877...12T, 2019MNRAS.486.5405G, 2020AA...640A...1C, 2020AJ....160..186L, 2020AJ....160..279K, 2020ApJ...904..140C, 2021AA...645A..84M, 2021AA...647A..19T, 2021ApJ...917...23K, 2021ApJS..254...20L, 2022AA...657L...3M, 2022AA...664A..70G, 2022AJ....164...88B, 2022ApJ...939...94M, 2023AA...673A.114H, 2023AA...677A..59R, 2023AA...678A..71R, 2023AJ....165...37L, 2023AJ....165...85W, 2023ApJ...952...68S, 2023ApJ...954..134K, 2023arXiv230503255S, 2024AJ....167...19L, 2024AJ....168...41T, 2024arXiv240605234T}

\clearpage
\pagebreak
\begin{deluxetable*}{lcccccc}
\tablewidth{0.985\textwidth}
\tablecaption{Age-calibrated TESS exoplanets candidates.\label{tab:exoplanet_candidates}}
\tablehead{
\noalign{\vskip 1.0ex}
\colhead{\shortstack{Host Star\\ Designation}} & \colhead{$N_{\rm exo}$} & \colhead{\texttt{moca\_aid}} & \colhead{\shortstack{Banyan~$\Sigma$\\ Prob. (\%)}} & \colhead{\shortstack{Mem.\\Ref.}} & \colhead{\shortstack{Age\\ (Myr)}} & \colhead{\shortstack{Age\\Ref.}}
}
\startdata
TIC 141141249 & 1 & ARG & 93.1 & 1 & 45-50 & 2 \\
TIC 278198753 & 1 & CAR & 81.3 & 1 & 33.7$^{+2.0}_{-1.9}$ & 3 \\
TIC 325468685 & 1 & CRIUS131 & 82.5 & 1 & 569.1 +619.1/-239.3 & -- \\
TIC 441420236 & 1 & BPMG & 98.1 & 4 & 26 $\pm$ 3 & 5 \\
TIC 31374837 & 1 & CRIUS197 & 93.0 & 1 & $\cdots$ & -- \\
TIC 192790476 & 1 & MEL5 & 97.2 & 6 & 210 $\pm$ 27 & 7 \\
TIC 299798795 & 1 & MEL5 & 99.8 & 6 & 210 $\pm$ 27 & 7 \\
TIC 51024887 & 1 & OCEMG & 90.1 & 1 & ~500-600 & 8 \\
TIC 27491137 & 1 & CRIUS224 & 99.5 & 9 & $\cdots$ & -- \\
TIC 180695581 & 1 & CRIUS224 & 99.6 & 9 & $\cdots$ & -- \\
TIC 437011608 & 1 & CRIUS197 & 92.3 & 1 & $\cdots$ & -- \\
TIC 4070275 & 1 & CHYA & 99.9 & 10 & 695$^{+85}_{-67}$ & 11 \\
TIC 434226736 & 1 & HYA & 99.9 & 12 & 695$^{+85}_{-67}$ & 11 \\
TIC 244161191 & 1 & COL & 92.7 & 1 & 33.7$^{+2.0}_{-1.9}$ & 3 \\
TIC 20182165 & 1 & OCTN & 97.5 & 13 & 30-100 & 13 \\
TIC 77951245 & 1 & COL & 99.9 & 12 & 33.7$^{+2.0}_{-1.9}$ & 3 \\
TIC 419957393 & 1 & OCTN & 96.3 & 1 & 30-100 & 13 \\
TIC 464646604 & 1 & ABDMG & 96.6 & 14 & 133$^{+15}_{-20}$ & 15 \\
TIC 18310799 & 1 & HYA & 99.9 & 16 & 695$^{+85}_{-67}$ & 11 \\
TIC 311183180 & 1 & CHYA & 86.5 & 1 & 695$^{+85}_{-67}$ & 11 \\
TIC 298981199 & 1 & OCTN & 99.2 & 1 & 30-100 & 13 \\
TIC 311271011 & 1 & CRIUS223 & 99.4 & 9 & 100-700 & -- \\
TIC 93125144 & 1 & ABDMG & 82.9 & 14 & 133$^{+15}_{-20}$ & 15 \\
TIC 39200363 & 1 & CRIUS205 & 99.9 & 9 & $\cdots$ & -- \\
TIC 391903064 & 1 & ABDMG & 99.3 & 1 & 133$^{+15}_{-20}$ & 15 \\
TIC 360630575 & 1 & MEL4 & 95.0 & 17 & 23-26 & -- \\
TIC 157081737 & 1 & THEIA301 & 97.9 & 18 & 195.0 & 19 \\
TIC 360156606 & 1 & HSC2523 & 99.9 & 1 & 7.0$^{+-4.2}_{-+4.1}$ & 20 \\
TIC 224225541 & 1 & SUN1 & 99.6 & 21 & 204 $\pm$ 45 & 22 \\
TIC 166053959 & 1 & GRX & 88.0 & 23 & 300 $\pm$ 60 & 24 \\
\enddata
\tablecomments{Only the first 30 rows are shown here; the full table is available in the online version of this journal.}
\tablerefs{(1)~This Paper, (2)~\cite{2019ApJ...870...27Z}, (3)~\cite{2024AJ....167...19L}, (4)~\cite{2001ApJ...562L..87Z}, (5)~\cite{2014ApJ...792...37M}, (6)~\cite{2024arXiv240605234T}, (7)~\cite{2024AJ....168...41T}, (8)~\cite{2022arXiv220800070G}, (9)~\cite{2022ApJ...939...94M}, (10)~\cite{2019AA...627A...4R}, (11)~\cite{2022AA...664A..70G}, (12)~\cite{2018ApJ...862..138G}, (13)~\cite{2013ApJ...778....5Z}, (14)~\cite{2010AA...520A..15M}, (15)~\cite{2018ApJ...861L..13G}, (16)~\cite{2019AA...621L...3M}, (17)~\cite{2023AJ....165...85W}, (18)~\cite{2019AJ....158..122K}, (19)~\cite{2020AJ....160..279K}, (20)~\cite{2023AA...673A.114H}, (21)~\cite{2023arXiv230503255S}, (22)~\cite{2023ApJ...952...68S}, (23)~\cite{2019ApJ...877...12T}, (24)~\cite{2022AA...657L...3M}.}
\end{deluxetable*}

\nocite{2001ApJ...562L..87Z, 2007ApJS..172..663S, 2008AJ....136..118F, 2009MNRAS.400L..20E, 2010AA...520A..15M, 2013ApJ...778....5Z, 2014ApJ...792...37M, 2015MNRAS.447.1267M, 2016AA...595A..22R, 2018AA...616A..10G, 2018AA...618A..93C, 2018AJ....156..165C, 2018ApJ...860...43G, 2018ApJ...861L..13G, 2018ApJ...862..138G, 2018ApJ...863...67G, 2019AA...621L...3M, 2019AA...623A.112D, 2019AA...627A...4R, 2019AJ....158..122K, 2019ApJ...870...27Z, 2019ApJ...877...12T, 2019MNRAS.486.5405G, 2020AA...640A...1C, 2020AJ....160..279K, 2020ApJ...903...96G, 2021AA...645A..84M, 2021AA...647A..19T, 2021ApJ...917...23K, 2022AA...657L...3M, 2022AA...664A..70G, 2022ApJ...939...94M, 2022ApJ...941..143K, 2022arXiv220800070G, 2023AA...673A.114H, 2023AJ....165...37L, 2023AJ....165...85W, 2023ApJ...952...68S, 2023ApJ...954..134K, 2023arXiv230503255S, 2024AJ....167...19L, 2024AJ....168...41T, 2024AJ....168..159L, 2024arXiv240605234T}

\clearpage
\pagebreak
{
\ifpdf
\pdfpageattr{/Rotate 90}
\fi
\movetabledown=60mm
\begin{rotatetable*}
\begin{deluxetable*}{lcccclcccl}
\tablewidth{0.985\textwidth}
\tablecaption{Candidate Age-Calibrated Substellar Objects.\label{tab:substellar}}
\tablehead{
\noalign{\vskip 1.0ex}
\colhead{Designation} & \colhead{\shortstack{Spectral\\ Type}} & \colhead{Cat.} & \colhead{\shortstack{BANYAN~$\Sigma$\\ \texttt{moca\_aid}}} & \colhead{\shortstack{Banyan~$\Sigma$\\ $P$ (\%)}} & \colhead{\shortstack{Known\\ Memberships}} & \colhead{\shortstack{Distance\\ (pc)}} & \colhead{\shortstack{RV\\(\kms)}} & \colhead{\shortstack{Mass\\ (\mjup)}} & \colhead{Refs\tablenotemark{a}}
}
\startdata
CWISE J053644.83-305539.5 & T9.5 & new & BPMG & 96.8 & $\cdots$ & $12.9 \pm 0.6$ & $\cdots$ & $1.7^{+1.4}_{-0.8}$ & 1;1;--;1;2 \\
WISEA J235402.79+024014.1 & Y1 & reclass. & HSC517 & 85.2 & BPMG,LM & $7.7 \pm 0.2$ & $\cdots$ & $2.5^{+2.0}_{-1.3}$ & 3;1;--;1;4 \\
CWISE J094005.45+523358.7 & $\geq$Y1 & new & CARN & 50.2 & $\cdots$ & $17 \pm 4$ & $\cdots$ & $2.8^{+3.0}_{-1.6}$ & 5;1;--;1;2 \\
WISE J113949.24-332425.1 & T7 & reclass. & TWA & 86.9 & BPMG,LM & $36 \pm 4$ & $\cdots$ & $3.0^{+2.2}_{-1.3}$ & 6;1;--;1;4 \\
CWISE J235644.87-481456.7 & Y0.5 & new & MEL5 & 93.5 & $\cdots$ & $\cdots$ & $\cdots$ & $3.1^{+2.3}_{-1.6}$ & 5;7;--;--;2 \\
WISEPC J225540.74-311841.8 & T8 & known & BPMG & 92.8 & BPMG,HM & $13.8 \pm 0.7$ & $\cdots$ & $3.4^{+2.3}_{-1.4}$ & 8;9;--;1;10 \\
CWISE J213249.12+690113.8 & T8.5 & new & ARG & 96.4 & $\cdots$ & $(23 \pm 3)$ & $\cdots$ & $3.6^{+2.2}_{-1.6}$ & 5;2;--;2;2 \\
ULAS J232600.40+020139.2 & T8 & new & CIC2391 & 77.4 & $\cdots$ & $22 \pm 2$ & $\cdots$ & $3.8^{+1.4}_{-0.6}$ & 11;1;--;1;2 \\
CWISE J201146.50-481259.8 & Y0 & new & MEL5 & 94.5 & $\cdots$ & $14.1 \pm 0.7$ & $\cdots$ & $3.8^{+2.6}_{-1.8}$ & 5;1;--;1;2 \\
WISE J035000.32-565830.2 & Y1 & new & CRIUS135 & 87.8 & $\cdots$ & $5.67 \pm 0.07$ & $\cdots$ & $3.9^{+3.7}_{-2.1}$ & 12;1;--;1;2 \\
ULAS J101721.40+011817.9 & T8pec & new & ARG & 92.6 & $\cdots$ & $(37 \pm 19)$ & $\cdots$ & $4.3^{+2.4}_{-1.7}$ & 13;14;--;2;2 \\
WISE J011952.75-450230.8 & T8 & new & ARG & 95.5 & $\cdots$ & $(21 \pm 4)$ & $\cdots$ & $4.3^{+2.4}_{-1.8}$ & 1;15;--;2;2 \\
WISE J064723.23-623235.5 & Y1 & new & OCEMG & 93.2 & $\cdots$ & $10.1 \pm 0.2$ & $\cdots$ & $4.6^{+3.0}_{-2.2}$ & 16;1;--;1;2 \\
CWISE J053512.02-773828.8 & T9.5 & new & CARN & 98.0 & $\cdots$ & $(26 \pm 9)$ & $\cdots$ & $4.6^{+3.0}_{-2.2}$ & 17;7;--;2;2 \\
CWISE J125720.14+715349.2 & Y1e & new & OCEMG & 95.7 & $\cdots$ & $\cdots$ & $\cdots$ & $4.7^{+3.0}_{-2.2}$ & 17;7;--;--;2 \\
WISEPC J205628.90+145953.3 & Y0 & new & CHYA & 55.7 & $\cdots$ & $7.1 \pm 0.1$ & $\cdots$ & $4.9^{+3.6}_{-2.5}$ & 8;1;--;1;2 \\
51 Eri b & T6.5: & known & BPMG & 95.3 & BPMG,HM & $29.8 \pm 0.1$ & $20 \pm 2$ & $5.0^{+3.0}_{-1.8}$ & 18;19;--;20;21 \\
WISE J233226.49-432510.6 & T9 & known & ABDMG & 99.6 & ABDMG,HM & $16.4 \pm 0.6$ & $\cdots$ & $5.0^{+3.0}_{-2.3}$ & 12;1;--;1;10 \\
S Ori J053804.65-021352.5 & T5 & new & CWNU1057 & 84.5 & $\cdots$ & $(92 \pm 23)$ & $\cdots$ & $5.1^{+3.0}_{-2.0}$ & 22;22;--;2;2 \\
PSO J159.2399-26.3885 & T1.5 & new & TWA & 89.6 & $\cdots$ & $70 \pm 16$ & $\cdots$ & $5.1^{+2.5}_{-1.8}$ & 23;24;--;24;2 \\
\enddata
\tablenotetext{a}{References for spectral types, proper motions, radial velocities, distances, and memberships, respectively.}
\tablecomments{Only the first 20 rows are shown here; the full table, which also contains the right ascension, declination, and proper motion columns, is available in the online version of this journal. Distances between parentheses are photometric rather than trigonometric.}
\tablerefs{(1)~\cite{2021ApJS..253....7K}, (2)~This Paper, (3)~\cite{2015ApJ...804...92S}, (4)~\cite{Best20US}, (5)~\cite{2020ApJ...889...74M}, (6)~\cite{2013PASP..125..809T}, (7)~\cite{2021ApJS..253....8M}, (8)~\cite{2011ApJS..197...19K}, (9)~\cite{2019ApJS..240...19K}, (10)~\cite{2021ApJ...911....7Z}, (11)~\cite{2013MNRAS.433..457B}, (12)~\cite{2012ApJ...753..156K}, (13)~\cite{2008MNRAS.391..320B}, (14)~\cite{2007MNRAS.379.1599L}, (15)~\cite{2021AJ....161..192N}, (16)~\cite{2013ApJ...776..128K}, (17)~\cite{2020ApJ...899..123M}, (18)~\cite{2017AJ....154...10R}, (19)~\cite{2023AA...674A...1G}, (20)~\cite{2018AA...616A...1G}, (21)~\cite{2001ApJ...562L..87Z}, (22)~\cite{2015AA...574A.118P}, (23)~\cite{2015ApJ...814..118B}, (24)~\cite{2020AJ....159..257B}.}
\end{deluxetable*}
\end{rotatetable*}
}

\nocite{1993yCat.3170....0M, 1996ApJ...469..706M, 1997AJ....113.1733H, 1998ApJ...499L.199S, 1999AA...343..477C, 1999AA...350..612N, 1999AJ....117..343R, 1999AJ....118..997G, 1999ApJ...510..266B, 1999ApJ...512L..63W, 1999ApJ...519..802K, 1999ApJ...527..219S, 2000AJ....120..447K, 2000AJ....120..479A, 2000AJ....120.1085G, 2000Sci...290..103Z, 2001AA...377L...9B, 2001ApJ...556..830B, 2001ApJ...562L..87Z, 2001MNRAS.328...45M_DUP, 2002AJ....123.2828C, 2002AJ....123.3409H, 2002AJ....124.1001C, 2002ApJ...575..484G, 2002ApJ...578..536Z, 2003AA...403..929K, 2003AA...404..171B, 2003AA...404..913S, 2003AA...406.1001C, 2003AJ....126.2421C, 2003ApJ...593.1093L, 2003ApJ...599..342S, 2004AA...417..583C, 2004AA...424..213B, 2004AJ....127..449M, 2004AJ....127.1131W, 2004AJ....127.2948V, 2004AJ....127.3553K, 2004AJ....128.2460H, 2004ApJ...602..816L, 2004ApJ...614..398L, 2005AA...430L..49S, 2005AA...438L..29C, 2005AA...440..139L, 2005AA...440.1061L, 2005AJ....129..829D, 2005AJ....129.1483L, 2006AA...446..485G, 2006AA...446..515P, 2006AA...458..805B, 2006AJ....131.2722C, 2006AJ....131.3016S, 2006AJ....132.2665S, 2006ApJ...637.1067B, 2006ApJ...645..676L, 2006ApJ...646.1215L, 2006ApJ...649..894L, 2006ApJ...651L..57A, 2006MNRAS.373...95L, 2006Natur.440..311S, 2007AA...474..653V, 2007AJ....133..439C, 2007AJ....133..971A, 2007AJ....133.2825R, 2007AJ....134..411M, 2007AJ....134.2340K, 2007ApJ...662..413K, 2007ApJ...669L..97L, 2007ApJS..172..663S, 2007ApJS..173..104L, 2007MNRAS.374..372L, 2007MNRAS.378L..24B, 2007MNRAS.379.1599L, 2007MNRAS.381.1077R, 2008AA...481..661B, 2008AA...485..155A, 2008AA...488..167S, 2008AJ....135..785W, 2008AJ....135..966F, 2008AJ....136.1290R, 2008ApJ...673L.185B, 2008ApJ...674..336G, 2008ApJ...676.1281M, 2008ApJ...684..654L, 2008ApJ...685..313P, 2008ApJ...687.1303G, 2008ApJ...688..377S, 2008ApJ...689.1295K, 2008ApJS..177..551M, 2008MNRAS.383..831P, 2008MNRAS.385L..53C, 2008MNRAS.391..320B, 2008yCat.2289....0W, 2009AA...494..949S, 2009AA...497..619Z, 2009AA...504..199B, 2009AA...504..461F, 2009AJ....137....1F, 2009AJ....137.3345C, 2009AJ....138..703C, 2009ApJ...699..649S, 2009ApJ...702..805S, 2009ApJ...703..399L, 2009ApJ...706.1484B, 2009ApJS..181..321E, 2009ApJS..184...18G, 2009MNRAS.397..258L, 2009MNRAS.400..603P, 2010AA...510A..27B, 2010AA...515A..75A, 2010AA...517A..53M, 2010AA...519A..93B, 2010AA...520A..15M, 2010AA...521L..54H, 2010AA...524A..38M, 2010AJ....139.1808S, 2010AJ....139.2184Z, 2010AJ....140..266W, 2010ApJ...710.1142B, 2010ApJ...715..671W, 2010ApJ...722.1226H, 2010ApJS..190..100K, 2010MNRAS.404.1817Z, 2010MNRAS.406.1885B, 2011AA...527A..24L, 2011AA...531A..92R, 2011AA...531A.164W, 2011AA...536A..63B, 2011AJ....141...97W, 2011AJ....141..203A, 2011AJ....142...77D, 2011ApJ...726...18K, 2011ApJ...729..139W, 2011ApJ...731...17B, 2011ApJ...732...61Z, 2011ApJ...738..122C, 2011ApJ...739...48A, 2011ApJ...740L..32L, 2011ApJS..197...19K, 2011MNRAS.416.3108R, 2011MNRAS.418.1231D, 2011PhDT.......245L, 2012AA...538A..64C, 2012AA...539A.151A, 2012AA...540A..85P, 2012AA...544A.111M, 2012AA...548A..53L, 2012AJ....143...80S, 2012ApJ...744....6S, 2012ApJ...752...56F, 2012ApJ...752...59H, 2012ApJ...753...19T, 2012ApJ...753..156K, 2012ApJ...758...31L, 2012ApJ...758...56S, 2012ApJ...758L...2J, 2012ApJS..201...19D, 2012MNRAS.419.3346G, 2012MNRAS.422.1495L, 2012MNRAS.424.3178B, 2012MNRAS.426.3403L, 2012MNRAS.426.3419B, 2012MNRAS.427.3280F, 2012yCat.1322....0Z, 2013AA...549A.123A, 2013AA...554A..43D, 2013AA...557A..43B, 2013AA...557L...8B, 2013AA...560A..52M, 2013AJ....145....2F, 2013AJ....146...85H, 2013AJ....146..161M, 2013ApJ...762...88M, 2013ApJ...772...79A, 2013ApJ...772..129B, 2013ApJ...776..128K, 2013ApJ...777...84B, 2013ApJ...777L..20L, 2013ApJS..205....6M, 2013MNRAS.430.1171D, 2013MNRAS.431.3222L, 2013MNRAS.433..457B, 2013MNRAS.433.2054S, 2013MNRAS.434..142B, 2013MNRAS.434..806B, 2013MNRAS.435.2474L, 2013PASP..125..809T, 2013yCat.2319....0L, 2014AA...562A.127B, 2014AA...565A..20S, 2014AA...568A..77Z, 2014AJ....147...20N, 2014AJ....147...34S, 2014AJ....147...94D, 2014ApJ...780L...4B, 2014ApJ...782....8I, 2014ApJ...783..121G, 2014ApJ...783..122K, 2014ApJ...784..126E, 2014ApJ...785L..14G, 2014ApJ...787....5N, 2014ApJ...787..104C, 2014ApJ...787..126L, 2014ApJ...788...81M, 2014ApJ...792..119D, 2014ApJ...794...36H, 2014ApJ...794..143B, 2014MNRAS.439..372M, 2014MNRAS.439.3890G, 2014MNRAS.442.1586D_DUP, 2014MNRAS.445.3694D, 2014MNRAS.445.3908L, 2015AA...574A.118P, 2015AA...577A.148B, 2015AA...579A..66M, 2015AJ....150..137Q, 2015ApJ...798...73G, 2015ApJ...801....4W, 2015ApJ...802...61L, 2015ApJ...804...92S, 2015ApJ...805L..10H, 2015ApJ...806...62B, 2015ApJ...808L..20G, 2015ApJ...812....3W, 2015ApJ...813...83C, 2015ApJ...814..118B, 2015ApJS..219...33G, 2015MNRAS.448.2737R, 2015MNRAS.449.3651M, 2015MNRAS.450.3490D, 2015MNRAS.453.2378M, 2015MNRAS.453L.103S, 2015MNRAS.454.4054B, 2016AA...587A..56M, 2016AA...593A.119M, 2016AA...594A..39F, 2016AJ....152...24W, 2016ApJ...816...78C, 2016ApJ...821..120A, 2016ApJ...822L...1S, 2016ApJ...826...73P, 2016ApJ...827...22F, 2016ApJ...830..144R, 2016ApJ...833...96L, 2016ApJS..224...36K, 2016ApJS..225...10F, 2016MNRAS.457.1028S, 2016PhDT.......189A, 2016arXiv161205560C, 2017AA...597A..90D, 2017AA...598A..48G, 2017AA...598A..92L, 2017AA...599A..78P, 2017AA...605L...9C, 2017AJ....153...18B, 2017AJ....153...46L, 2017AJ....153...92T, 2017AJ....153..165T, 2017AJ....153..188F, 2017AJ....153..196S, 2017AJ....153..257O, 2017AJ....154...10R, 2017AJ....154...29K, 2017AJ....154...46E, 2017AJ....154...69S, 2017AJ....154..112K, 2017AJ....154..134E, 2017AJ....154..147D, 2017AJ....154..165B, 2017AJ....154..224R, 2017ApJ...837...95B, 2017ApJ...838...64N, 2017ApJ...838..150K, 2017ApJ...841L...1G, 2017ApJS..228...18G, 2017MNRAS.465.4723K, 2017RNAAS...1...42B, 2018AA...615A.160C, 2018AA...616A...1G, 2018AA...616A..10G, 2018AA...618A..93C, 2018AA...620A.130P, 2018AAS...23143601F, 2018AJ....155...34C, 2018AJ....156...75E, 2018AJ....156...76L, 2018AJ....156..271L, 2018ApJ...852...55D, 2018ApJ...854..101B, 2018ApJ...854L..27G, 2018ApJ...858...41Z, 2018ApJ...862..138G, 2018ApJ...867..109M, 2018ApJ...869L..33O, 2018ApJS..234....1B, 2018ApJS..236...28T, 2018MNRAS.473.2020L, 2018MNRAS.480.5447Z, 2018MNRAS.481.3548S, 2018RNAAS...2...33S, 2019AA...623A..35L, 2019AA...623A.112D, 2019AA...624L..11F, 2019AA...626A..80C, 2019AA...627A...4R, 2019AA...629A.114C, 2019AJ....157...85B, 2019AJ....157..231K, 2019AJ....157..234S, 2019AJ....157..247R, 2019AJ....158...54E, 2019AJ....158...75H, 2019AJ....158..122K, 2019AJ....158..182G, 2019ApJ...877...12T, 2019ApJ...883L..16C, 2019ApJ...887...87Z, 2019ApJS..240...19K, 2019MNRAS.485.2167G, 2019MNRAS.486.5405G, 2019MNRAS.487.2937O, 2020AA...644A.169M, 2020AJ....159..257B, 2020AJ....159..282E, 2020AJ....160...44L, 2020AJ....160..156S, 2020AJ....160..186L, 2020AJ....160..279K, 2020ApJ...889...74M, 2020ApJ...890..106S, 2020ApJ...892...31B, 2020ApJ...898...77S, 2020ApJ...898L..16B, 2020ApJ...899..123M, 2020ApJ...902L...6W, 2020ApJ...904..146F, 2020ApJS..251....6M, 2020PASP..132j4401A, 2021AA...647A..19T, 2021AA...647A.137J, 2021AJ....161..192N, 2021ApJ...911....7Z, 2021ApJ...917...23K, 2021ApJS..253....7K, 2021ApJS..253....8M, 2021ApJS..254...20L, 2022AA...660A..38W, 2022AA...664A.175P, 2022AJ....163...24L, 2022AJ....163...64E, 2022AJ....163..242S, 2022AJ....164..151L, 2022ApJ...924...68V, 2022ApJ...935...15Z, 2022ApJ...939...94M, 2022ApJ...941..143K, 2022MNRAS.517..161K, 2022arXiv220800070G, 2023AA...671A..46M, 2023AA...672A..94D, 2023AA...673A.114H, 2023AA...674A...1G, 2023AA...677A..59R, 2023AA...679A..82M, 2023AJ....165...37L, 2023AJ....165...85W, 2023AJ....165..269L, 2023AJ....166..103S, 2023ApJ...943L..16S, 2023ApJ...954..134K, 2023ApJ...959...63S, 2023ApJS..267....7Z, 2023Sci...380..198C, 2023arXiv230711882S, 2024AJ....167...19L, 2024AJ....167..253R, 2024AJ....168..159L, 2024AJ....168..165S, 2024ApJ...961..121H, 2024ApJ...967..115B, 2024ApJS..271...55K, 2024arXiv240605234T, 2025AJ....170...86S, Best20US, Schn23b}

\clearpage
\pagebreak
{
\ifpdf
\pdfpageattr{/Rotate 90}
\fi
\movetabledown=40mm
\begin{rotatetable*}
\begin{deluxetable*}{lcclccl}
\tablewidth{0.985\textwidth}
\tablecaption{Literature Planetary-Mass Objects which Membership is not Corroborated by BANYAN~$\Sigma$.\label{tab:rej_planemos}}
\tablehead{
\noalign{\vskip 1.0ex}
\colhead{Designation} & \colhead{\shortstack{Spectral\\ Type}} & \colhead{\shortstack{Banyan~$\Sigma$\\ $P$ (\%)}} & \colhead{\shortstack{Lit.\\ Memberships}} & \colhead{\shortstack{Distance\\ (pc)}} & \colhead{\shortstack{RV\\(\kms)}} & \colhead{Refs\tablenotemark{a}}
}
\startdata
$[$AKC2006$]$ 17 & L0 & 1.2 & LUP,CM & $305 \pm 99$ & $\cdots$ & 1;2;--;2;3 \\
2MASS J00464841+0715177 & L0$\delta$ & 0.0 & BPMG,CM & $37.8 \pm 0.4$ & $-2.8 \pm 0.3$ & 4;2;--;2;4 \\
2MASS J01531463-6744181 & L3$\beta$ & 2.2 & THA,CM & $(25 \pm 5)$ & $\cdots$ & 4;5;--;6;4 \\
2MASS J02411151-0326587 & L0$\gamma$ & 1.2 & THA,CM & $56 \pm 7$ & $6 \pm 8$ & 7;8;--;8;4 \\
2MASS J04215450+2652315 & M8.5 & 4.9 & TAU,CM & $\cdots$ & $\cdots$ & 9;10;--;--;11 \\
2MASS J04311907+2335047 & M7.75 & 4.6 & TAU,HM & $168 \pm 8$ & $32.3 \pm 0.6$ & 9;2;--;2;12 \\
2MASS J04373705+2331080 & L0 & 5.1 & TAU,CM & $(31 \pm 7)$ & $\cdots$ & 9;13;--;6;14 \\
2MASS J06085283-2753583 & L0 $\delta$ & 10.0 & COL,LM & $44.2 \pm 0.3$ & $26.7 \pm 0.6$ & 4;2;--;2;4 \\
2MASS J09553336-0208403 & L7pec(red) & 0.0 & TWA,CM & $(29 \pm 6)$ & $-20 \pm 4$ & 15;16;--;6;15 \\
2MASS J10212570-2830427 & L5$\beta$ & 0.0 & TWA,CM & $(10 \pm 2)$ & $\cdots$ & 15;17;--;6;4 \\
2MASS J11472421-2040204 & L7pec(red) & 6.0 & TWA,CM & $38 \pm 3$ & $7 \pm 3$ & 15;18;--;18;19 \\
2MASS J12271545-0636458 & M8.5$\beta$ & 0.1 & TWA,CM & $39.1 \pm 0.5$ & $\cdots$ & 4;2;--;2;4 \\
2MASS J12563961-2718455 & L4$\beta$ & 2.5 & TWA,CM & $(32 \pm 8)$ & $-19 \pm 4$ & 15;20;--;6;4 \\
2MASS J16072342-3905099 & L1 & 0.0 & LUPIII,CM & $3,161 \pm 1,628$ & $\cdots$ & 21;2;--;2;21 \\
2MASS J16262218-2423523 & M8.5 & 0.0 & ROPH,CM & $\cdots$ & $\cdots$ & --;16;--;--;22 \\
2MASS J16263252-2426354 & M8 & 0.0 & ROPH,CM & $\cdots$ & $\cdots$ & --;16;--;--;22 \\
2MASS J16263991-2422334 & M8.3 & 0.0 & ROPH,CM & $\cdots$ & $\cdots$ & 23;16;--;--;22 \\
2MASS J16270592-2418402 & L3 & 5.1 & ROPH,CM & $(22 \pm 4)$ & $\cdots$ & 24;16;--;6;22 \\
2MASS J19350976-6200473 & L1$\gamma$ & 2.7 & THA,CM & $(31 \pm 7)$ & $\cdots$ & 4;25;--;6;4 \\
2MASS J20113196-5048112 & L3$\gamma$ & 9.2 & THA,CM & $2,042 \pm 1,315$ & $\cdots$ & 4;17;--;2;4 \\
2MASS J22134491-2136079 & L0$\gamma$ & 0.0 & BPMG,LM & $48 \pm 2$ & $-5 \pm 4$ & 7;2;--;2;4 \\
2MASS J22163082+1953167 & T3 & 0.0 & BPMG,CM & $1,495 \pm 569$ & $\cdots$ & 26;2;--;2;26 \\
2MASS J22351658-3844154 & L1.5$\gamma$ & 0.0 & THA,CM & $41 \pm 2$ & $\cdots$ & 4;2;--;2;4 \\
\enddata
\tablenotetext{a}{References for spectral types, proper motions, radial velocities, distances, and memberships, respectively.}
\tablecomments{The right ascension, declination, and proper motion columns are available in the online version of this journal. Distances between parentheses are photometric rather than trigonometric.}
\tablerefs{(1)~\cite{2006ApJ...647L.167J}, (2)~\cite{2023AA...674A...1G}, (3)~\cite{2008ApJS..177..551M}, (4)~\cite{2015ApJS..219...33G}, (5)~\cite{2015ApJ...798...73G}, (6)~This Paper, (7)~\cite{2009AJ....137.3345C}, (8)~\cite{2016ApJ...833...96L}, (9)~\cite{2023AJ....165...37L}, (10)~\cite{2018ApJ...858...41Z}, (11)~\cite{2009ApJ...703..399L}, (12)~\cite{2006AJ....132.2665S}, (13)~\cite{2025AJ....170...86S}, (14)~\cite{2017ApJ...837...95B}, (15)~\cite{2017ApJS..228...18G}, (16)~\cite{2007MNRAS.379.1599L}, (17)~\cite{2021AJ....161..192N}, (18)~\cite{2020AJ....159..257B}, (19)~\cite{2016ApJ...822L...1S}, (20)~\cite{2018ApJS..234....1B}, (21)~\cite{2005AA...440..139L}, (22)~\cite{2008hsf2.book..351W}, (23)~\cite{2011ApJ...726...23G}, (24)~\cite{2012AA...539A.151A}, (25)~\cite{2021ApJS..253....8M}, (26)~\cite{2015ApJ...814..118B}.}
\end{deluxetable*}
\end{rotatetable*}
}

\nocite{2005AA...440..139L, 2006AJ....131.2722C, 2006AJ....132.2665S, 2006ApJ...647L.167J, 2007ApJ...657..511A, 2007ApJ...662..413K, 2007ApJS..173..104L, 2007MNRAS.374..372L, 2007MNRAS.379.1599L, 2008ApJ...676..427A, 2008ApJ...680.1295S, 2008ApJS..177..551M, 2008hsf2.book..351W, 2009AJ....137.3345C, 2009ApJ...703..399L, 2010AA...515A..75A, 2011AA...527A..24L, 2011ApJ...726...23G, 2012AA...539A.151A, 2012AA...548A..26D, 2013MNRAS.435.2474L, 2015ApJ...798...73G, 2015ApJ...814..118B, 2015ApJS..219...33G, 2016ApJ...822L...1S, 2016ApJ...833...96L, 2017AA...602A..82D, 2017AJ....154...46E, 2017ApJ...837...95B, 2017ApJS..228...18G, 2018AJ....156...76L, 2018ApJ...858...41Z, 2018ApJ...869...72Y, 2018ApJS..234....1B, 2019AJ....158...54E, 2020AA...633A.152C, 2020AJ....159..257B, 2020AJ....160...44L, 2021AJ....161..192N, 2021ApJS..253....8M, 2022AJ....163..242S, 2023AA...674A...1G, 2023AJ....165...37L, 2023ApJS..267....7Z, 2024AJ....168..159L, 2025AJ....170...86S, Best20US}

\end{document}